\documentclass[12pt,twocolumn]{aastex631}

\usepackage{natbib}
\usepackage{color}
\usepackage[english]{babel}
\usepackage[normalem]{ulem}
\usepackage{blindtext}
\usepackage{textgreek}
\usepackage{threeparttablex}
\usepackage{tabularx}
\usepackage{tabulary}
\usepackage{longtable}
\usepackage{rotating}
\usepackage{graphicx}
\usepackage{appendix}

\extrafloats{100}

\newcommand{\lsim}{\raise0.3ex\hbox{$<$}\kern-0.75em{\lower0.65ex\hbox{$\sim$}}}
\newcommand{\gsim}{\raise0.3ex\hbox{$>$}\kern-0.75em{\lower0.65ex\hbox{$\sim$}}}

\usepackage{listings}
\usepackage{xcolor}
\definecolor{codegreen}{rgb}{0,0.6,0}
\definecolor{codegray}{rgb}{0.5,0.5,0.5}
\definecolor{codepurple}{rgb}{0.58,0,0.82}
\definecolor{backcolour}{rgb}{0.95,0.95,0.92}
\lstdefinestyle{mystyle}{
    backgroundcolor=\color{backcolour},   
    commentstyle=\color{codegreen},
    keywordstyle=\color{magenta},
    numberstyle=\tiny\color{codegray},
    stringstyle=\color{codepurple},
    basicstyle=\ttfamily\footnotesize,
    breakatwhitespace=false,         
    breaklines=true,                 
    captionpos=b,                    
    keepspaces=true,                 
    numbers=left,                    
    numbersep=5pt,                  
    showspaces=false,                
    showstringspaces=false,
    showtabs=false,                  
    tabsize=2
}
\begin{document}

\title{The Delay Time Distribution of Tidal Disruption Events}

\author[0009-0005-1158-1896]{Margaret Shepherd}
\affil{Department of Astronomy, University of Illinois at Urbana-Champaign, 1002 W. Green Street, Urbana, IL 61801, USA}
\email{ms169@illinois.edu}

\author[0000-0002-4235-7337]{K. Decker French}
\affil{Department of Astronomy, University of Illinois at Urbana-Champaign, 1002 W. Green Street, Urbana, IL 61801, USA}

\author[0000-0002-4337-9458]{Nicholas C. Stone}
\affil{Racah Institute of Physics, The Hebrew University, 91904 Jerusalem, Israel}
\affil{Department of Astronomy, University of Wisconsin, Madison, WI 53706, USA}

\author[0000-0003-1714-7415]{Nicholas Earl}
\affil{Department of Astronomy, University of Illinois at Urbana-Champaign, 1002 W. Green Street, Urbana, IL 61801, USA}

\author[0000-0002-7854-1953]{Denyz Melchor}
\affil{Department of Physics and Astronomy, University of California, Los Angeles, CA 90095, USA}
\affil{Mani L. Bhaumik Institute for Theoretical Physics, Department of Physics and Astronomy, UCLA, Los Angeles, CA 90095, USA}

\author{Teddy R. Smith}
\affil{Department of Physics and Astronomy, Carleton College, 300 N. College Street Northfield, MN 55057, USA}

\author[0000-0001-8426-5732]{Jean Somalwar}
\affil{Department of Astronomy, University of California, Berkeley, CA 94720, USA}
\affil{Kavli Institute for Particle Astrophysics \& Cosmology, P.O. Box 2450, Stanford University, Stanford, CA 94305, USA}

\author[0009-0006-1177-7466]{Odelia Teboul}
\affil{Department of Particle Physics \& Astrophysics, Weizmann Institute of Science, Rehovot 76100, Israel}

\author[0000-0003-1535-4277]{Margaret E. Verrico}
\affil{Department of Astronomy, University of Illinois at Urbana-Champaign, 1002 W. Green Street, Urbana, IL 61801, USA}

\begin{abstract}

Tidal disruption events (TDEs) can be observed when stars get too close to supermassive black holes and are torn apart and accreted. The delay time distribution of TDEs, or rate of TDEs as a function of time since a burst of star formation, can be used to determine what mechanisms influence the TDE rate. We compile a catalog of 41 TDE host galaxies with optical spectra, model the stellar populations with \textsc{Bagpipes}, and retrieve the age of the most recent burst of star formation to construct the delay time distribution of TDEs. TDEs occur more frequently in post-starburst galaxies than in other types of galaxies, though the mechanism causing this rate enhancement is unknown. We find that the TDE rate increases with post-burst age to reach a peak at $\sim$1 Gyr relative to a control sample. We compare the observational TDE delay time distribution to theoretical models, which propose overdense stellar nuclei, radial anisotropies in stellar orbits, supermassive black hole binaries, and AGN disks as potential mechanisms that may enhance the TDE rate in post-starburst galaxies. Most models predict a TDE rate that declines with post-burst age, in contrast to our observational results, though some models are still feasible at certain ages (e.g., the black hole binary model matches at old burst ages and the stellar overdensity model matches at intermediate burst ages).  

\end{abstract}	

\section{Introduction}
\label{sec:intro}

A tidal disruption event (TDE) is a transient that occurs when a star is torn apart by a supermassive black hole (SMBH). If the star's orbit causes it to approach the black hole too closely, the gravitational tidal forces from the black hole can overcome the star's self-gravity and tidally disrupt the star \citep{Rees_1988, Phinney_1989, Evans_1989}. During a TDE, about half of the star's material is flung out of the system and about half returns and is gravitationally bound \citep{Gezari_2021}. The resulting flare can be seen in the radio \citep{vanVelzen_2016, Alexander_2016, Somalwar_2025}, IR \citep{vanVelzen_2016b, Masterson_2024}, optical/UV \citep{Gezari_2006, vanVelzen_2011, Yao_2023}, and X-ray \citep{Komossa_1999, Sazonov_2021} regions of the electromagnetic spectrum. However, the majority of TDEs discovered thus far have been selected based on their UV/optical emission.

One pressing discrepancy in TDE rate studies is that the rate of TDEs in post-starburst (PSB) and quiescent Balmer-strong (QBS) galaxies has been shown to be enhanced compared to the TDE rate in other types of galaxies \citep{Arcavi_2014, French_2016, LawSmith_2017, Graur_2018, French_2020}. PSB galaxies have undergone a burst of star formation within the past gigayear and are now approaching quiescence. PSB galaxies make up roughly 0.2\% of the local universe \citep{French_2016} but make up 17\% of the TDE host galaxies presented in this study. The reason for this rate enhancement in PSB galaxies remains unknown. 

The phase space of stellar positions and velocities that result in a TDE is called the loss cone \citep{Frank_1976, Lightman_1977, Cohn_1978, Wang_2004, Stone_2020, Teboul_2024a}. A star is said to be “in the loss cone” if the star’s trajectory will result in the disruption of the star. Whatever mechanism drives the TDE rate enhancement in PSB galaxies should (re)fill the loss cone more efficiently than “normal” nuclear dynamics. There are a number of proposed mechanisms that would increase the rate of loss cone replenishment for galaxies that have experienced a recent starburst and/or galaxy merger. 

One hypothesis is that the central stellar environment of PSB galaxies is overdense, leading to more stars entering the loss cone and thus more disruptions \citep{Stone_2016}. \cite{Stone_2018}, \cite{Bortolas_2022}, and \cite{Teboul_2025} explore this possibility. \cite{Stone_2018} find that the TDE rate is enhanced by a factor of $\sim$10--1000 in this scenario. \cite{Bortolas_2022} proposes that a galactic nucleus is best described by a top-heavy initial mass function (IMF). \cite{Bortolas_2022} implements a complete IMF into their TDE rate calculations and re-calculates the rates from \cite{Stone_2018}, which results in a TDE rate that is enhanced by a factor of $\sim$10--100 before declining. In contrast, \cite{Teboul_2025} find that rate enhancements are reduced by a factor of $\sim$10 due to ejections of stars caused by strong scattering interactions, and turn into net rate reductions at later times. 

If stars created in a starburst are biased towards having radial orbits as opposed to tangential orbits, then TDE rates can be enhanced by a factor $\sim$2--200 compared to galaxies with no radial velocity bias (with an isotropic velocity distribution) \citep{Stone_2018}. When accounting for strong scatterings, \cite{Teboul_2025} find that the rate enhancement in this scenario is at most $\sim$50 at early times and turns into a rate reduction at late times.  

SMBH binaries can produce dynamical effects that temporarily increase the TDE rate around one of the black holes. \cite{Ivanov_2005}, \cite{Chen_2009}, and \cite{Li_2019} find that the TDE rate can be briefly enhanced by multiple orders of magnitude above the single-black hole TDE rate, depending on the size of the central stellar cusp and the black hole binary mass ratio. This period of enhancement typically lasts $\sim10^{5}$ years \citep{Chen_2009, Chen_2011}, and is driven by a combination of the Kozai-Lidov mechanism \citep{Kozai_1962, Lidov_1962} and strong scatterings. \cite{Mockler_2023} and \cite{Melchor_2024} propose that a TDE rate enhancement can result from the combination of the two-body interactions between stars and the eccentric Kozai-Lidov (EKL) mechanism. The EKL mechanism increases the eccentricities of stellar orbits so that TDEs become more likely. \cite{Melchor_2024} show that the enhanced TDE rates achieved through this mechanism can match the observed TDE rate in PSB galaxies.

The TDE rate may also be affected by the presence and disappearance of an AGN disk. \cite{Wang_2024} and \cite{Kaur_2025} propose that an evolving AGN disk may increase the TDE rate in PSB galaxies. Under the assumption that AGNs tend to reside in galaxies having a recent starburst \citep{Hopkins_2008}, they show that star-disk interactions as well as in-situ star formation in the outer AGN disk result in an enhancement of stars on circular orbits near the SMBH. As the AGN transitions to quiescence, these stars are rapidly scattered into the loss cone. \cite{Wang_2024} find that AGN disk evolution results in a TDE rate enhanced by a factor of 5--320 over a period of 1 Gyr compared to non-active galaxies. 

There are other hypotheses for the PSB overrepresentation not tested in this paper.  \cite{Madigan_2018} propose that eccentric nuclear stellar disks briefly increase the TDE rate in PSB galaxies to rates as high as 0.1--1 TDEs per year per galaxy. \cite{Merritt_2004} investigate the effect of triaxial gravitational potentials around central black holes and finds that TDE rates may reach a few $\times 10^{-3}$ TDEs per year per galaxy. Additionally, \cite{Hamers_2017} find that perturbations of stellar orbits due to nuclear spiral arms can also enhance the TDE rate, but not high enough to match the observed TDE rate in PSB galaxies.

We consider here theoretical models that make predictions for the TDE rate as a function of time, which is known as a delay time distribution (DTD), and construct an observational DTD to compare to these models. We first compile a sample of 41 TDE host galaxies with optical spectra from the literature and archival data, described in Section~\ref{sec:data}. We then use a stellar population synthesis code called Bayesian Analysis of Galaxies for Physical Inference and Parameter EStimation (\textsc{Bagpipes}, \citealt{Carnall_2018, Carnall_2019}) to fit the spectra and recreate the host galaxies’ star formation history (SFH). Section~\ref{sec:methods} describes the methods of using \textsc{Bagpipes} and determining the host galaxies' galaxy type. Section~\ref{sec:results} describes the results obtained from \textsc{Bagpipes} and presents PSB overrepresentation values. We use the information from \textsc{Bagpipes} to compute the DTD for TDEs in PSB galaxies and compare to theoretical models in Section~\ref{sec:analysis}. Section~\ref{sec:discussion} compares the methods in this paper to previous attempts to model the SFH of PSB galaxies, discusses caveats and limitations of the sample, investigates potential biases in the results, and considers the potential effects of other parameters (dust, mass, redshift). Finally, Section~\ref{sec:conclusion} provides a summary of the findings. Appendix~\ref{appendix:a} presents more information on alternative methods to modeling SFHs, Appendix~\ref{appendix:b} presents the \textsc{Bagpipes} fitting results, Appendix~\ref{appendix:c} presents a DTD for the host galaxies of TDEs that displayed broad lines, and Appendix~\ref{appendix:d} presents information about the host galaxies and their spectra.

\section{Data}
\label{sec:data}

To compile a large sample of TDE host galaxy spectra, we searched for TDEs in the Transient Name Server (TNS), the Weizmann Interactive Supernova Data Repository\footnote{https://wiserep.org} (WISeREP, \citealt{wiserep}), compilations of TDE hosts in the literature \citep{French_2020, Graur_2018, Hammerstein_2021}, and the MOST Hosts Survey \citep{Soumagnac_2024}. We selected TDEs from any TDE subclass, and removed cases that were later re-classified as non-TDEs\footnote{AT 2022ablq was reclassified as a SN Ibn \citep{Charalampopoulos_2023}.}. Next, using the coordinates from the previous search, we collected optical host galaxy spectra for as many TDE host galaxies as possible from the Sloan Digital Sky Survey (SDSS, Section~\ref{sec:sdss_spectra}), WISeREP (Section~\ref{sec:wiserep_spectra}), \citet{Graur_2018} (Section~\ref{sec:graur_spectra}), \citet{Hammerstein_2021} (Section~\ref{sec:kcwi_spectra}), \citet{Tadhunter_2021} (Section~\ref{sec:F01004_spectra}), the MOST Hosts Survey (\citealt{Soumagnac_2024}, Section~\ref{sec:2020nov_spectra}), and private communication from other individuals (Section~\ref{sec:muse_spectra}). 

These host galaxy spectra were either taken before the discovery of the TDE or at least 365 days after the discovery of the TDE to avoid contamination from the TDE itself. In Section~\ref{sec:tdeplateauemission} we consider the possible contribution of long-lasting contamination from the TDE disk. We also impose a signal-to-noise ratio (SNR) per pixel minimum of 15 to ensure all spectra were of sufficient quality for stellar population fitting. The SNRs of SDSS galaxy spectra were pre-calculated. We calculated the SNRs of galaxy spectra from other sources in the wavelength band of $5200$ {\AA} $< \lambda < 5900$ {\AA} (rest wavelengths). In total, 41 TDEs had host galaxy spectra that passed our selection cuts. We also gathered the galaxies' redshifts, redshift uncertainties, and stellar masses. Spectra for comparison samples were drawn from SDSS (Section~\ref{sec:control_samples}).

\subsection{TDE Host Galaxy Spectra and Ancillary Data}
\label{sec:spectra}

The following subsections describe information about the TDE host galaxy spectra and associated information, summarized in Table~\ref{table:finalsample}. Column 6 in this table describes the sources of the stellar masses that were used in \textsc{Bagpipes} fitting (from SDSS, from WISeREP, calculated from the 2MASS K-band magnitude\footnote{To calculate the stellar mass of a galaxy using its 2MASS K-band apparent magnitude, we begin by calculating the distance with the redshift and the \texttt{astropy} \texttt{cosmo.distmod} method. Then, we use the distance and the apparent magnitude to calculate the absolute magnitude, which we convert to luminosity. Finally, using the mass-to-light ratio in \cite{Just_2015}, we calculate the stellar mass of the host galaxy.}, and from individual papers: \citealt{Yao_2023, French_2020, Wevers_2019, Graur_2018, Hammerstein_2021, Hammerstein_2023}). The stellar mass is later used to estimate the metallicity, which we provide to \textsc{Bagpipes}. Column 8 in this table provides information on the slit width of the instrument taking the spectrum.

Redshifts were available from various databases for all host galaxies: from SDSS for galaxies with SDSS spectra; from \cite{Yao_2023}, \cite{Hammerstein_2023}, and WISeREP for galaxies with WISeREP spectra; and from \cite{Graur_2018} for galaxies with \cite{Graur_2018} spectra. When a redshift uncertainty was not available (WISeREP spectra, spectra from \citealt{Graur_2018}, spectra from \citealt{Hammerstein_2021}, some MUSE spectra, and the MOST spectrum), we instead estimate the redshift uncertainty based on the significant figures reported in the redshift. For example, for a redshift of 0.13, we set a redshift uncertainty of 0.005; for a redshift of 0.083, we set a redshift uncertainty of 0.0005.

Given the inhomogeneity of the data, flux uncertainties were not always provided with the spectra. When available, we provided \textsc{Bagpipes} with the reported flux uncertainties for the spectra. If unavailable, we assume a flux uncertainty per wavelength step of 10\%.

The final sample of 41 host galaxies can be found in Table~\ref{table:finalsample} in Appendix~\ref{appendix:d}. When necessary, we masked telluric features and trimmed noisy red/blue ends of the spectra to ensure \textsc{Bagpipes} worked well and did not mistake noisy regions for signal. In the analysis going forward, the TDE host galaxy sample is treated as if it has 42 galaxies, because we include F01004 twice (see Section~\ref{sec:F01004_spectra}). We have further considered how the results would change if F01004 were only included once (see Section~\ref{sec:F01004_bias}).

\subsubsection{SDSS Spectra}
\label{sec:sdss_spectra}

19 TDEs had host galaxy spectra in SDSS that had SNR $>$ 15. ASASSN-14li, ASASSN-14ae, and AT 2018hyz had spectra in both the SDSS and MUSE samples. In the case of ASASSN-14ae, the MUSE and SDSS SNRs were quite similar (the MUSE spectrum had a SNR of 18.3 while the SDSS spectrum had a SNR of 19.7), but the standard deviation of the age of the burst calculated from the MUSE spectrum was smaller ($\sigma_t$ for the MUSE spectrum was 0.025 Gyr while $\sigma_t$ for the SDSS spectrum was 0.058 Gyr), so the MUSE spectrum was chosen and ASASSN-14ae was removed from the SDSS sample. Thus, a total of 18 SDSS galaxies are in the final sample. Flux uncertainties were available for all galaxies in this sample.

\subsubsection{WISeREP Spectra}
\label{sec:wiserep_spectra}

The objects with spectra selected from WISeREP often had multiple spectra associated with them. In order to select which spectra to use, we filtered out any spectra that had SNR $<$ 15. Then, we removed any spectra that had been taken in the 365 days after the TDE discovery date. After these cuts, seven unique galaxies that had SNR $>$ 15 remained. For galaxies with multiple spectra, the spectrum that was taken furthest from the date of TDE discovery was used. 

Redshift values were drawn from WISeREP, except for two galaxies where redshift from \citet{Yao_2023} was used (as we prefer to use values from a published scientific paper when available). AT 2018zr was removed from the WISeREP sample because the spectrum had an unusual shape that \textsc{Bagpipes} was unable to fit after repeated attempts, possibly due to spectral calibration. AT 2021lo was removed from the WISeREP sample because there is still evidence of TDE emission in the spectrum, notably a broad H$\alpha$ line and HeII line. AT 2020acka's host galaxy spectrum was removed from the WISeREP sample because \citet{Soraisam_2022} showed that there is re-brightening evident in the TDE, potentially contaminating the host galaxy spectrum. After these removals, a total of four WISeREP galaxies remained in the final sample. Flux uncertainties were available for one of these galaxies.

\subsubsection{\cite{Graur_2018} Spectra}
\label{sec:graur_spectra}
 
To select good-quality spectra, we exclude host spectra for which the resolution is poor ($>$ 500 km/s), which led to the exclusion of the galaxy containing the TDE 2MASXJ0249. We also removed iPTF16fnl from the \cite{Graur_2018} sample because the spectrum was still visibly contaminated by the broad TDE emission. A total of eight galaxies from \cite{Graur_2018} remain in the final sample. Flux uncertainties were available for one of these galaxies.

\subsubsection{MUSE Spectra}
\label{sec:muse_spectra}

The galaxy spectra in this subsample came from the MUSE spectrograph (Thomas Wevers, priv. comm.). AT 2019qiz had spectra in both the MUSE and \cite{Hammerstein_2021} samples, but the \cite{Hammerstein_2021} spectra was chosen because the SNR was higher than that of the MUSE spectrum, and the standard deviation of the age of the burst was lower. ASASSN-14li and AT 2018hyz had spectra in both the SDSS and MUSE samples; the SDSS spectra of both were chosen because the SDSS spectrum's SNR is higher than the MUSE spectrum's SNR, and the SDSS spectrum contains Balmer break information, which is indicative of recent star formation and useful for \textsc{Bagpipes}. After these removals, a total of six galaxies with MUSE spectra remained in the final sample. Flux uncertainties were available for all galaxies in this sample. \cite{Pursiainen_2025} use the MUSE sample of TDE host galaxies to do stellar population fitting, which we compare to in Section~\ref{sec:comparison_french2018}.

\subsubsection{\cite{Hammerstein_2021} Spectra}
\label{sec:kcwi_spectra}
 
After implementing the minimum SNR cut, a total of three galaxies from \cite{Hammerstein_2021} remain in the final sample. Flux uncertainties were available for none of the galaxies in this sample.

\subsubsection{\cite{Tadhunter_2021} Spectrum}
\label{sec:F01004_spectra}

An optical spectrum for the host galaxy of the TDE F01004 was provided by \citet{Tadhunter_2021}, which included flux uncertainties. We did not use the F01004 spectra from \citet{Graur_2018} because the pre-flare spectrum contains broad lines that \textsc{Bagpipes} was unable to fit. 

Evidence from \citet{Sun_2024} shows that there may have been two independent TDEs in the host galaxy for F01004, though observations can also be plausibly explained by a double TDE or a repeating partial TDE. More observations are needed to test these hypotheses. We have included this galaxy twice in our sample to account for two possible TDEs. 

\subsubsection{MOST Hosts Spectrum}
\label{sec:2020nov_spectra}

After searching the MOST Hosts database \citep{Soumagnac_2024} for high-quality spectra of TDE host galaxies, only one candidate was found: the host galaxy spectra for the TDE AT 2020nov. The redshift of this host galaxy came from \cite{Earl_2025}. Flux uncertainties were available for this galaxy spectra.

\subsection{Control Samples}
\label{sec:control_samples}

\begin{figure}[h]
\begin{center}
    \includegraphics[width=0.95\linewidth]{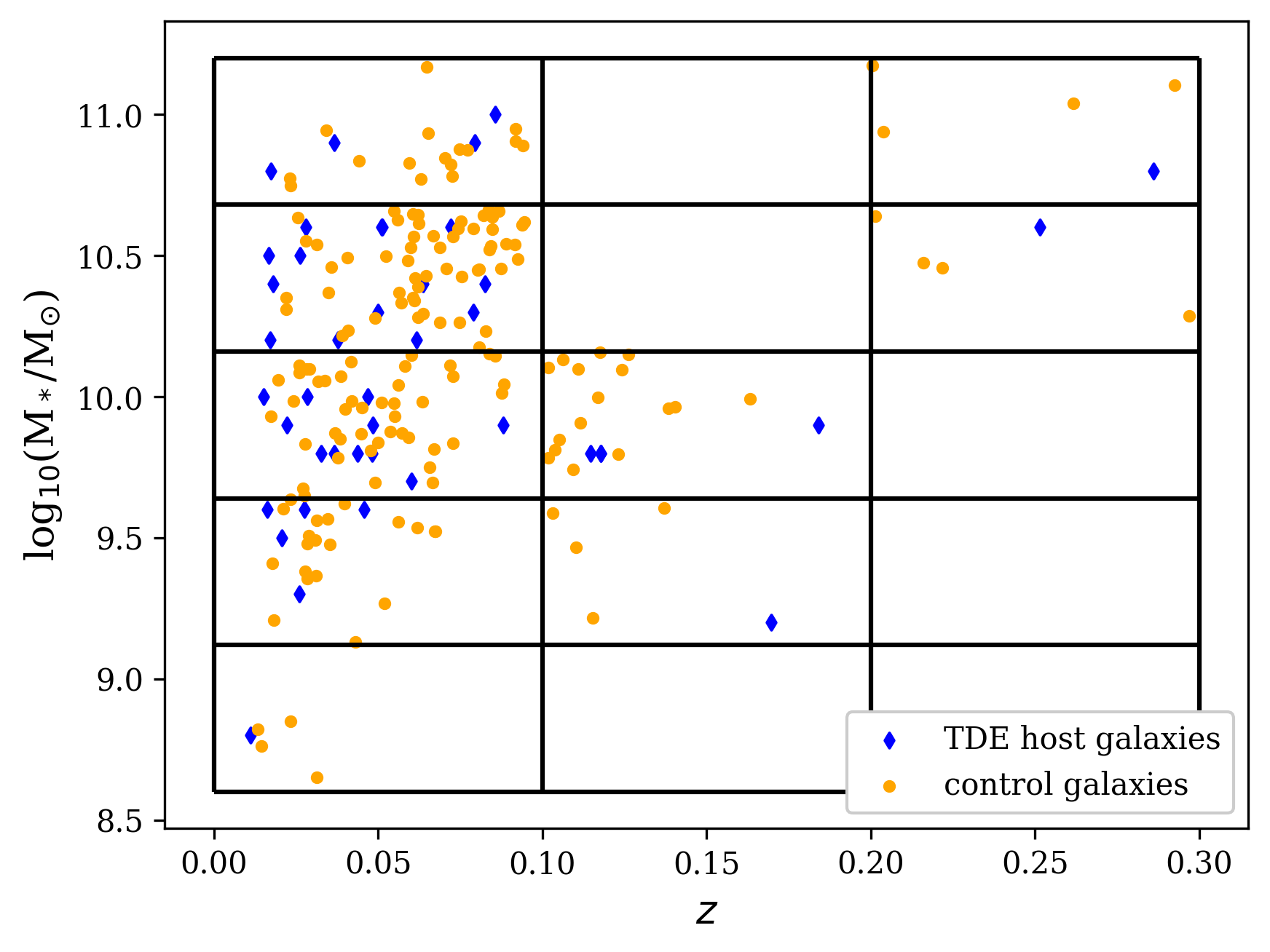}
\end{center}
\caption{Stellar mass versus redshift of the TDE host galaxy sample and the control sample. The control sample was selected to match the redshift and stellar mass distribution of the TDE host galaxy sample based on how many TDE hosts fell within “cells” of this parameter space.}
\label{fig:stelmass_vs_z_TDE_control}
\end{figure}

We constructed a “control sample” composed of galaxies in SDSS that are matched in stellar mass space and redshift space to the final TDE host galaxy sample, but which are not TDE hosts themselves. We require the control sample galaxies to have SNR $>15$, consistent with the TDE host galaxies. We placed a 5x3 grid on the TDE host galaxies plotted in stellar mass space and redshift space, then for each TDE host in each grid cell, we randomly selected four SDSS galaxies that also fell into that grid cell (see Figure~\ref{fig:stelmass_vs_z_TDE_control}). Thus, the control sample's distribution in stellar mass space and redshift space roughly matches the TDE sample's respective distribution. We chose to have more cells in stellar mass space than redshift space because stellar mass has a strong correlation with stellar age, and thus is likely more important in affecting the TDE rate. TDEs become less common as the central black hole mass approaches $10^8 M_{\odot}$ and will eventually reach a cut-off due to event horizon suppression \citep{Hills_1975, Kesden_2012, Yao_2023}. For the TDE host galaxies, we find there is no systematic trend in the burst ages across stellar mass bins. We do find that burst age of the TDE host galaxies increases with redshift, but because the redshift distribution of our TDE host galaxy sample does not match that of SDSS, this motivates the necessity of having cells in the first place. Because we expect to find systematic differences in stellar types in galaxies across a range of stellar masses, it is important that the control sample matches the mass (and thus related properties) of the TDE host galaxies. Our matched selection provides a control sample for which the stellar mass distribution is statistically indistinguishable from that of the TDE host galaxy sample (Kolmogorov–Smirnov (KS) test p-value = 0.375). SDSS is not a volume-complete survey, so the binning method is necessary to ensure that the control sample is matched in important characteristics. We test a volume-limited subset of the TDE host galaxies and the control sample in Section~\ref{sec:nonuniformsample} to make sure that any outliers in redshift or stellar mass space are not biasing our results.

We used this control sample to establish the distribution of ages since a burst of star formation occurred in non-TDE host galaxies and determine how the same distribution is different in TDE host galaxies. There are 168 (42 $\times$ 4) galaxies in the control sample. This comparison sample balances the need for an accurate characterization of normal galaxies with the computational expense of fitting a large number of galaxies with \textsc{Bagpipes}. 

When calculating the PSB overrepresentation (Section~\ref{sec:psb_overrepresentation}), we are not limited by the computational expense of running \textsc{Bagpipes}, so we select a larger comparison sample. This “expanded control sample” was created with the same method as the control sample explained above, but 53 SDSS galaxies were randomly selected per TDE host galaxy in each cell, for a total of 2,226 (42 $\times$ 53) galaxies in the expanded control sample. 53 control galaxies per TDE host galaxy is the maximum number of control galaxies that could be assigned to a TDE host galaxy in the least populated bin, so 2,226 galaxies is the maximum possible size for this control sample. 

\begin{figure}[h]
\begin{center}
    \includegraphics[width=0.95\linewidth]{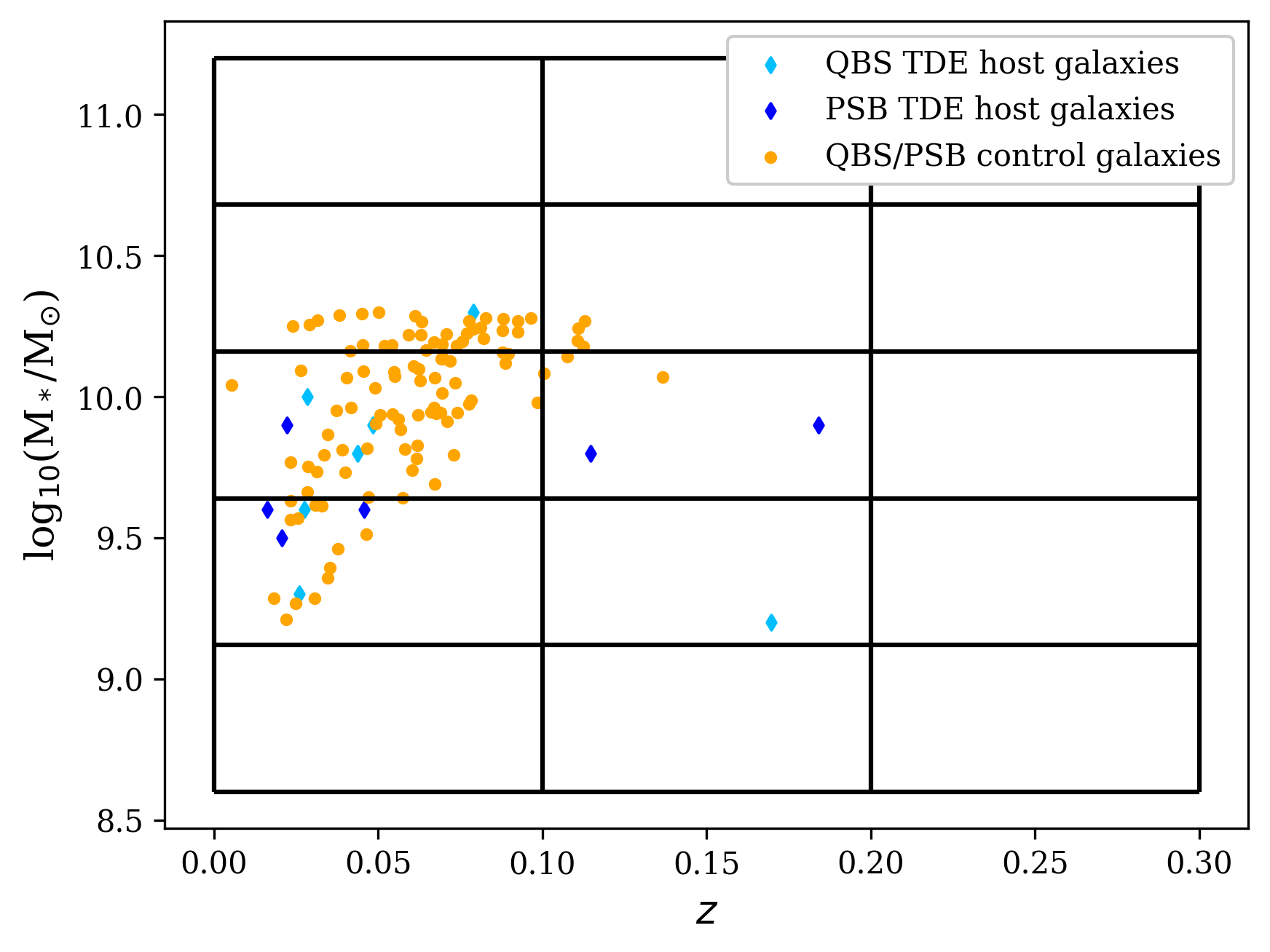}
    \includegraphics[width=0.95\linewidth]{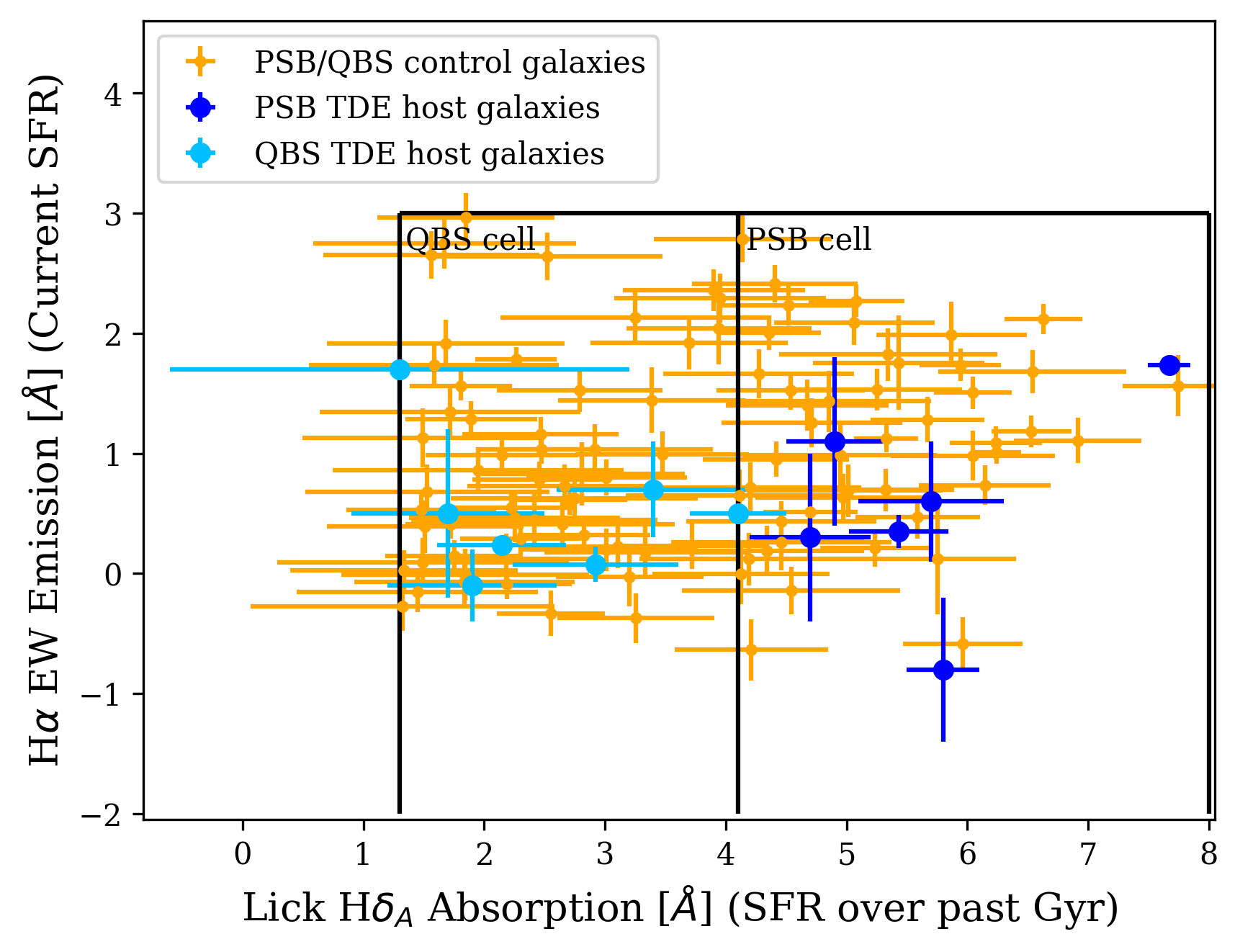}
\end{center}
\caption{{\it Top:} Stellar mass versus redshift of the PSB/QBS subset of the TDE host galaxy sample and the PSB/QBS control sample. The PSB/QBS control sample was selected to roughly match the redshift and stellar mass distribution of the PSB/QBS TDE hosts. The cell method used in Figure~\ref{fig:stelmass_vs_z_TDE_control} could not be used exactly because there were no SDSS PSB/QBS galaxies matching our criteria where $0.1 < z < 0.2$ and $9.12<\log_{10}(M_{*}/M_{\odot})<9.64$. Instead, PSB/QBS non-TDE hosts within a redshift range of $0<z<0.2$ and a stellar mass range of $9.12<\log_{10}(M_{*}/M_{\odot})<10.3$ were chosen to be in the PSB/QBS control sample. {\it Bottom:} H$\alpha$ equivalent width emission and Lick H$\delta$ absorption of the PSB/QBS subset of the TDE host galaxy sample and the PSB/QBS control sample. The PSB/QBS control sample was selected to roughly match the H$\alpha$ and H$\delta$ distribution of the PSB/QBS TDE hosts.}
\label{fig:PSBQBScontrol}
\end{figure}

To compare the PSB/QBS galaxies that host TDEs to the larger population of PSB/QBS galaxies, we create a “PSB/QBS control sample” matched to the properties of the PSB/QBS TDE hosts (see Section~\ref{sec:galaxy_classification_methods} for how we assigned galaxy type labels). Galaxies for this control sample were drawn from SDSS and had SNR $>$ 15. They were required to be PSB or QBS based on the definition presented in Section~\ref{sec:galaxy_classification_methods} with values of the H$\alpha$ equivalent widths and H$\delta$ indices from SDSS {\tt galspec} \citep{Kauffmann_2003, Brinchmann_2004, Tremonti_2004} and roughly matched the redshift and stellar mass distribution of the PSB/QBS TDE hosts in the sample ($0 < z < 0.2$, $9.12 < \log(M_{\star}/M_{\odot}) < 10.3 $). For each PSB and QBS TDE host galaxy in the sample, 8 non-TDE hosts of the same classification were drawn from SDSS. There are 104 (13 $\times$ 8) galaxies in the PSB/QBS control sample. The distribution of the PSB/QBS control sample in redshift and stellar mass space as well as H$\alpha$ equivalent width and H$\delta$ index space is shown in Figure~\ref{fig:PSBQBScontrol}. 

\section{Methods}
\label{sec:methods}

\subsection{\textsc{Bagpipes} Methods}
\label{sec:bagpipesmethods}

\textsc{Bagpipes} \citep{Carnall_2018, Carnall_2019} is a stellar population synthesis code that uses the \textsc{MultiNest} nested sampling algorithm \citep{Feroz_2019} to estimate best-fit values for various galaxy parameters from user-provided spectra. We use \textsc{Bagpipes} to model a galaxy's SFH given the spectra and related priors. \textsc{Bagpipes} utilizes stellar population models from \cite{Bruzual_2003}.

\subsubsection{SFH Model}
\label{sec:sfhmodel} 

Modeling galaxy spectra using stellar population synthesis requires that we assume either a parametric \citep[e.g.,][]{Wild_2020} or non-parametric \citep[e.g.,][]{Suess_2022} SFH. This choice will vary depending on the goal of spectral modeling (for example, whether deducing the SFH is the final goal or if it is marginalized over). In order to compare the observational DTD of the TDE hosts to theoretical models, we must determine the time elapsed since the most recent strong burst of star formation. Our method must accurately measure both the presence and the age of a burst, while also allowing us to exclude galaxies in which no recent burst has occurred. 

We enforce a two-component SFH: an old stellar component modeled by a delayed exponential function ($SFR(t) \propto t \times e^{-t/\tau}$, \citealt{Simha_2014}) and a new stellar component modeled by a double power law function ($SFR(t) \propto [(t/\tau)^{\alpha} + (t/\tau)^{-\beta}]^{-1}$, \citealt{Wild_2020}) representing the burst (see Table~\ref{tab:freeparameters} for more details, and Appendix~\ref{appendix:a} for a discussion of our choice of functional form). This approach forces \textsc{Bagpipes} to assume that there was a period of recent star formation, but it can make that period as small, large, short/bursty, or long/extended as it wants. This two-component strategy allows \textsc{Bagpipes} to fit galaxies with a recent burst as well as older galaxies that formed the majority of their stellar mass significantly in the past. Choosing to use a parametric SFH places emphasis on accurately recovering the time since the burst occurred.

We set the burst to occur between the time of observation and 3 Gyr before the time of observation. We fix the rising slope index $\beta$ \citep{Wild_2020} and allow the falling slope index $\alpha$ to vary \citep{Carnall_2019}. We calculated the time of observation to be the age of the universe given the redshift of the galaxy, using a flat cosmology defined by $H_0 = 70$ km s$^{-1}$ Mpc$^{-1}$ and $\Omega_M = 0.3$.

\subsubsection{Other Priors}
\label{sec:otherpriors}

Other parameters are shown in Table \ref{tab:freeparameters}, more information on our SFH fitting process is in Appendix~\ref{appendix:a}, and the full fit instructions are shown in Appendix~\ref{appendix:b}. We use a Gaussian prior on the redshift, centered on the values given in Table~\ref{table:finalsample}. Because \textsc{Bagpipes} is extremely sensitive to mismatches in the redshift solution, we use a Gaussian width of 40$\sigma_z$ for the SDSS spectra. The redshift uncertainties we assumed for the WISeREP, \cite{Graur_2018}, \cite{Hammerstein_2021}, and some MUSE galaxies were quite large, so we set the redshift uncertainty prior equal to the redshift uncertainty in these cases. This assumption leads to a roughly equivalent redshift uncertainty prior between the samples. Both our data and the \textsc{Bagpipes} model spectra use vacuum wavelengths, though any small mismatch is accounted for by \textsc{Bagpipes} when fitting the redshift. No extinction correction was applied. To address any inaccurate flux calibration in the spectra or template mismatch between the models and the data, a Chebyshev polynomial perturbation is fit to the data in the “Calibration” parameter. The metallicity priors for each galaxy (median and variance, with a Gaussian shape) were determined from the galaxy's stellar mass by using the stellar mass--metallicity trend shown in Table 2 of \cite{Gallazzi_2005}. These were applied to both the new and old star formation components.

\begin{table*}[t]
  \centering
  \begin{tabular}{lll}
    \hline
    \hline 
    Parameter and Unit & Value or Allowed Range & Shape of Prior \\
    \hline
    \hline 
    Nebular continuum emission ($\log_{10}$ of ionization parameter) & -3 & -- \\
    \hline
    Dust attenuation, Calzetti ($V$ band magnitudes) & 0, 2 & Flat \\
    \hline
    White noise, scaled & 1, 10 & $\log_{10}$ \\
    \hline
    Polynomial Bayesian calibration, 0th order & 0.5, 1.5 & Gaussian, $\mu = 1.0$, $\sigma = 0.25$ \\
    \hline
    Polynomial Bayesian calibration, 1st order & -0.5, 0.5 & Gaussian, $\mu = 0.0$, $\sigma = 0.25$ \\
    \hline
    Polynomial Bayesian calibration, 2nd order & -0.5, 0.5 & Gaussian, $\mu = 0.0$, $\sigma = 0.25$ \\
    \hline
    Velocity dispersion (km/s) & 30, 300 & $\log_{10}$ \\
    \hline
    Time since SF began in old SF component (Gyr) & 3, 15 & $\log_{10}$ \\
    \hline
    Timescale of SF decrease $\tau$ in old SF component (Gyr) & 1 & -- \\
    \hline
    Old SF component mass formed ($\log_{10}(M_{\star}/M_{\odot})$) & 1, 15 & $\log_{10}$ \\
    \hline
    Old SF component metallicity ($Z/Z_{\odot}$) & 0.1, 2 & Gaussian \\
    \hline
    Falling slope index of new SF component $\alpha$ & 0.1, 1000 & $\log_{10}$ \\
    \hline
    Rising slope index of new SF component $\beta$ & 250 & -- \\
    \hline
    Age of universe $\tau$ at turnover of new SF component (Gyr) & $T_0-3$, $T_0$\footnote{$T_0$ is the age of the universe at time of observation} & $\log_{10}$ \\
    \hline
    New SF component mass formed ($\log_{10}(M_{\star}/M_{\odot})$) & 1, 15 & log$_{10}$ \\
    \hline
    New SF component metallicity ($Z/Z_{\odot}$) & 0.1, 2 & Gaussian \\
    \hline
    Redshift $z$ & 0.0001, 0.9 & Gaussian \\
    \hline
  \end{tabular}
  \caption{Parameters used by \textsc{Bagpipes} to fit all host galaxy spectra (see Appendix~\ref{appendix:b} to see these parameters implemented in the fit instructions to \textsc{Bagpipes}). $\mu$ and $\sigma$ for the metallicity priors were determined by $M_{\star}$.}
  \label{tab:freeparameters}
\end{table*}

\subsection{Measuring H$\alpha$ Emission and H$\delta$ Absorption}
\label{sec:HalpHdelmethods}

Another way to assess the current and recent SFH of a galaxy is to use the equivalent width of the H$\alpha$ emission line and the H$\delta$ index in the galaxy spectra. For some of the galaxies in the TDE host sample, these values had already been calculated \citep{French_2016, Graur_2018, French_2020}. For the rest of the galaxies, we performed these calculations ourselves (Table~\ref{table:broadlinestable}). For the galaxies in the expanded control sample and PSB/QBS control sample, we used the H$\alpha$ equivalent width and H$\delta$ index available in SDSS {\tt galspec} \citep{Kauffmann_2003, Brinchmann_2004, Tremonti_2004}.

To compute the H$\alpha$ emission line equivalent width, we extracted the H$\alpha$ line flux, as calculated by \textsc{Bagpipes} accounting for stellar absorption at H$\alpha$. Then, we calculated the average continuum level on both sides of the line, using a region free of other strong emission lines\footnote{The left continuum region was at $\lambda_{rest} = 6507.81, 6542.81$. The right continuum region was at $\lambda_{rest} = 6597.81, 6637.81.$}. The H$\alpha$ emission line's equivalent width is the summed flux (across the entire line) divided by the average continuum in the region. We used \textsc{pyLick} \citep{pyLick} to obtain the Lick H$\delta_{\rm{A}}$ absorption line equivalent width. We are unable to calculate the H$\delta$ equivalent width for three galaxies (AT 2018fyk, AT 2019ahk, and AT 2019dsg) because the spectra did not extend to blue enough wavelengths.

\subsection{Assigning Galaxy Type Labels}
\label{sec:galaxy_classification_methods}

To calculate the PSB overrepresentation in our TDE host galaxy sample, we assigned galaxy type labels to the hosts. These labels are quiescent, quiescent Balmer-strong (QBS), star forming (SF), and post-starburst (PSB). We used a classification scheme based on H$\alpha$ equivalent widths and H$\delta$ indices. Star forming galaxies must have H$\alpha > 3$ {\AA}, quiescent galaxies must have H$\alpha < 3$ {\AA} and H$\delta < 1.3$ {\AA}, QBS galaxies must have H$\alpha < 3$ {\AA} and $1.3$ {\AA} $\leq$ H$\delta < 4$ {\AA}, and PSB galaxies must have H$\alpha < 3$ {\AA} and H$\delta \geq 4$ {\AA} \citep{French_2016}. We plot the H$\alpha$ equivalent widths and Lick H$\delta_{\rm A}$ indices of 38 out of 41 TDE host galaxies in Section~\ref{sec:psb_overrepresentation}. We cannot plot three TDE host galaxies (AT 2018fyk, AT 2019ahk, and AT 2019dsg) because their MUSE spectra does not extend to blue enough wavelengths for us to calculate their H$\delta$ index and the indices had not been reported elsewhere \citep{Graur_2018}. Despite this limitation, we are still able to classify two of these galaxies because their H$\alpha$ equivalent widths place them in the star forming region (see Table~\ref{table:broadlinestable}). Thus, we have classifications for 41 TDE host galaxies (40 unique hosts plus the doubly-counted F01004 host). The degree to which PSB and QBS galaxies are overrepresented in the TDE host galaxy sample is explored in Section~\ref{sec:psb_overrepresentation}.

Some of the galaxies in our sample had already been labeled in \citet{French_2020}. For the galaxies which also had classifications in \cite{French_2020}, all of our classifications agree except for the host galaxy of RBS 1032. This was classified as PSB under our classification scheme, but was considered QBS in \cite{French_2020} because \cite{French_2020} required a galaxy's H$\delta$ index to be at least 1$\sigma$ more than 4 {\AA} to be labeled as PSB.  

\section{Results}
\label{sec:results}

\subsection{TDE Rates versus Burst Age}
\label{sec:tderate_age}

\begin{figure}[h]
\begin{center}
    \includegraphics[width=0.95\linewidth]{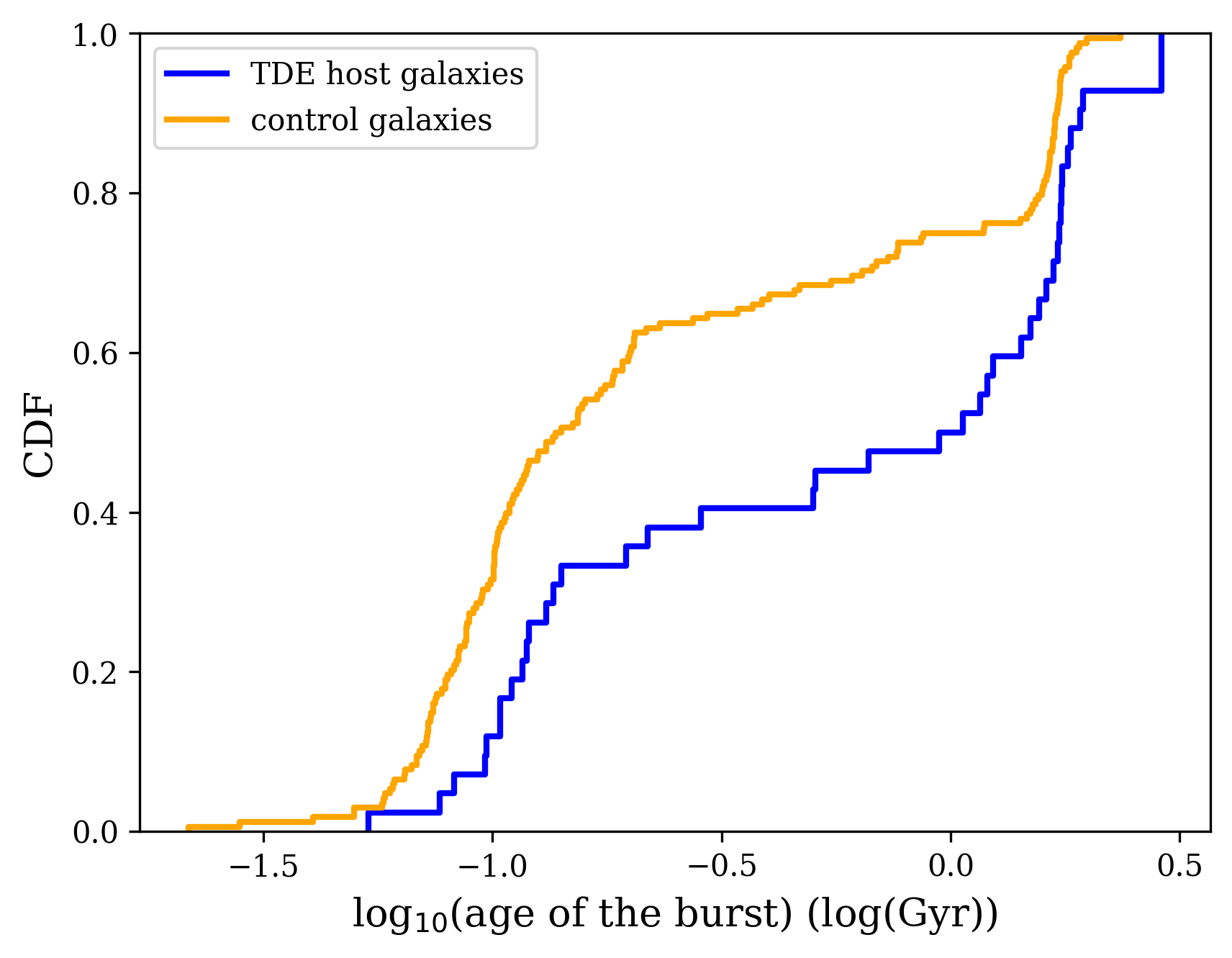}
\end{center}
\caption{Cumulative distribution function of the time elapsed since a burst of star formation in the entire TDE host galaxy sample and the control sample. The Anderson-Darling 2-sample test shows with a significant result that the two samples are not drawn from the same population (p-value $< 0.001$).}
\label{fig:burstage_compare}
\end{figure}

After using \textsc{Bagpipes} to model the galaxy spectra of the TDE host galaxy sample, the control sample, and the PSB/QBS control sample, we extract the posterior distributions of all free parameters in the \textsc{Bagpipes} fits, including information about the SFHs (Table~\ref{table:BagpipesOp4}). We plot a cumulative distribution function of the age of the burst for all of the TDE host galaxies and the control sample galaxies in Figure~\ref{fig:burstage_compare}. There are differences between the two samples; the control sample galaxies are more likely to have young burst ages and the TDE host galaxies are more likely to have older burst ages. The Anderson-Darling 2-sample test comparing the two samples returns a p-value $< 0.001$, indicating that the difference in their distributions of burst age is statistically significant. 

\begin{figure*}
\begin{center}
\includegraphics[width=0.49\linewidth]{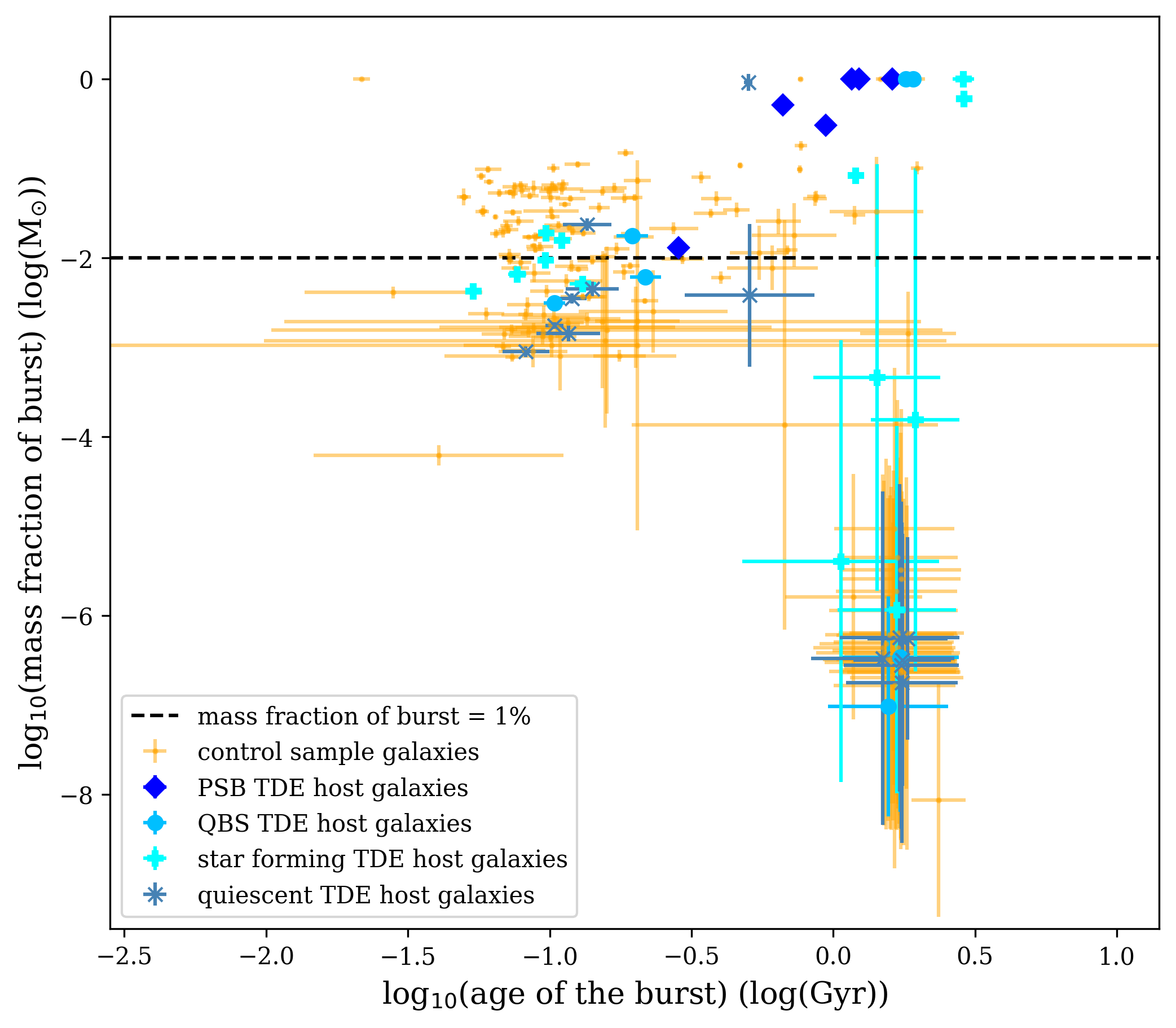}
\includegraphics[width=0.49\linewidth]{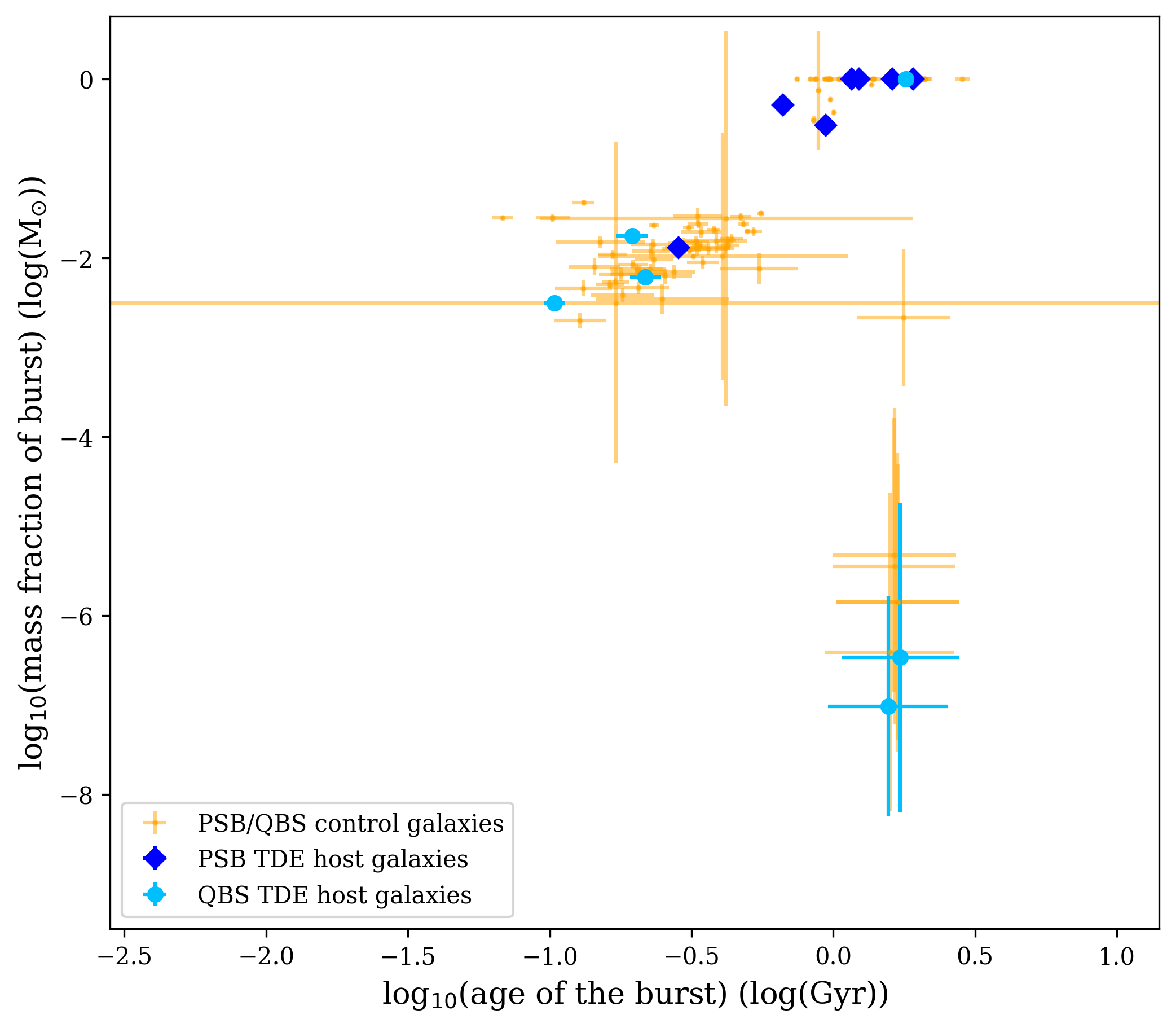}
\end{center}
\caption{Burst mass fraction versus age of the burst. {\it Left:} The control sample galaxies are shown in orange dots and the TDE host galaxies are shown in shades of blue, with different symbols and colors for different galaxy type labels. The dashed line shows the minimum burst mass fraction needed to be considered having a “high” burst mass fraction (1\%), above which all the PSB TDE host galaxies and most of the QBS TDE host galaxies fall. {\it Right:} The PSB/QBS control sample galaxies are shown in orange dots and the PSB/QBS subset of the TDE host galaxies are shown in shades of blue. Not shown in this figure is the falling slope index $\alpha$; the galaxies with low $\alpha$, corresponding to a very slowly declining “burst”, tend to fall in the lower right corner alongside galaxies with older stellar populations. Of note are the large error bars on galaxies with burst mass fractions less than 1\%. In these cases, \textsc{Bagpipes'} results indicate that these galaxies have not had a burst. \textsc{Bagpipes} is forced to return values for such a “burst” anyway, but because the burst is so small/did not happen, the error bars on the burst age and burst mass fraction are very large.}
\label{fig:age_vs_massfrac}
\end{figure*}

In order to meaningfully compare the post-burst ages, we must select a subset of galaxies that have experienced recent bursts, excluding galaxies with low-level fluctuations of on-going star formation or a lack of significant star formation within the past $\sim$3 Gyr, for which we cannot accurately measure a post-burst age. To determine which galaxies have experienced a true burst, we plot the galaxies' burst mass fractions versus the burst ages, with the points labeled by their galaxy type (Figure~\ref{fig:age_vs_massfrac}). Several distinct populations of galaxies can be seen. For star forming galaxies with young burst ages, we see a population of galaxies with burst fractions $<1\%$, for which low-level fluctuations in their SFR cause them to have small repeated bursts. Additionally, we can see that the burst mass fraction becomes poorly constrained below 1\%, and the burst age of galaxies below this threshold is also poorly constrained.

We impose a minimum burst mass fraction (stellar mass formed in the burst divided by total stellar mass formed) of 1\% for a galaxy to have a “high” burst mass fraction. \cite{French_2018} find that burst mass fractions are typically above 3\% for the PSB galaxies in their sample. The threshold of 1\% is selected to include PSB SFHs while excluding low-level fluctuations in star formation from main sequence galaxies. When we select for galaxies with a high burst mass fraction, we are able to recover all of the PSB galaxies.\footnote{Increasing the threshold for high burst mass fraction from 1\% to 3\% removes TDE host galaxies with younger burst ages, making comparison to models more difficult but does not affect the overall result that TDE host galaxies have older burst ages.} 

Additionally, we implement a minimum steepness of the burst's decline, represented by $\alpha$ in \textsc{Bagpipes}, of $\alpha > 10$ for a galaxy to be considered having a significant burst. This cut removes galaxies with extremely slowly-declining “burst” components with $\alpha < 1$, which much more closely resemble constant star formation, from consideration as galaxies with a significant burst\footnote{This cut on $\alpha$ removes one TDE host (host of F01004) from consideration as a galaxy that has experienced a significant burst.}. The subset of the TDE host galaxy sample that has experienced a significant burst of star formation contains 15 galaxies.

For galaxies with a burst mass fraction of $>1$\%, there is generally an upward trend between burst mass fraction and burst age (see both panels of Figure~\ref{fig:age_vs_massfrac}). Galaxies with smaller bursts (still above 1\%) can only be identified as PSB for a short period of time, as the spectroscopic signatures of a galaxy being PSB will fade faster. Galaxies with larger bursts can be identified as PSB for a longer period of time \citep{French_2018}, but if the larger burst is enabled by a longer period of star formation that fades more slowly, they will be observable as PSBs after a longer delay. Despite the selection effects discussed above, the cut of $>1\%$ nonetheless selects both normal and PSB host galaxies over a full range of burst ages.

To calculate the TDE rate as a function of time since burst, we use two different methods: (1) using the TDE host galaxies that our stellar population analysis determines have experienced a significant burst of star formation, normalized by the control sample, and (2) using the TDE host galaxies that are spectroscopically confirmed as QBS or PSB, normalized by the PSB/QBS control sample. For both of these methods, we calculate the TDE rate and TDE rate enhancement.
 
To infer the TDE rate per age bin for TDE host galaxies that have experienced a significant burst of star formation, we must estimate the number of galaxies and the timescale over which the astronomical community would have been sensitive to TDEs. In practice, we have combined TDEs from multiple discovery surveys. We estimate a survey duration ($T_{survey} = 5$ years) and survey volume ($N_{gal} = 500,000$ galaxies) using ZTF as a guide. The survey volume estimate comes from the approximate number of galaxies with spectra in SDSS Data Release 8 \citep{Aihara_2011} that have stellar masses above $10^{8.6} M_{\odot}$ (the lower bound of stellar masses for TDE host galaxies in our sample) and that have a median SNR $> 10$. We find that the average TDE rate over all age bins is close to the measured TDE rate per year per galaxy \citep{Yao_2023}. This validates our assumptions, but we note that in our comparisons below, our measurements are more capable of measuring the TDE rate \textit{enhancement} over time than the absolute TDE rate over time. If we chose a different survey volume and duration, the effect would be a DTD whose shape stays the same but whose normalization can increase or decrease.

For the DTD of TDE host galaxies that have experienced a significant burst, we calculate the TDE rate per post-burst age as 
\begin{equation} 
R_{TDE,burst} =\frac{N_{TDE}}{T_{survey} \times N_{gal}} \times \frac{f_{TDE,burst}}{f_{con}},
\end{equation}
where $R_{TDE,burst}$ is the TDE rate per age bin in units of TDEs per year per galaxy. $N_{TDE}$ is the number of TDEs in our TDE host galaxy sample. $T_{survey}$ and $N_{gal}$ are the survey duration and survey volume, respectively, that we estimate as described above. $f_{TDE,burst}$ is the fraction of TDEs whose host galaxies have experienced a significant burst of star formation that fall into each age bin, and $f_{con}$ is the fraction of control sample galaxies per age bin. To convert our results to units of rate enhancement, we divide the TDE rate by the average optical TDE rate calculated in \cite{Yao_2023} in units of TDEs per year per galaxy, $R_{opt} = 3.2 \times 10^{-5}$ yr$^{-1}$ gal$^{-1}$, so that 

\begin{equation}
\Gamma_{TDE,burst} = \frac{R_{TDE,burst}}{R_{opt}},
\end{equation}
where $\Gamma_{TDE,burst}$ is a multiplicative factor above the fiducial TDE rate, or the “rate enhancement” per age.

For the case of the PSB/QBS TDE host galaxies, we use the PSB/QBS control sample of non-TDE host galaxies to perform a similar estimate of $R_{TDE,burst}$ and $\Gamma_{TDE,burst}$ as a function of post-burst age. For this analysis, instead of using the estimated survey volume to set the TDE rate normalization, we use the QBS overrepresentation factor as calculated in Section~\ref{sec:psb_overrepresentation}. Including the PSB galaxies as a subset of the QBS sample, we find the QBS overrepresentation factor to be $O_{QBS} = 8.61$, averaged across all age bins. We can then calculate the rate enhancement per age bin to be

\begin{equation}
\Gamma_{TDE,QBS} =  O_{QBS} \times \frac{f_{TDE,QBS}}{f_{con,QBS}},
\end{equation}

where $f_{TDE,QBS}$ is the fraction of TDEs whose host galaxies are PSB/QBS that fall into each age bin and $f_{con,QBS}$ is the fraction of PSB/QBS control galaxies per age bin. Multiplying by the average optical TDE rate from \citet{Yao_2023}, we can also estimate the TDE rate per age bin as

\begin{equation}
R_{TDE,QBS} = O_{QBS} \times \frac{f_{TDE,QBS}}{f_{con,QBS}} \times R_{opt},
\end{equation}
where $R_{TDE,QBS}$ is the TDE rate per age bin.

\begin{figure*}
\begin{center}
\includegraphics[width=0.49\linewidth]{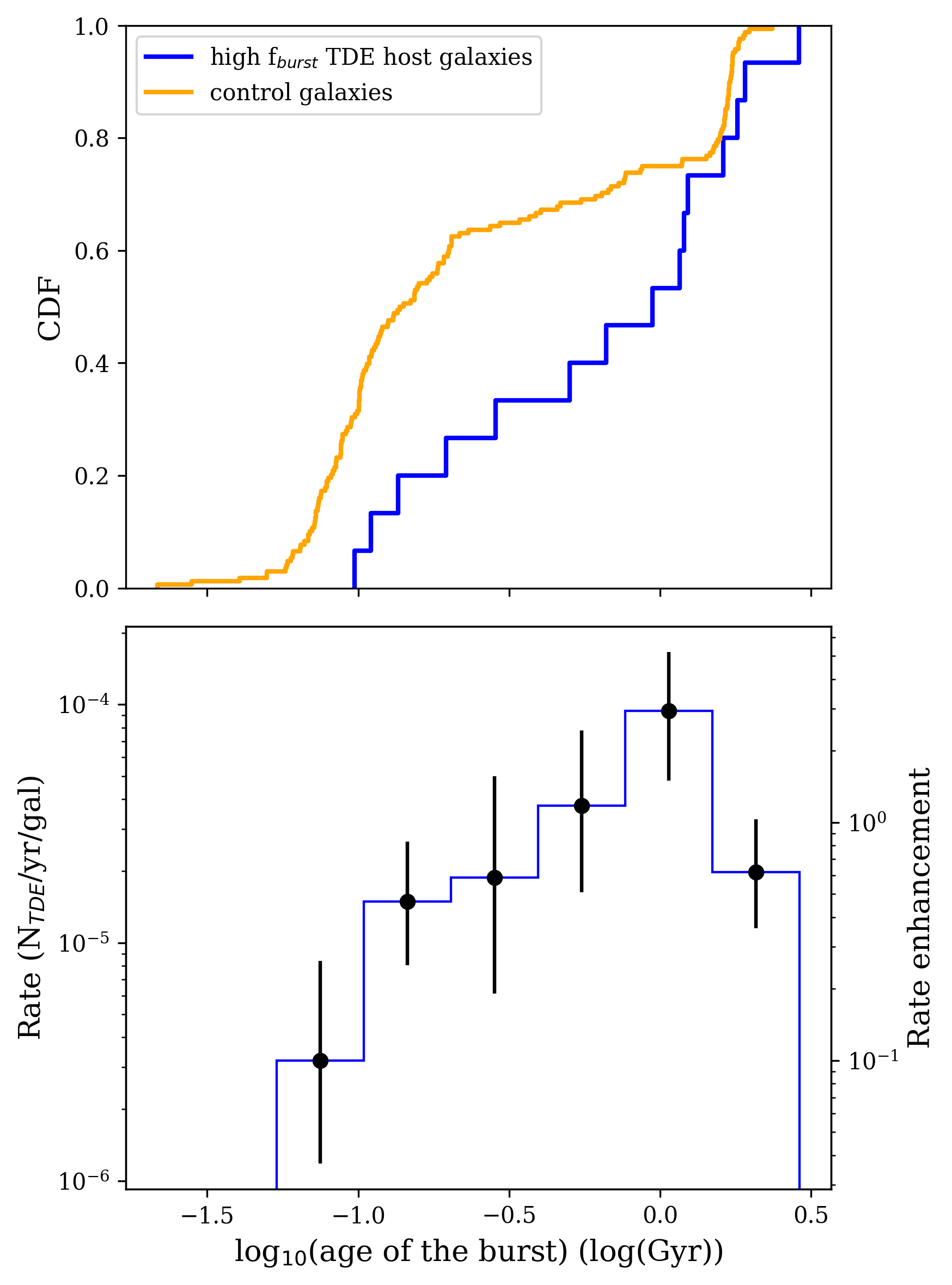}
\includegraphics[width=0.49\linewidth]{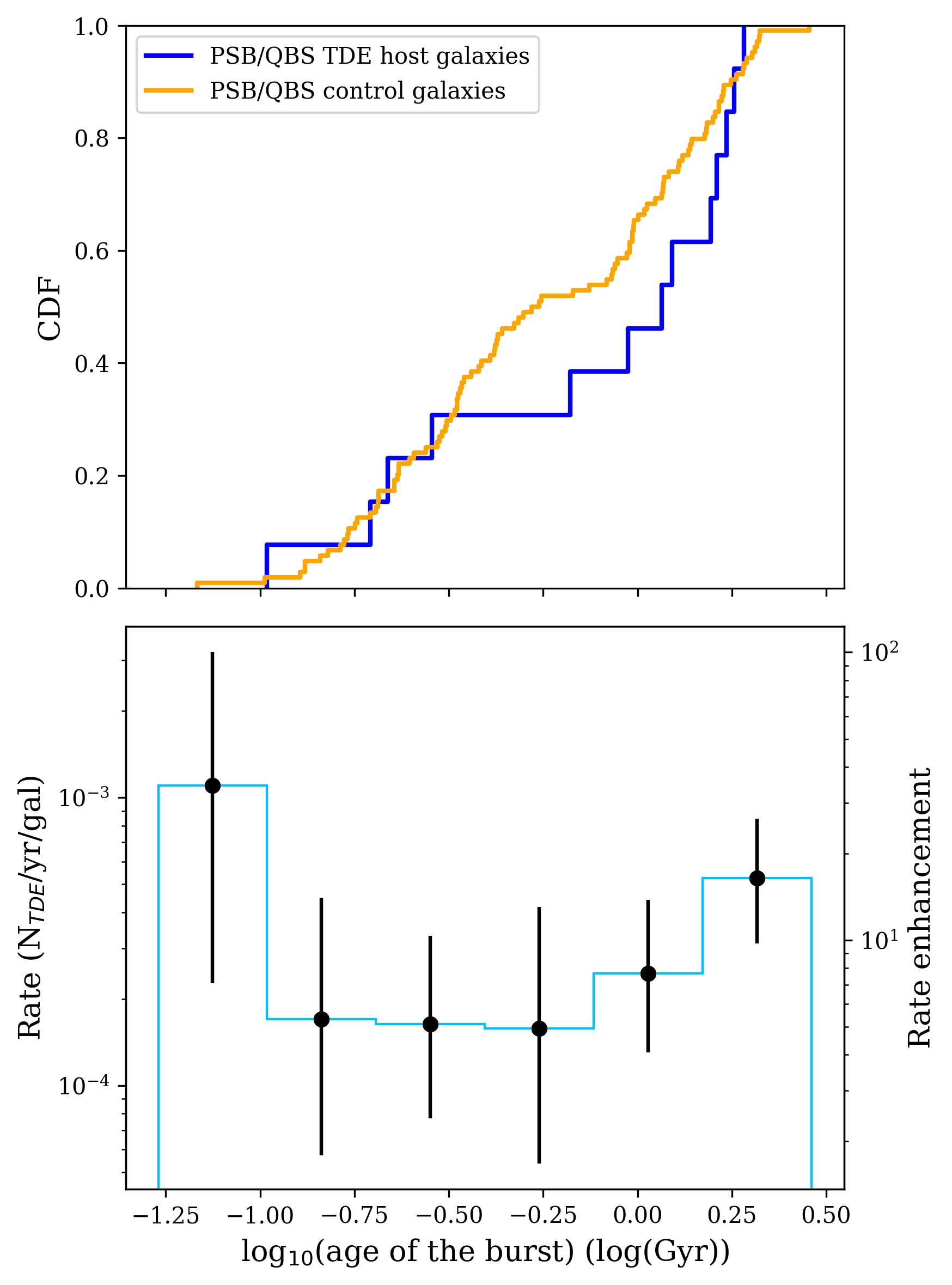}
\end{center}
\caption{{\it Top left:} A cumulative distribution function showing the burst ages for the TDE host galaxies that have had a significant burst (using the criteria described in Section~\ref{sec:tderate_age}) and for the control sample galaxies. {\it Bottom left:} The rate and rate enhancement of TDEs in galaxies that have recently had a significant burst as a function of burst age. There is a peak in the TDE rate at $\sim$1 Gyr because very few control galaxies have a burst that old, while many TDEs in galaxies with a significant burst did have a burst at that age. We find a significant result using the Anderson-Darling 2-sample test that the high f$_{burst}$ TDE host burst ages and the control galaxy burst ages are not drawn from the same population (p-value $=0.00791$). {\it Top right:} A cumulative distribution function showing the burst ages for the PSB/QBS subset of TDE host galaxies and the PSB/QBS control sample. {\it Bottom right:} The rate and rate enhancement of TDEs in PSB/QBS galaxies as a function of burst age. This DTD is flatter than the high f$_{burst}$ DTD to the left, as the burst age distributions of the PSB/QBS TDE hosts and the PSB/QBS control galaxies are not statistically distinct.}
\label{fig:finalrate_results}
\end{figure*}

The cumulative distribution functions of burst ages for each subset of TDE host galaxies and their corresponding control samples are shown in the top panels of Figure~\ref{fig:finalrate_results}. We use the Anderson-Darling 2-sample test to determine if each of these sample pairs (high f$_{burst}$ TDE host galaxies and control galaxies for the upper left panel, and PSB/QBS TDE host galaxies and PSB/QBS control galaxies for the upper right panel) are drawn from the same population. For high f$_{burst}$ TDE hosts, we find a significant result that the two samples are not drawn from the same population (p-value $=0.00791$). For the PSB/QBS TDE hosts, we do not find a significant result (p-value $> 0.25$), which means the PSB/QBS TDE host burst age distribution cannot be distinguished from the PSB/QBS control galaxy burst age distribution.

The resulting TDE rates and rate enhancements are shown in the bottom panels. These are our “fiducial” DTDs. The error bars in the lower panels are Poisson errors using the \texttt{Pearson} method in the \texttt{astropy} function \texttt{poisson\_conf\_interval}, which are propagated through the rate and rate enhancement equations. We ensure the bin size is larger than the typical age and set the bin size by calculating $\frac{3\bar{\sigma_t}}{ln(10)\bar{t}}$, where $\bar{\sigma_t}$ is the average of the standard deviation values of the burst ages and $\bar{t}$ is the average of the burst ages.

\begin{figure*}
\begin{center}
\includegraphics[width=0.49\linewidth]{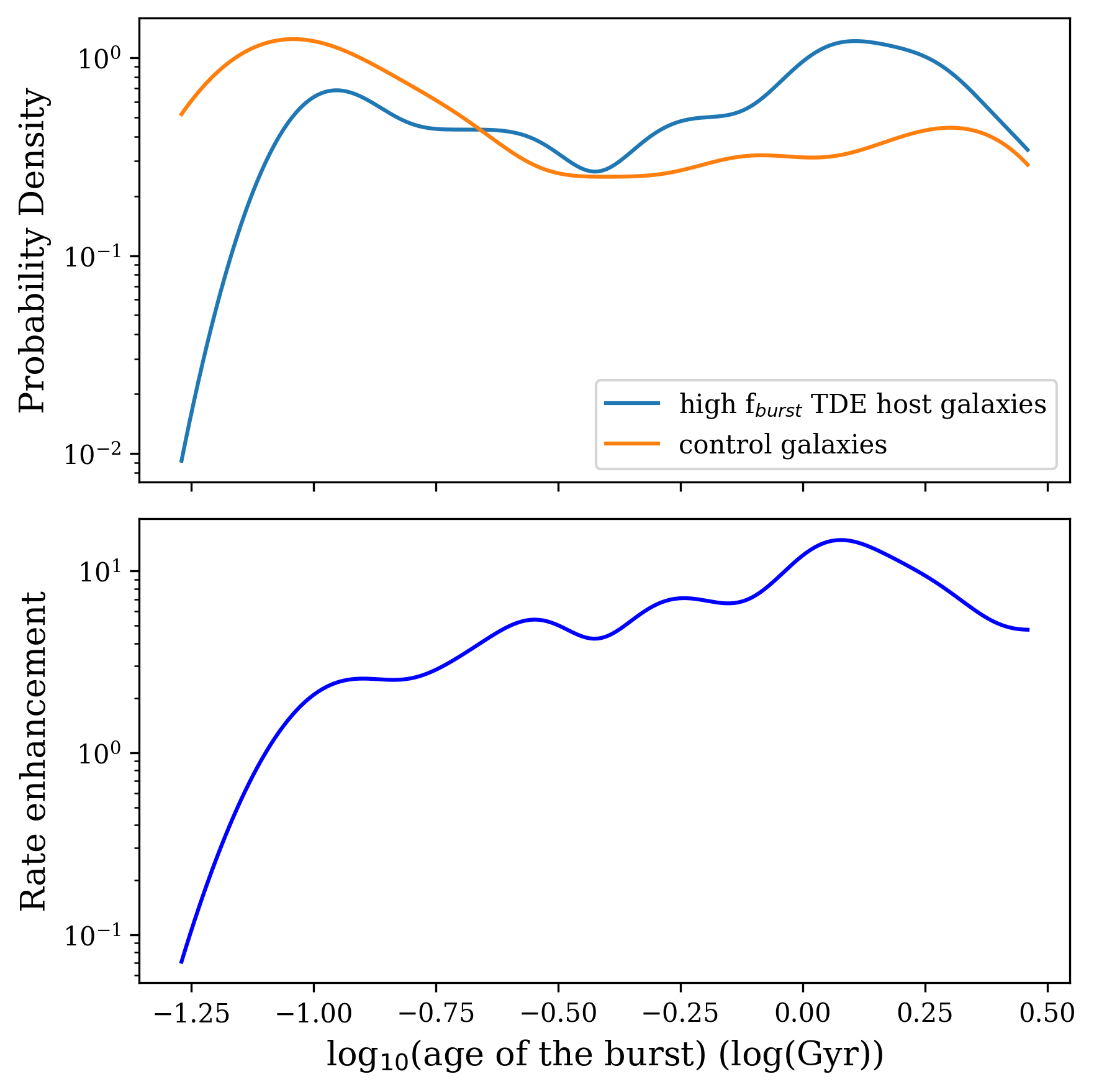}
\includegraphics[width=0.49\linewidth]{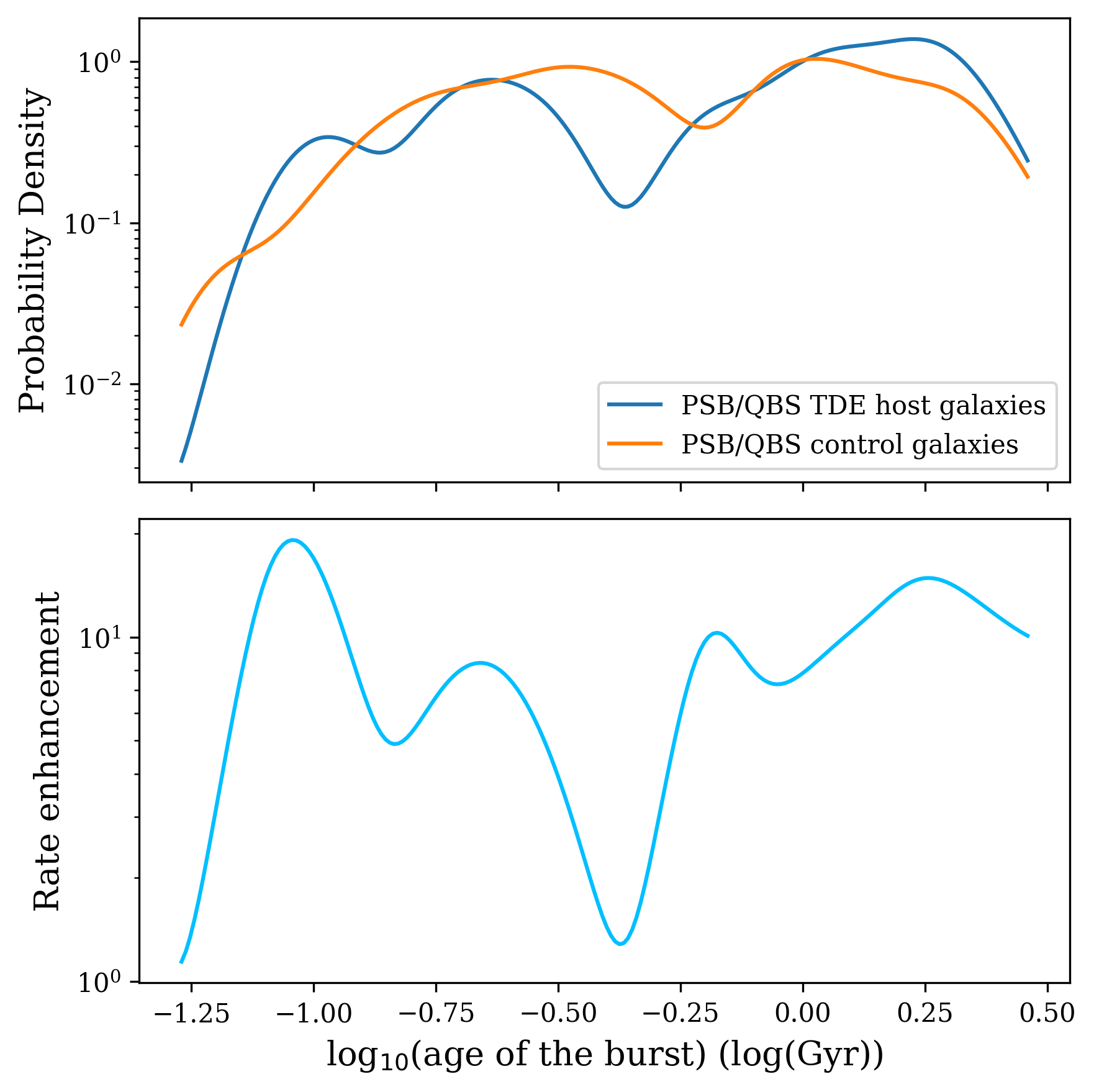}
\end{center}
\caption{This plot shows the same data as Figure~\ref{fig:finalrate_results}, but uses kernel density estimation to produce a smoothed histogram. This incorporates the entire posterior returned by \textsc{Bagpipes}, which is important to take into account the large error bars on the burst age for the galaxies with lower burst mass fractions, some of which are included in the subset of PSB/QBS galaxies. {\it Top left:} A probability density plot showing the distribution of burst ages for the TDE host galaxies that have had a significant burst of star formation and for the control sample. {\it Bottom left:} The rate enhancement of TDEs in galaxies that have recently had a significant burst as a function of burst age. The control galaxies are used to normalize this histogram, which leads to a peak in the TDE rate at a burst age just after 1 Gyr. {\it Top right:} A probability density plot showing the distribution of burst ages for the PSB/QBS subset of TDE host galaxies and for the PSB/QBS control sample. {\it Bottom right:} The rate enhancement of TDEs in PSB/QBS galaxies as a function of burst age. The PSB/QBS control sample galaxies are used to normalize this histogram. The distribution has more variation, with peaks in the TDE rate at the youngest and oldest burst ages and a valley at an intermediate age. The broad trends shown in this plot mirror those shown in Figure~\ref{fig:finalrate_results}, indicating that the variable errors on burst age returned by \textsc{Bagpipes} are not influencing our results.}
\label{fig:finalkde_results}
\end{figure*}

While we use the median value of the posteriors that \textsc{Bagpipes} returns to collect the burst ages of the TDE host galaxies and the control sample galaxies, we can also make use of the full distributions. We retrieved the full probability distribution of the time since burst for all galaxies in all samples and used kernel density estimation (KDE, bandwidth = 0.2 dex) to achieve a smoothed histogram. Figure~\ref{fig:finalkde_results} shows these distributions where the bottom panels are analogous to the bottom panels of Figure~\ref{fig:finalrate_results}. The non-binned nature of this graph allows us to see more variation than may be present in the binned DTDs. However, the broad trends are reproduced, indicating that any galaxies with large errors on their burst ages are not unduly affecting the DTDs.

The upper left panel of Figure~\ref{fig:finalkde_results} shows the probability density distributions for the burst ages of the TDE host galaxies that have experienced a significant burst of star formation and of the control sample galaxies. To show the TDE rate enhancement as a function of time since burst (lower left panel), we calculate $\frac{KDE_{TDE}}{N_{TDE}} \times \frac{N_{con}}{KDE_{con}}$, where $N_{TDE}$ is the number of TDE host galaxies and $N_{con}$ is the number of control sample galaxies. The shape of the KDE version of this DTD matches that of the histogram version in Figure~\ref{fig:finalrate_results}, where the TDE rate increases with burst age until a peak is reached at a burst age of $\sim$1 Gyr.

The upper right panel of Figure~\ref{fig:finalkde_results} shows the probability density distributions for the burst ages of the PSB/QBS TDE host galaxies and of the PSB/QBS control sample galaxies. The rate enhancement was calculated in the same manner as above. The shape of the KDE version of this DTD roughly matches that of the histogram version in Figure~\ref{fig:finalrate_results}, where there is variation in the rate enhancement as a function of time, but the overall trend in both panels does not show a clear increase in the rate enhancement from young burst ages to old burst ages.

\subsection{TDE Rates versus Burst Strength}
\label{sec:tderate_strength}

\begin{figure}
\begin{center}
    \includegraphics[width=0.95\linewidth]{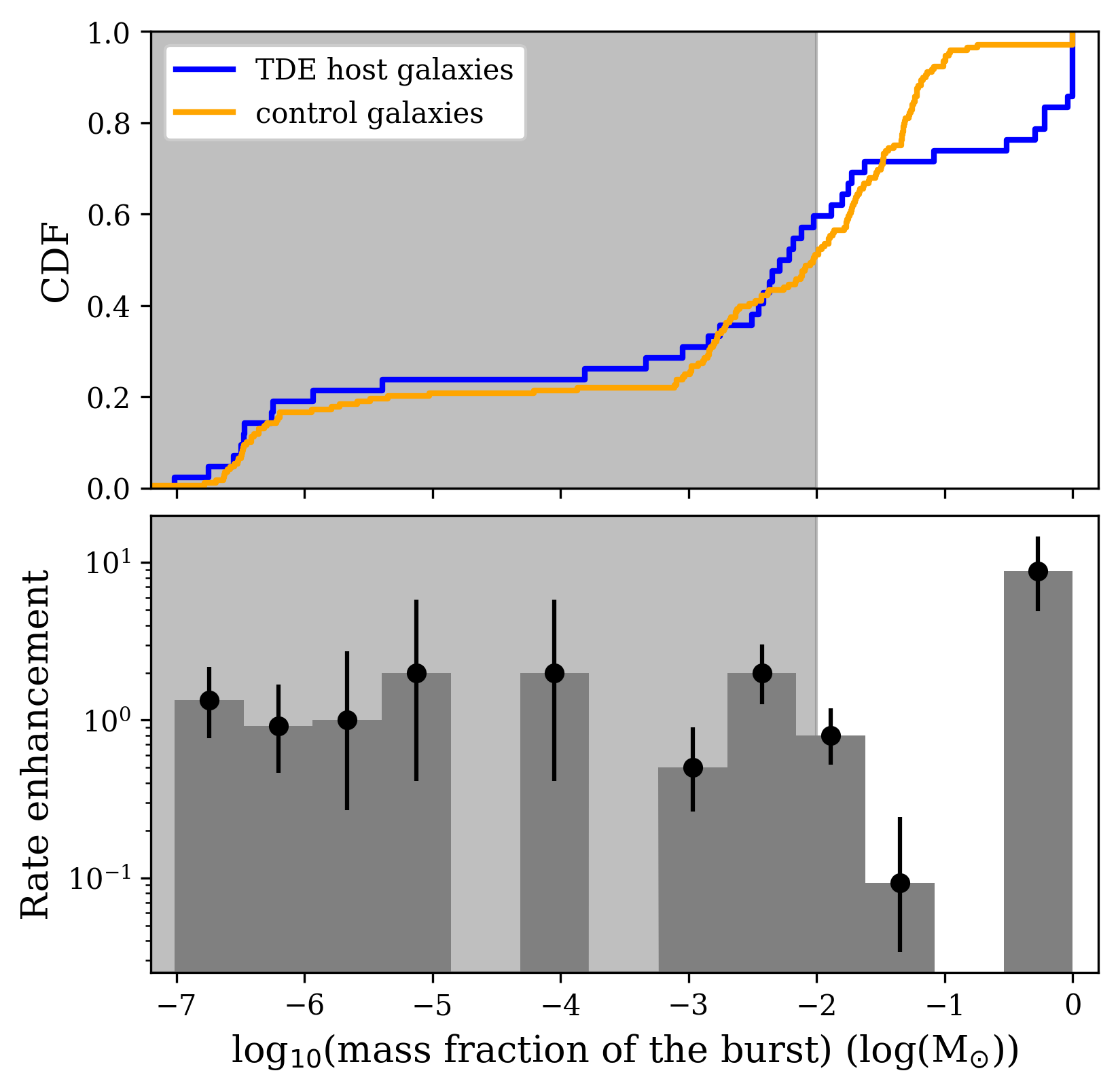}
\end{center}
\caption{{\it Top:} A cumulative distribution function showing the burst mass fractions for the entire TDE host galaxy sample and for the control sample. {\it Bottom:} The rate enhancement of TDEs as a function of burst mass fraction. The full TDE sample is used and it is normalized by the control sample. The gray shaded region denotes burst mass fractions below $10^{-2}$, which have quite large error bars, so conclusions should not be drawn about these values. At large burst mass fractions (greater than 10\%), we see that there is a large rate enhancement. As there is no rate enhancement greater than 1 for burst mass fractions between 1\% and a few $\times$ 10\%, we see that the observed rate enhancement in our TDE host galaxy sample is entirely concentrated at the highest burst mass fractions.}
\label{fig:massfrac}
\end{figure}

Another way to measure the TDE overrepresentation in galaxies that have recently experienced a significant burst of star formation is to calculate the TDE rate enhancement as a function of burst mass fraction. This is shown in Figure~\ref{fig:massfrac}, which was constructed in a similar fashion to Figure~\ref{fig:finalrate_results}. The x-axis shows the log burst mass fraction. The y-axis shows the rate enhancement in log spacing. The number of TDE host galaxies in each bin of burst mass fraction is normalized by the fraction of control sample galaxies that also fall into that bin. However, galaxies with very small burst mass fractions often have large errors on their burst age (see horizontal error bars on the points below the dotted line in Figure~\ref{fig:age_vs_massfrac}). For this reason, we have grayed out the region of the graph where the burst mass fraction is less than 1\%. 

We use the Anderson-Darling 2-sample test to determine if these two samples are drawn from the same population. When considering the full TDE host galaxy sample and control sample across the full range of burst mass fractions, we do not find a significant result (p-value $=0.0721$). However, when only considering TDE host galaxies and control galaxies with burst mass fractions $> 1\%$, the Anderson-Darling 2-sample test returns a significant result (p-value $< 0.001$). This indicates that when considering galaxies with a significant burst of star formation, TDE host galaxies are likely to have greater burst mass fractions than non-TDE host galaxies.

TDE host galaxies with burst mass fractions between 1\% and 10\% have rate enhancements between 0.1 and 1 (effectively reductions in the rate). No TDEs in our sample have burst mass fractions at or slightly above 10\%, so there is no bar in the second-highest burst mass fraction bin. Importantly, the TDE rate enhancement of galaxies with the highest burst mass fraction (greater than 10\% and approaching unity) is almost a factor of 10. This is consistent with the expectation that galaxies that have experienced a significant burst of star formation have high TDE rates, and suggests that a combination of high burst mass fractions and bursts of star formation that are $\sim$1 Gyr old may help to boost the TDE rate.

\subsection{Post-Starburst Overrepresentation}
\label{sec:psb_overrepresentation}

\cite{Arcavi_2014} found a strong preference for E+A galaxies, a classification similar to PSB, among a survey of TDEs observed with the Palomar Transient Factory. \cite{French_2016} quantify the overrepresentation for PSB and QBS galaxies, finding an overrepresentation in PSB galaxies by a factor of $190^{+115}_{-100}$ and an overrepresentation in QBS galaxies by a factor of $33^{+7}_{-11}$. Follow-up studies have further refined these quantities. \cite{LawSmith_2017} control for a variety of selection effects, which on net favor selection of PSB hosts, but still finds a PSB overrepresentation factor of $\sim$25-38 and a QBS overrepresentation factor of $\sim$10-13. \cite{Graur_2018} find a PSB overrepresentation factor of $35^{+21}_{-17}$ and a QBS overrepresentation factor of $18^{+8}_{-7}$ in a sample of optical and X-ray selected TDEs. Both \cite{Hammerstein_2021} and \cite{Roth_2021} find that the PSB overrepresentation persists even when controlling for selection effects. \cite{Hammerstein_2021} find a PSB overrepresentation factor of 22-29 and a QBS overrepresentation factor of 16-17. \cite{Roth_2021} find that selection effects, including dust obscuration that prevents TDEs in star forming galaxies from being detected, may revise the overrepresentation factor lower but cannot explain the overrepresentation entirely.

\begin{figure}
\begin{center}
    \includegraphics[width=0.95\linewidth]{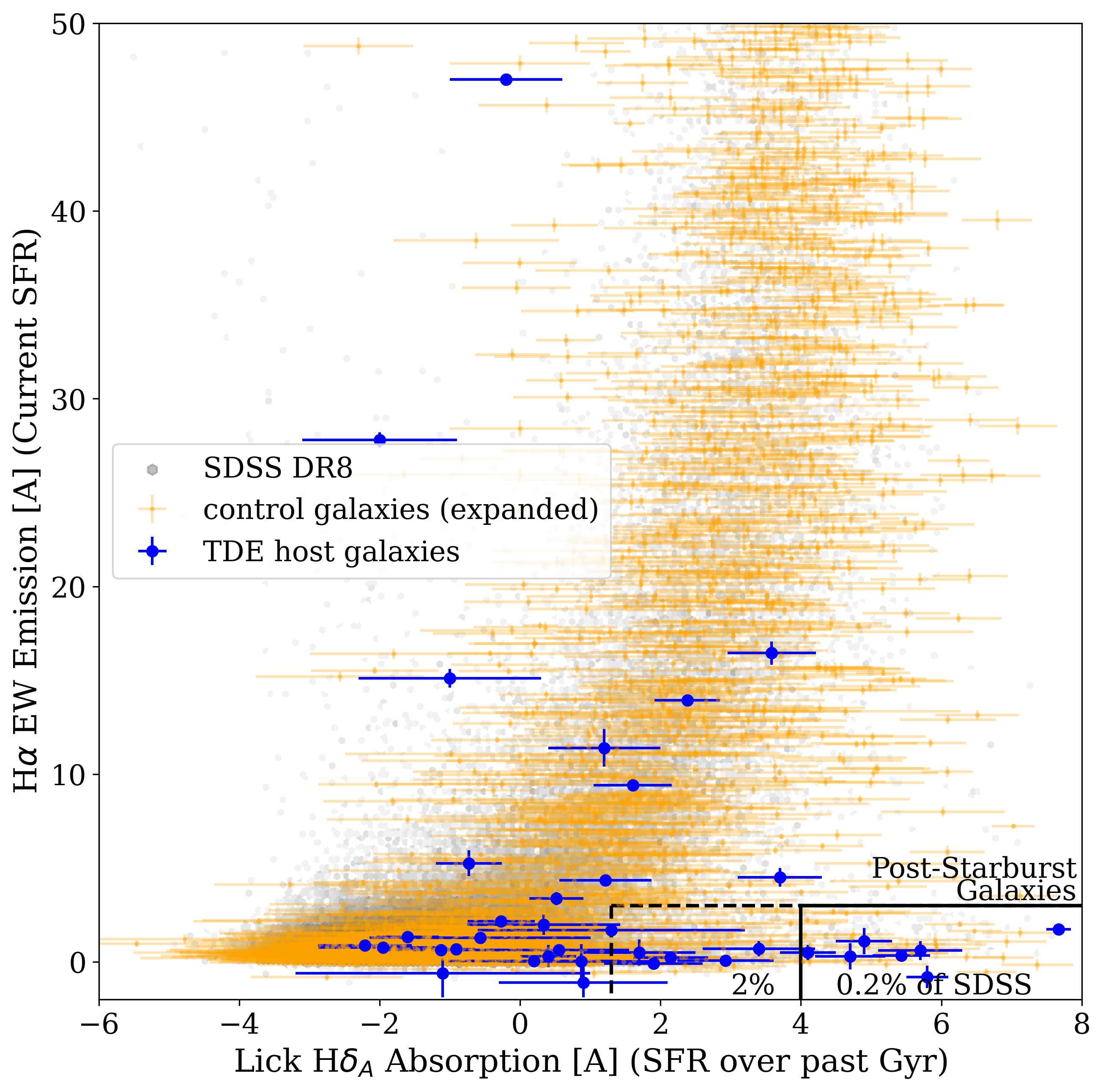}
\end{center}
\caption{The H$\alpha$ emission equivalent width (proxy for current star formation) and Lick H$\delta$ absorption (proxy for star formation in the past Gyr) in the optical spectra of TDE host galaxies (blue points), galaxies from the expanded control sample (orange points), and SDSS DR8 (gray background). The box in the lower right corner bounded by solid black lines represents the region of parameter space where PSB galaxies reside, encompassing 0.2\% of galaxies in SDSS DR8. Extending the box towards lower H$\delta$ values (dashed black lines) encompasses 2\% of galaxies in SDSS DR8, representing QBS galaxies. Not all galaxies from the expanded control sample are shown on this plot; the bounds of the axes are chosen for ease of viewing the TDE host galaxies. PSB and QBS galaxies are overrepresented in our TDE host galaxy sample compared to both SDSS DR8 and the expanded control sample.}
\label{fig:psb_overrepresentation}
\end{figure}

The overrepresentation of PSB and QBS TDE hosts in our sample can be seen in Figure~\ref{fig:psb_overrepresentation}, where we plot the H$\alpha$ equivalent widths and Lick H$\delta_{\rm A}$ indices of the TDE host galaxy sample. The gray points in the background come from spectra in SDSS DR8 \citep{Aihara_2011}, which represent the typical distribution of H$\alpha$ equivalent widths and H$\delta$ indices for local galaxies. The TDE hosts do not show the same distribution as SDSS DR8 or the expanded control sample. The TDE host galaxies are likely to have low H$\alpha$ equivalent widths, and a significant number of them are gathered in the lower right corner with high Balmer absorption. 

Using the galaxy classifications of TDE hosts and comparison galaxies, we can measure the degree to which TDEs are overrepresented in PSB or QBS hosts. We use several methods to account for selection biases, and the resulting overrepresentation factors are summarized in Table \ref{tab:overrepresentation}. Galaxies within H$\alpha < 3$ {\AA} and H$\delta \geq 4$ {\AA} are PSB galaxies because the high H$\delta$ index means that recent star formation is high, while the low H$\alpha$ equivalent width means that current star formation is low. Only 0.204$\pm$0.006\% of galaxies in SDSS DR8 fall in this regime \citep{French_2016}. However, 7 out of the 40 TDE host galaxies for which we have galaxy classifications are labeled as PSB galaxies. We calculate the percentage of PSB galaxies in our sample to be 7/41 = 17$\pm$6\%. The number of TDE host galaxies that we have galaxy classifications for increased to 41 because we are double counting the star forming host of TDE F01004 in our sample (see Section~\ref{sec:F01004_spectra}). The errors quoted are binomial errors. This results in an overrepresentation factor of 17/0.204 = 83$\pm$29 with respect to the background SDSS population.

In recent years, some nuclear transients have been followed up on \textit{because} they were located in PSB galaxies, potentially biasing observed TDE samples towards PSB hosts. To estimate the lower limit on PSB overrepresentation accounting for this search selection effect, we calculate the overrepresentation factor for TDEs in PSB galaxies discovered prior to 2018. Five PSB galaxies in our sample hosted TDEs discovered prior to 2018, which account for 5/41 = 12$\pm$5\% of the sample and result in an overrepresentation factor of $>$ 12/0.204 = 59$\pm$25 with respect to the background SDSS population. 

The expanded control sample of galaxies matched in stellar mass and redshift to the TDE hosts described in Section~\ref{sec:control_samples} can be used to measure the overrepresentation factor accounting for broader selection biases. PSBs make up 31/2,226 = 1.39$\pm$0.25\% of the expanded control sample. This results in an overrepresentation factor of 17/1.39 = 12$\pm$5 with respect to the expanded control sample. Removing the TDEs in PSB galaxies discovered 2018 and later as described above, we measure an PSB overrepresentation factor of $>$ 12/1.39 = 8.6$\pm$3.9. The main source of the difference between the expanded control sample and SDSS DR8 measurements is the lack of higher mass galaxies in the expanded control sample, for which black hole masses are likely $>10^8 M_{\odot}$ and TDEs would not be observed. Though high mass galaxies, especially those in the exponential tail of the Schechter luminosity function \citep{Schechter_1976}, have lower number density throughout the universe, the turnover of the luminosity function (converted to a mass function) is close to the upper stellar mass limit of the expanded control sample. \cite{Weigel_2016} find the turnover to be at $M^{\star} = 10^{10.79} M_{\odot}$, while the expanded control sample has an upper limit of $M_{\star} = 10^{11.2} M_{\odot}$. Thus, there is still a significant number density of high mass galaxies at the turnover of the Schechter luminosity function present in SDSS that we eliminate by using the expanded control sample.

Galaxies within the bounds of H$\alpha < 3$ {\AA} and H$\delta$ $\geq$ $1.3$ {\AA} are labeled as QBS galaxies. These galaxies have still had recent starbursts, but with smaller burst mass fractions, less recent bursts, or longer duration bursts than galaxies with H$\delta > 4$ {\AA}. Only 2.32$\pm$0.02\% of SDSS galaxies fall in this regime \citep{French_2016}. However, 13 out of the 40 TDE host galaxies for which we have galaxy classifications are labeled as QBS galaxies, including 7 PSB galaxies. Double counting the star forming host of F01004 as before, the percentage of QBS galaxies is 13/41 = 32$\pm$7\%, with an overrepresentation factor of 32/2.32 = 14$\pm$3 relative to the background SDSS population. QBS galaxies make up 82/2,226 = 3.7$\pm$0.4\% of the expanded control sample. We calculate an overrepresentation value of 32/3.7 = 8.6$\pm$2.2 for the QBS galaxies with respect to the expanded control sample. 

\begin{table*}[t]
\caption{PSB and QBS overrepresentation factors with respect to different comparison samples and removing TDEs discovered in PSB galaxies in 2018 and later (“adjusted” method). The expanded control sample is matched in $M_{\star}$ and $z$ to the TDE host galaxy sample. Uncertainties come from the binomial error on the number of TDE host galaxies and comparison sample galaxies that are classified as PSB or QBS.}
\label{tab:overrepresentation}
\centering
\begin{tabular}{llc}
\hline
\hline
Population & Method & Overrepresentation Factor \\
\hline
\hline
PSB & SDSS DR8 & $83 \pm 29$ \\
\hline
PSB & SDSS DR8, adjusted & $>59 \pm 25$ \\
\hline
PSB & expanded control sample & $12 \pm 5$ \\
\hline
PSB & expanded control sample, adjusted & $>8.6 \pm 3.9$ \\
\hline
QBS & SDSS DR8 & $14 \pm 3$ \\
\hline
QBS & expanded control sample & $8.6 \pm 2.2$ \\
\hline
\end{tabular}
\end{table*}

Though the galaxies in our sample were not selected for galaxy type (only for the availability of high quality optical spectra), we find that PSB and QBS galaxies are overrepresented among TDE host galaxies. The overrepresentation factors calculated here are lower when the expanded control sample is used as the comparison and when TDEs discovered in PSB galaxies after 2018 are removed in an attempt to account for observational biases. However, our results are broadly consistent to previous work when considering analogous comparison samples \citep{LawSmith_2017, Graur_2018, Hammerstein_2021}. This analysis is limited by the uncertainties due to small number statistics, and unbiased spectroscopic follow-up studies will be an important tool in refining these estimates. 

\section{Analysis}
\label{sec:analysis}

We compare the TDE DTDs calculated in Section~\ref{sec:tderate_age} (Figure~\ref{fig:finalrate_results}) to theoretical DTDs modeling different dynamical and environmental effects in galaxy nuclei. This comparison allows us to discern which effect(s) is responsible for the enhanced TDE rate in PSB galaxies. The observational DTDs displayed in this section are the TDE rate enhancement as a function of burst age for (1) TDE host galaxies that have experienced a significant burst of star formation (left panel of Figure~\ref{fig:finalrate_results}) and for (2) TDE host galaxies that are spectroscopically classified as PSB or QBS (right panel of Figure~\ref{fig:finalrate_results}). The former DTD was normalized with the general control sample and the latter was normalized with the PSB/QBS control sample. The error bars on the observational DTDs in the plots in this section are Poisson errors using the \texttt{Pearson}  method in the \texttt{astropy} function \texttt{poisson\_conf\_interval}. See Section~\ref{sec:caveats_limits} for a discussion of the various caveats related to constructing a DTD.

\subsection{Stellar Overdensities}
\label{sec:overdensity}

\cite{Stone_2018}, \cite{Bortolas_2022}, and \cite{Teboul_2025} propose models where stellar overdensities in galactic nuclei can drive the TDE rate enhancement in PSB galaxies. PSB galaxies \citep{Quintero_2004, Yang_2008} as well as TDE host galaxies \citep{LawSmith_2017, Graur_2018} have been shown to have high central stellar concentrations on scales of the effective radius ($\sim 1$ kpc).  Stars entering the loss cone originate on the far smaller scales of the SMBH radius of influence ($\sim 1-10$ pc), which are harder to probe observationally, although similar overdensities have been seen on $\sim 10-100$ pc scales in PSB TDE hosts \citep{French_2020b}, and even on $\sim 1$ pc scales in the nearest E+A galaxy \citep{Stone_2016b}. If a burst of stars creates an unusually high stellar density in or near the radius of influence, the TDE rate can be enhanced over shorter relaxation timescales, but will decline over time as the overdense nuclear star cluster relaxes and expands. 

\begin{figure*}
\begin{center}
\includegraphics[width=0.49\linewidth]{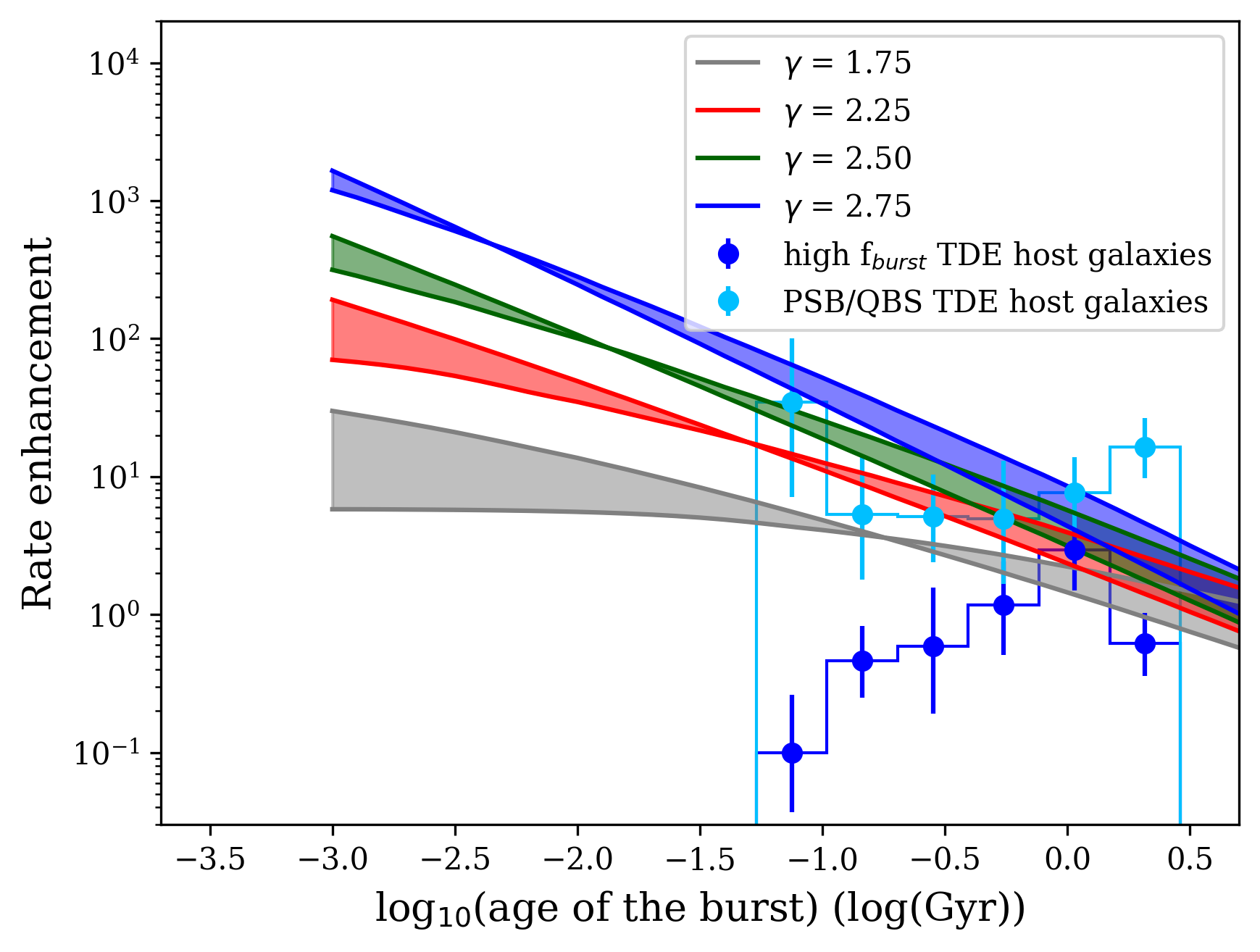}
\includegraphics[width=0.49\linewidth]{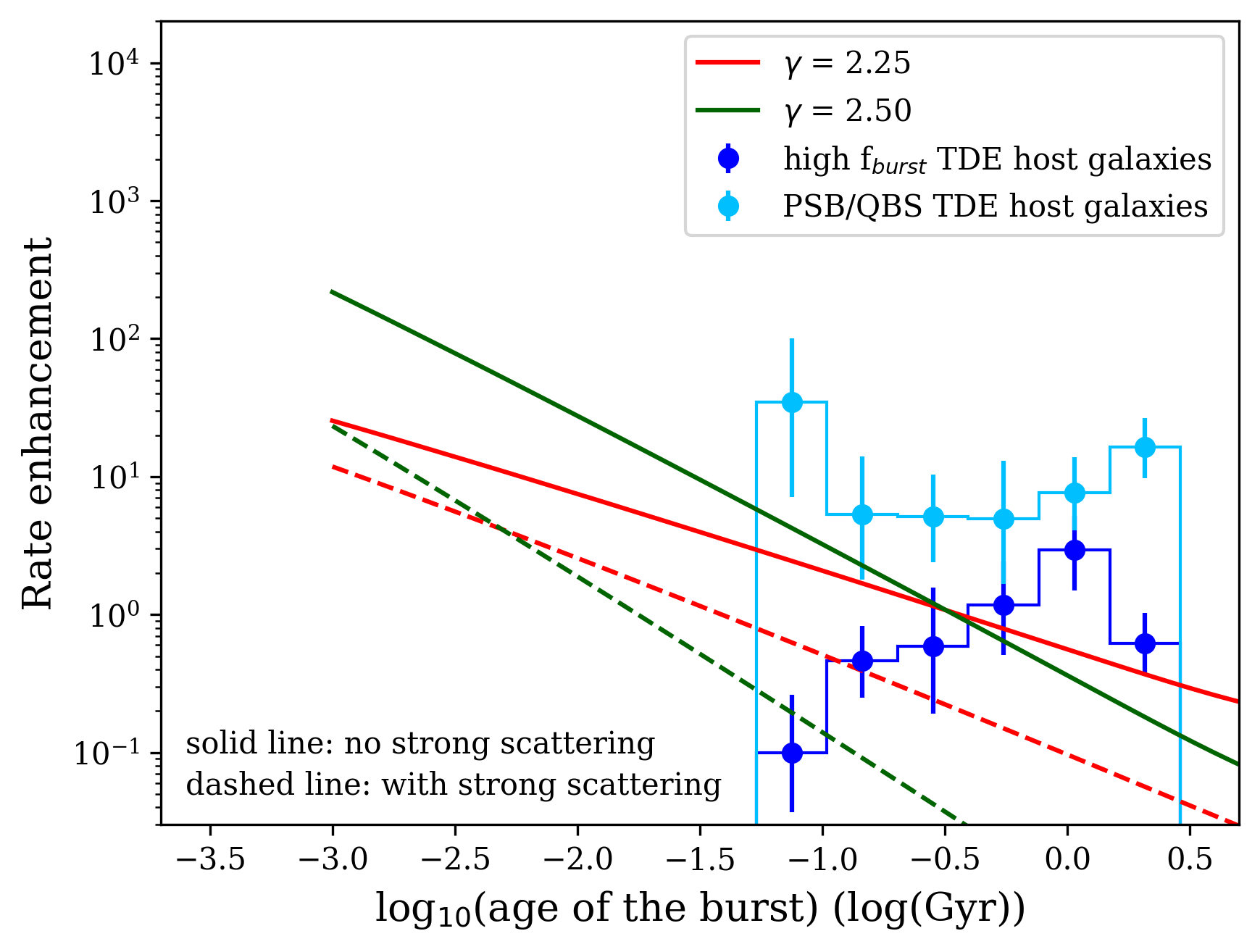}
\includegraphics[width=0.49\linewidth]{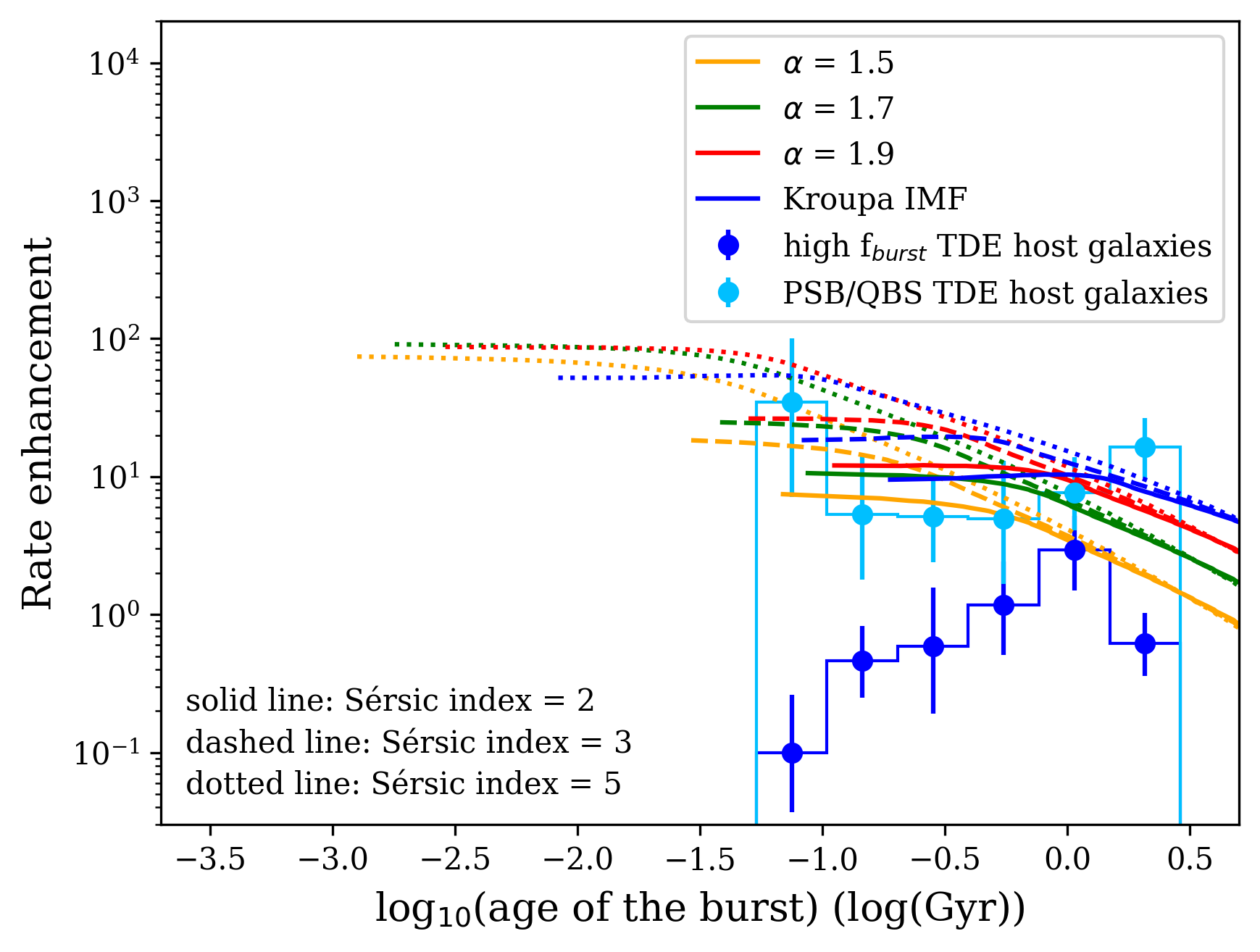}
\end{center}
\caption{Comparisons between our measured DTDs and stellar overdensity models from \citet{Stone_2018} ({\it top left}), \citet{Teboul_2025} ({\it top right}), and \citet{Bortolas_2022} ({\it bottom}) for the TDE rate as a function of burst age. {\it Top left:} \cite{Stone_2018} model a stellar density profile with a range of initial power law slopes. {\it Top right:} \cite{Teboul_2025} model a stellar density profile with different initial power law slopes, though their model incorporates the evolution of the slope with time using \textsc{PhaseFlow} \citep{Vasiliev_2017}. {\it Bottom:} \cite{Bortolas_2022} models a stellar nucleus that includes the effects of mass segregation and a complete IMF, where the blue lines represent a Kroupa IMF, and the orange, green, and red lines represent a top-heavy IMF having a variety of different slopes. The \cite{Stone_2018} and \cite{Teboul_2025} models do not match the shape of the data, given that the models decline as the burst age increases while the data do not present such a decline. The \cite{Bortolas_2022} models have a plateau at intermediate burst ages that matches the plateau in the DTD for TDEs in PSB/QBS galaxies, but the observational DTD increases at old burst ages while the \cite{Bortolas_2022} models decline. This indicates that stellar nuclear overdensities may be a contributor to the TDE rate enhancement in PSB galaxies, especially at early or intermediate burst ages, but cannot explain the complete trend shown in the data.}
\label{fig:stellaroverdensityplot}
\end{figure*}

\cite{Stone_2018} consider two different types of nuclear overdensities: “overconcentrated” nuclei with standard density profile slopes but exceptionally small influence radii, and “ultrasteep” nuclei with standard influence radii but much steeper profiles than the classic Bahcall-Wolf stationary state solution, with 3D density power-law index $\gamma=1.75$ \citep{Bahcall_1976}. They model the DTD resulting from a range of stellar density profiles with initial power-law slopes of $\gamma =$ 1.75, 2.25, 2.5, and 2.75, for a single-mass population of 1 $M_\odot$ stars. These theoretical TDE rates are shown in the top left panel of Figure~\ref{fig:stellaroverdensityplot} on top of the observationally inferred DTDs. The shape of these models, which decline as the burst age increases, does not match the shape of our data, which rises as burst age increases, though the models generally do overlap with the data at a burst age of 1 Gyr. In order to reconcile this model with our data, there would need to be many TDEs in young systems that remain undetected, a possibility which we discuss in Section~\ref{sec:bias}.

\citet{Bortolas_2022} argues that time-dependent models of post-burst TDE rates must account for the disruption of high mass stars sampled from a complete IMF, rather than assuming that the effect of a stellar mass distribution can be accounted for with time-independent correction factors, as is common in steady-state loss cone calculations (e.g. \citealt{Stone_2016}). In particular, \citet{Bortolas_2022} shows that mass segregation will bring higher mass stars closer to the black hole in the nuclear star cluster, enhancing TDE rates in a way that (unlike in the steady state limit) cannot be accounted for with a single-mass loss cone calculation. This may be even more important for starbursting systems, which generally have top-heavy IMFs \citep{Toyouchi_2022}. \citet{Bortolas_2022} simulates the TDE rate for a Milky Way-like galaxy with a Kroupa IMF \citep{Kroupa_2001} and three different top-heavy IMFs having varying slopes. Because these models formally show TDE rate, not TDE rate enhancement, we have converted their predictions to units of TDE rate enhancement by dividing their rates by the average optical TDE rate, as used above \citep{Yao_2023}. The \citet{Bortolas_2022} models are plotted on top of our data in the bottom panel of Figure~\ref{fig:stellaroverdensityplot}. These models feature a plateau in the TDE rate at young/intermediate burst ages, which agrees with the middle bins of the PSB/QBS DTD. However, the models' decline at older burst ages does not agree with our data. 

Stars in nuclear star clusters have a large number of weak (small-angle) scatterings with distant stars that slowly modify each others' orbits. The resulting diffusion through angular momentum space forms the basis of standard loss cone theory \citep{Frank_1976, Lightman_1977, Cohn_1978}. Strong scatterings are traditionally neglected because of their subdominant effect on the bulk transport of stars through angular momentum and energy space \citep{Cohn_1978}. Recently, \citet{Teboul_2024a} have shown that strong scatterings, i.e., pairwise interactions between stars or compact objects that are powerful enough to eject a star, have an outsized impact on highly eccentric stars. Under certain circumstances, ejections resulting from strong scattering interactions can “shield” the loss cone from the incoming diffusive flux of stars, reducing the TDE rate significantly. Loss cone shielding is most relevant when either stars or stellar mass black holes achieve ultrasteep density profiles, with $\gamma \ge 9/4$ \citep{Teboul_2024a}.

\citet{Teboul_2025} construct a series of DTD models with and without strong scattering, using a Kroupa IMF. After implementing strong scattering into their stellar overdensity model, they find that TDE rates are briefly enhanced after the production of an ultrasteep cusp in a burst, but then drop so much that the rates fall below the average rate. These models are shown in top right panel of Figure~\ref{fig:stellaroverdensityplot}, where differently-colored lines display the TDE rate modeled using ultra-steep central profiles. As shown in \cite{Teboul_2025}, when including strong scattering in the model, the magnitude of the rate enhancement is smaller and does not reproduce the overall level of rate enhancement seen in PSB/QBS galaxies. Additionally, the shape of the models (declining as burst age increases) does not match the shape of our data. If there were many TDEs in young systems that we cannot detect (Section~\ref{sec:bias}), the shape of the DTD might agree more with the model, but the overall magnitude of this model's rate is too low.

\subsection{Radial Anisotropies}
\label{sec:anisotropy}

\begin{figure*}
\begin{center}
\includegraphics[width=0.49\linewidth]{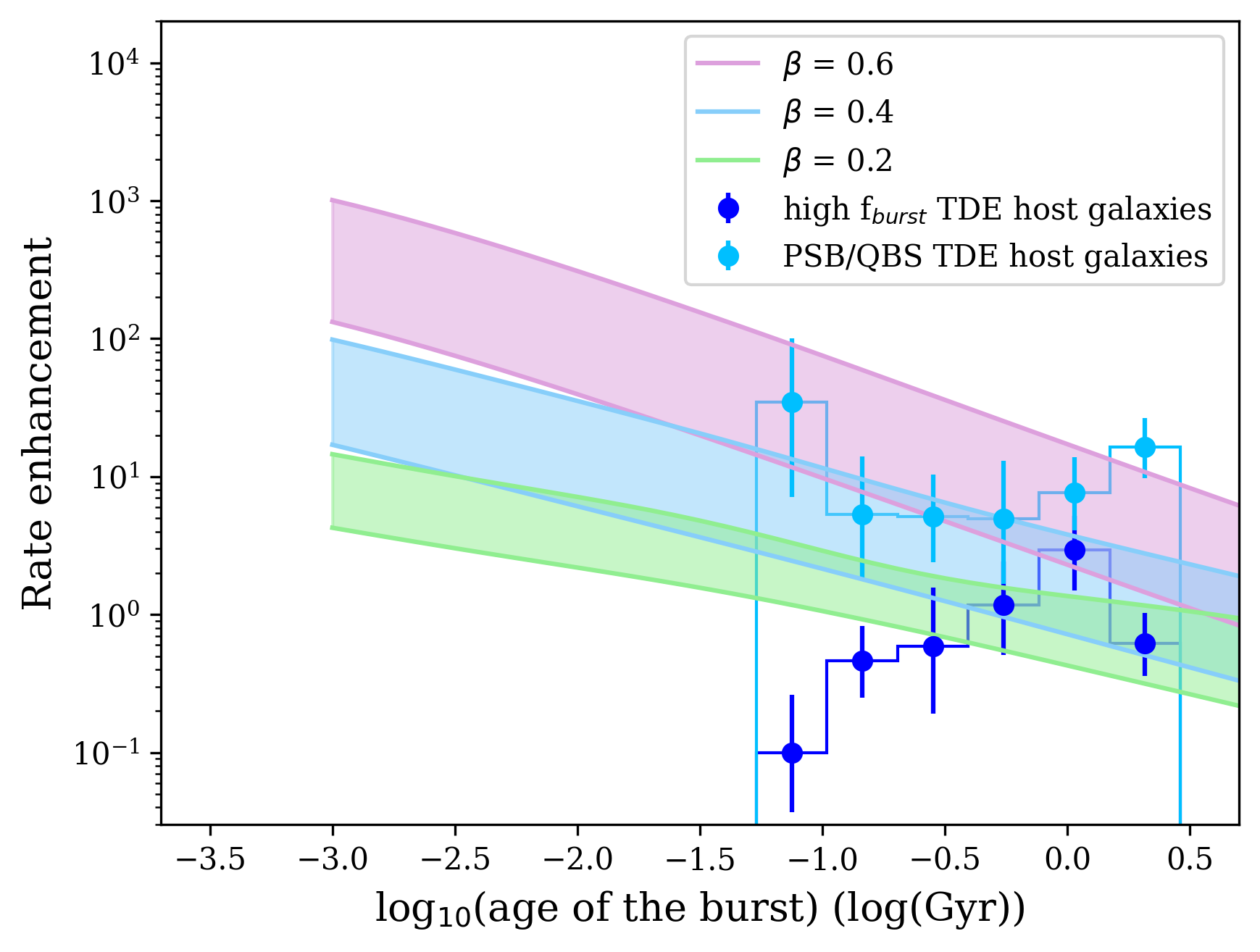}
\includegraphics[width=0.95\linewidth]{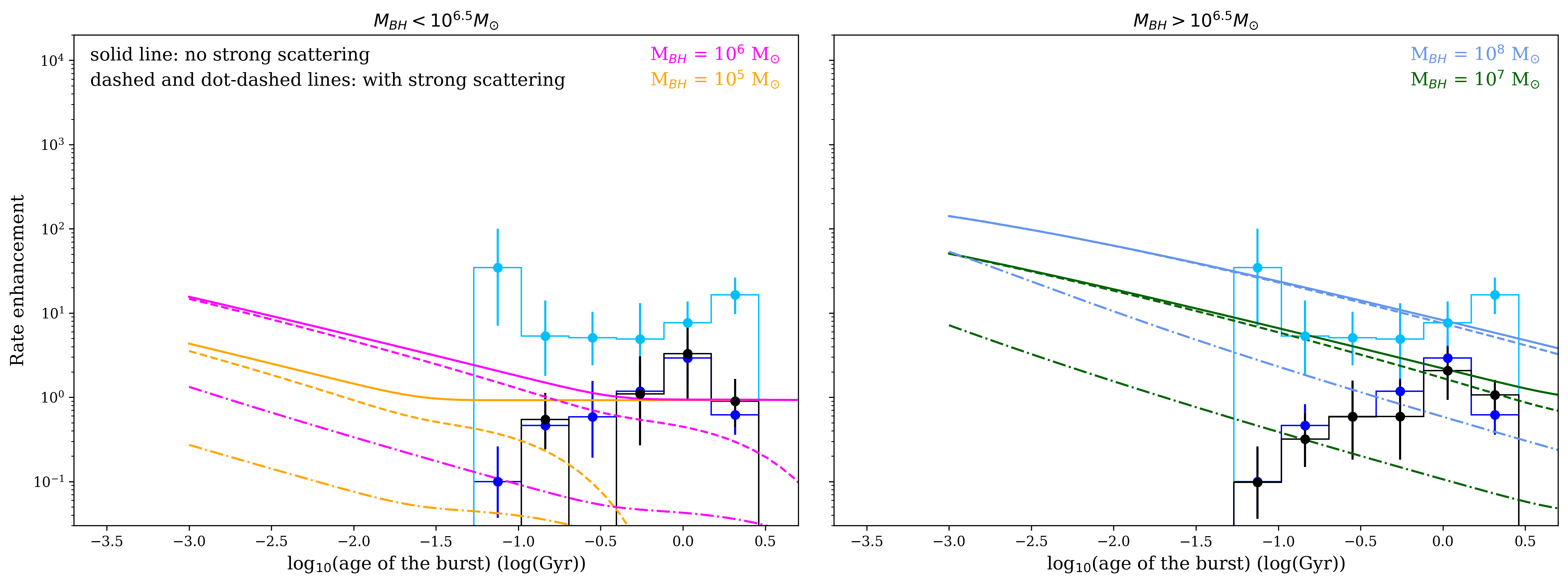}
\includegraphics[width=0.95\linewidth]{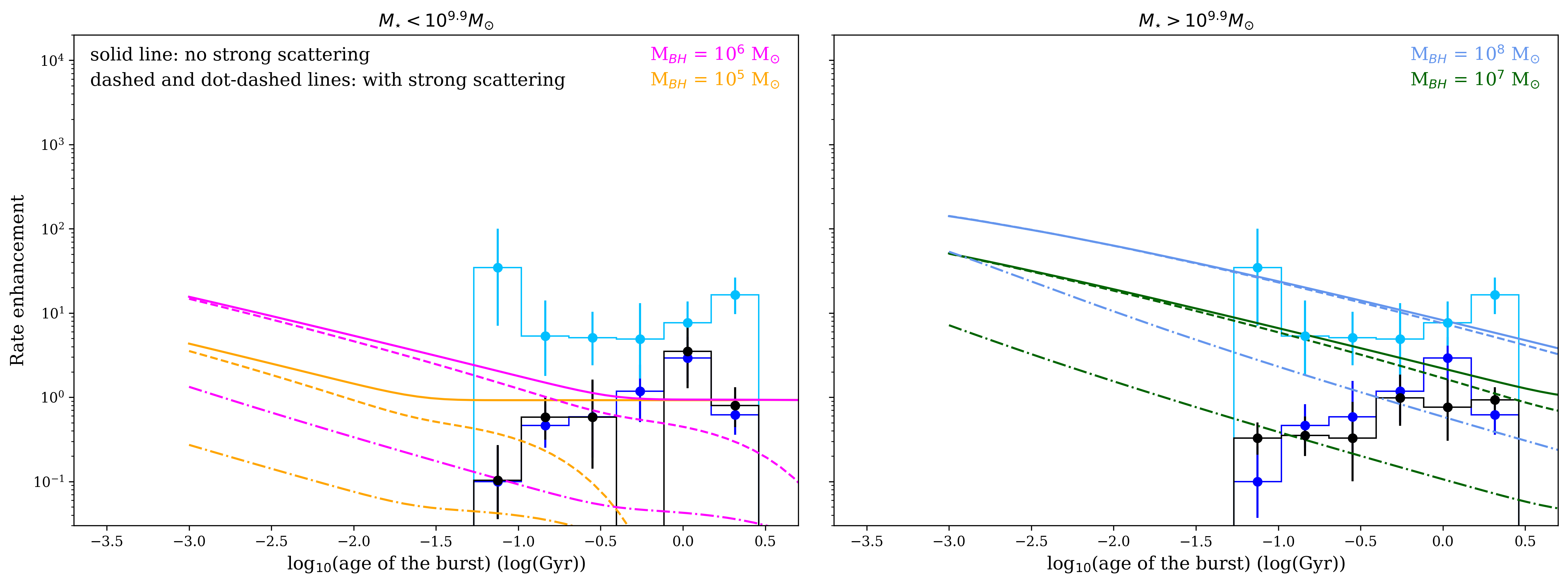}
\end{center}
\caption{Comparisons between our measured DTDs and radial anisotropy models from \citet{Stone_2018} ({\it top}) and \citet{Teboul_2025} ({\it middle, bottom}) for the TDE rate as a function of burst age. {\it Top:} \cite{Stone_2018} model a galaxy nucleus where the stellar orbits have a range of anisotropy values. {\it Middle:} The TDE rate as a function of burst age, using the subset of 23 galaxies with measured black hole masses, split into low ($M_{BH} < 10^{6.5}$ M$_{\odot}$) and high ($M_{BH} > 10^{6.5}$ M$_{\odot}$) black hole mass subsets, is shown in black. {\it Bottom:} The TDE rate as a function of burst age, split into low ($M_{\star} < 10^{9.9}$ M$_{\odot}$) and high ($M_{\star} > 10^{9.9}$ M$_{\odot}$) stellar mass subsets, is shown in black. Theoretical TDE rates from \cite{Teboul_2025} are shown in different colors for disruption around different black hole masses where the anisotropy factor $\beta_0 = 0.5$. In all three panels, the shape of the models (declining as burst ages increase) do not fit the shape of the data (increasing or U-shaped as burst ages increase), though one of the models from \cite{Stone_2018} is wide enough to agree within errors with the PSB/QBS DTD and models from \cite{Teboul_2025} at low black hole masses showing a plateau agrees with the high $f_{burst}$ DTD at intermediate ages. The general disagreement indicates that radial anisotropies cannot completely explain the PSB overrepresentation.}
\label{fig:anisotropyplot}
\end{figure*}

Models from \cite{Stone_2018} and \cite{Teboul_2025} display the TDE rate in a post-burst nucleus with strong radial anisotropies. \cite{Stone_2018} model initial anisotropies of $\beta_0 =$ 0.2, 0.4, and 0.6. These theoretical TDE rates are shown in the top panel of  Figure~\ref{fig:anisotropyplot}. The shape of these models (declining as burst age increases) does not match the shape of our data, though the model with $\beta_0 = 0.6$ agrees within errors with the PSB/QBS DTD as it is quite wide.

\citet{Teboul_2025} revisit the radial anisotropy scenario proposed in \citet{Stone_2018} and additionally include the effects of strong scattering, which can decrease TDE rates if an ultrasteep density component (stars or black holes) exists. These models are shown in the middle and bottom panels of Figure~\ref{fig:anisotropyplot}. Because the models are differentiated based on the SMBH mass, we split our sample into low mass and high mass subsets to draw a better comparison.

We obtain black hole mass measurements for 23 of the TDE host galaxies from \cite{Mummery_2024}, using the peak $g$-band luminosity method. Black hole masses calculated from the plateau luminosity of the TDE light curve at $\nu = 10^{15}$ Hz are also available from \cite{Mummery_2024}, but for fewer galaxies, and the distribution of black hole masses from the peak $g$-band luminosity and black hole masses from the plateau luminosity are not statistically significantly different. Thus, the black hole masses from the peak $g$-band luminosity were used. We split the host galaxies into low mass and high mass black hole subsets at $10^{6.5} M_{\odot}$. 

In order to include \emph{all} galaxies in this subset analysis, we recreate the split into low mass and high mass black hole subsets using the stellar masses available for all the TDE host galaxies in our sample. Thus, we create low mass and high mass stellar mass subsets as an analog for black hole masses. We use the “all, limits” black hole-stellar mass scaling relation from \cite{Greene_2020}, calculated using both early and late type galaxies and including upper limits, to determine which stellar mass best predicts a black hole mass of $10^{6.5} M_{\odot}$. We then split the host galaxies into low and high stellar mass subsets at $10^{9.9} M_{\odot}$.

The DTDs based on TDE host galaxies with low and high black hole (stellar) masses are shown in the middle (bottom) panel of Figure~\ref{fig:anisotropyplot}. These DTDs are both normalized by a subset of the control sample galaxies with low and high stellar masses split at $10^{9.9} M_{\odot}$. The similarity between the shapes of the low mass DTDs and high mass DTDs indicates that the mass of the black hole does not have a strong impact on the TDE rates over this mass range. One difference between the low and high stellar mass DTDs is their predicted TDE rate at 1 Gyr, where low stellar mass TDE hosts have a higher TDE rate at 1 Gyr than high stellar mass TDE hosts. The models from \cite{Teboul_2025}, which decline as burst age increases followed by either a plateau or further decline, broadly do not agree with our data. The models that take strong scattering into account also display a rate enhancement that is too small to be in agreement with our data, especially for the galaxies with lower mass black holes and stellar masses. However, increasing the anisotropy factor would increase the rate enhancement (see Figure 2 in \citealt{Teboul_2025}), so this issue is not insurmountable. Additionally, models that display a plateau in the TDE rate (in particular, the solid pink and yellow lines representing the TDE rate without strong scattering around black holes with $M_{BH} = 10^6 M_{\odot}$, $10^5 M_{\odot}$ respectively) agree with the DTD for the high $f_{burst}$ TDE host galaxies and the TDE host galaxies with low black hole masses at the intermediate burst ages.

\subsection{Supermassive Black Hole Binaries}
\label{sec:smbhb}

\cite{Mockler_2023} and \citet{Melchor_2024} propose that the EKL mechanism \citep{Kozai_1962, Lidov_1962, Naoz_2016} from a SMBH companion, in conjunction with two-body relaxation, may increase the TDE rate in PSB galaxies up to observed levels. Because starbursts can occur after a galaxy merger, the presence of a SMBH companion in PSB galaxies is reasonable. The EKL mechanism increases the eccentricity of stellar orbits to funnel more stars toward the central black hole. \citet{Melchor_2024} calculate TDE rates for different values of the eccentricity of the black hole binary, the mass ratio of the black hole binary, and the central stellar density profile. 

\begin{figure}[t]
\begin{center}
\includegraphics[width=0.95\linewidth]{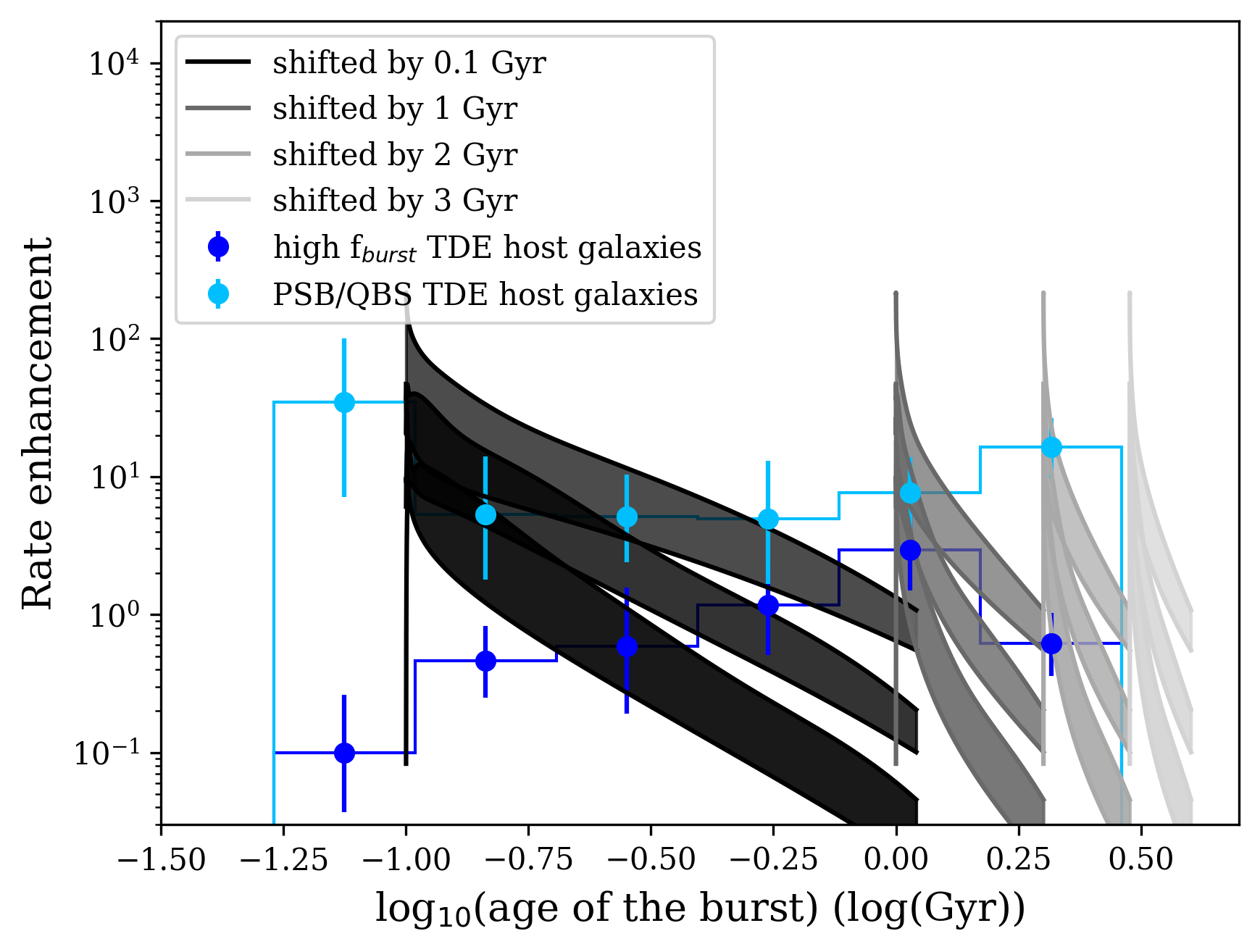}
\end{center}
\caption{Comparison between our measured DTDs and SMBH binary models from \citet{Melchor_2024} for the TDE rate as a function of burst age. These models predict the TDE rate in a system where a SMBH binary is increasing the stars' eccentricities via the EKL mechanism. A selection of models with a black hole binary mass ratio of 0.1, a core-like central stellar distribution, and three different eccentricity values ($e = 0.3, 0.5, 0.7$ from highest to lowest rate enhancement) are shown. To shift the burst ages reported in the simulations from \cite{Melchor_2024} to more physical ages, the black hole binary merger time needs to be considered. The models are shifted by a reasonable range of merger times, where merger times of 1-2 Gyr show better agreement with the data at older burst ages \citep{Kelley_2017, Taffoni_2003, BoylanKolchin_2008, HolleyBockelmann_2025}. Because we expect a range of coalescence times, the shape of the DTD integrated over many systems might more closely reflect the range of coalescence times rather than the DTD for a particular system.}
\label{fig:melchorplot}
\end{figure}

A selection of rates from \citet{Melchor_2024} are shown on top of our data in Figure~\ref{fig:melchorplot}. Because the \citet{Melchor_2024} models formally show TDE rate, not TDE rate enhancement, we have converted their models to units of TDE rate enhancement by dividing their rates by the average optical TDE rate \citep{Yao_2023}. We display these models with a shift in the x-axis representing a range of reasonable coalescence times to account for the time it takes for the black hole binary to merge. This is necessary to map the simulation onto physical timescales, as the \citet{Melchor_2024} simulations do not begin until the black holes are $\sim~$1-2 pc apart. The values chosen for the coalescence times are meant as a qualitative illustration of the delayed coalescence effects, as well as a test of the shape of the modeled DTD. The timescale for black hole binaries to reach this close separation after the initial coalescence of the galaxies and starburst depends on the mass ratio of the two black holes. More unequal mass ratio binaries will have a longer dynamical friction timescale \citep{BoylanKolchin_2008}, and for mass ratios beyond a threshold $\lesssim 0.01$, the dynamical friction time can exceed the Hubble time \citep{Stone_2018}. Using these shifted models, we can see that the models shifted by 1 or 2 Gyr come into agreement with the second-oldest or oldest age bin. This leaves no models that we can use to compare to the younger burst ages, but a range of mass ratios could produce rate enhancements over a range of ages due to the different coalescence timescales \citep{Kelley_2017, BoylanKolchin_2008, HolleyBockelmann_2025}.

Previous research has shown that the mass ratio of the black hole binary should impact the degree of the rate enhancement. \cite{Chen_2011}, though not explicitly calculating the impact of the binary mass ratio on the TDE rates, note that the TDE rate in a merger with an extremely small mass ratio ($q < 0.01$) may be quenched due to precession dominating the Kozai-Lidov effect. Mergers with a more equal mass ratio ($0.01 < q < 0.1$) allow the Kozai-Lidov effect to dominate and the TDE rate can be increased by orders of magnitude during the merger. \cite{Li_2019} use a numerical simulation to determine the effect of the binary mass ratio on the TDE rate after a galaxy merge, and find that the TDE rate increases as the mass ratio becomes more equal, though this effect is more pronounced for TDEs around the secondary black hole than the primary black hole. \cite{Melchor_2024} investigate TDE rates at two different mass ratios ($q=0.1$ and $q=0.01$), but the rates do not appear to have any significant increase or decrease as a function of the mass ratio. Instead, the choice of stellar density profile (core versus cusp) has a greater impact on the rate, with models using the cusp distribution having higher overall rates (also seen in \citealt{Mockler_2023}).

\subsection{Disk Interactions}
\label{sec:agndisk}

Past works have proposed that TDE rates may be enhanced in galactic nuclei hosting AGN. These rate enhancements can occur due to the gravitational quadrupole moment of a massive AGN disk \citep{Karas_2007, Kaur_2025}, the capture of pre-existing stars onto embedded and inspiraling orbits \citep{Wang_2024b}, or the production of a tightly bound, dense stellar population via Toomre instability in the AGN gas \citep{Wang_2024}. In this section we focus on the predictions of the \citet{Wang_2024} model in particular, as these are tailored to explain the PBS/QBS association of observed TDE populations, though other effects may operate during the AGN phase itself. 

\begin{figure}
\begin{center}
    \includegraphics[width=0.95\linewidth]{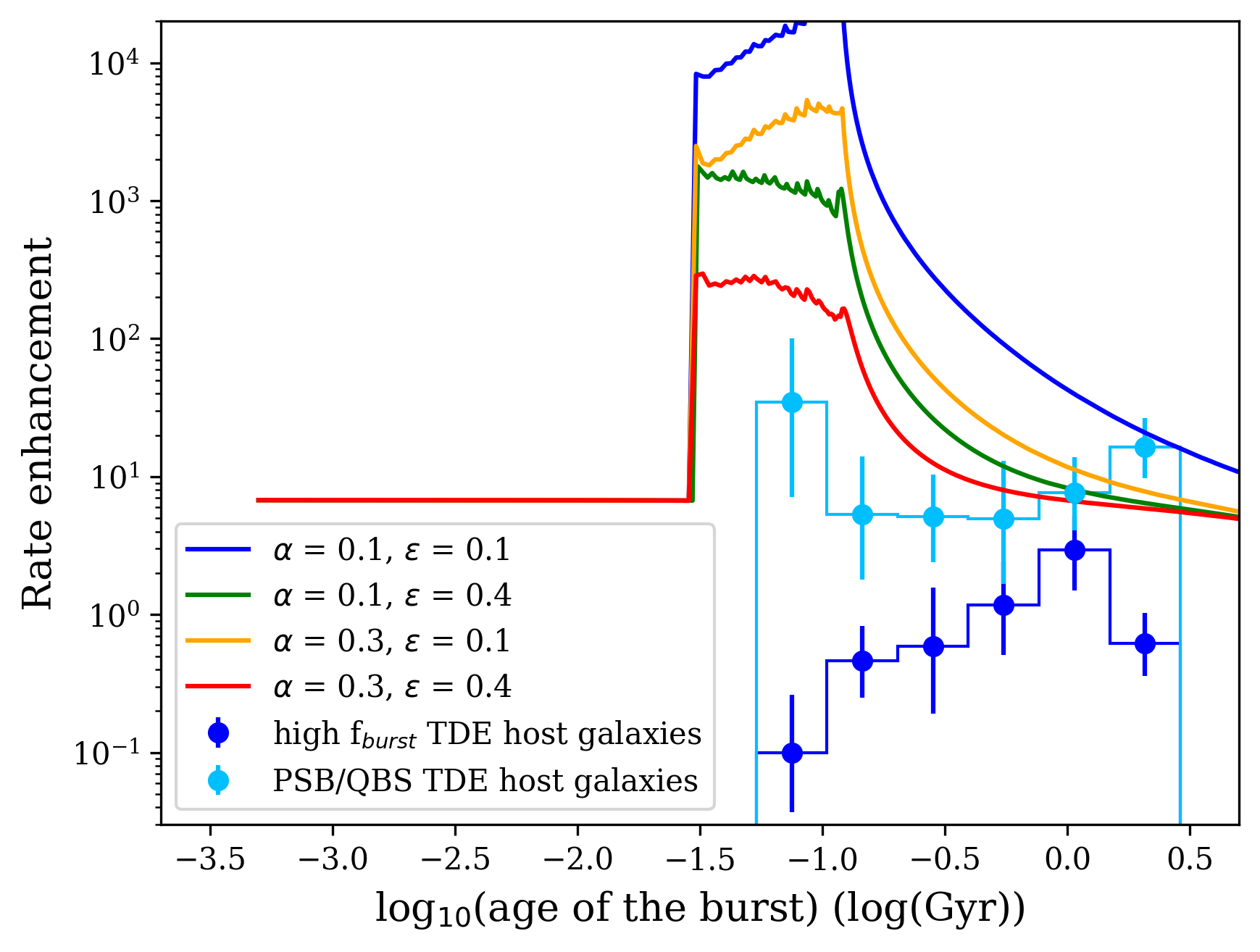}
\end{center}
\caption{Comparison between our measured DTDs and AGN disk models from \citet{Wang_2024} for the TDE rate as a function of burst age. The theoretical models predict the TDE rate in a system where interactions with an AGN disk briefly increase the TDE rate. \cite{Wang_2024} vary the viscosity parameter $\alpha$ and the efficiency parameter $\epsilon$. The models do not match either DTD, but they do show a brief peak in the TDE rate at the same time as the PSB/QBS DTD. However, the magnitude of the models' peak is too high compared to the data and the duration of the peak is too long. Additionally, the shape of the models do not match either DTD at older burst ages, suggesting that a disappearing AGN disk may be a contributor to the TDE rate enhancement in PSB galaxies at early burst ages, but cannot completely explain the trend shown in the data.}
\label{fig:wangplot}
\end{figure}

More specifically, \citet{Wang_2024} propose that TDE rates are enhanced in PSB galaxies via the appearance and disappearance of an AGN disk, which may be more likely to be found in galaxies that have experienced a recent starburst \citep{Hopkins_2008, Schawinski_2009, Kaviraj_2011}.  Fragmentation in such disks via Toomre instability creates a dense population of stars that may scatter into the loss cone at elevated rates. The authors find that TDE rates are enhanced by 2-3 orders of magnitude during the stage when the AGN disk is transitioning into quiescence, with models specifically examining a black hole mass of 10$^{6.75} M_{\odot}$. Because the \citet{Wang_2024} models formally show TDE rate, not TDE rate enhancement, we have converted their models to units of TDE rate enhancement by dividing their rates by the average optical TDE rate \citep{Yao_2023}. Because the baseline rate used in \cite{Wang_2024} is rather high ($\sim2 \times 10^{-4}$ yr$^{-1}$ gal$^{-1}$), their baseline rate enhancement is also high. Figure~\ref{fig:wangplot} plots the models from \cite{Wang_2024} on top of our data for the TDE rate enhancement as a function of time since a burst of star formation. The four models shown correspond to different combinations of two disk parameters, the viscosity parameter $\alpha$ and the efficiency parameter $\epsilon$, governing the conversion from rest mass energy from star formation to radiation. Though their models show a peak in the TDE rate enhancement at the same time as the youngest age bin in the PSB/QBS DTD, the overall trend in the models (peak and then decline) does not match the overall trend in either DTD (increase with burst age). However, it is possible that adjusting the time of the burst of star formation with respect to the transition of the AGN disk to quiescence could push the peak in the TDE rate towards older bursts. We explore the potential biases that AGN may introduce in sample selection in Section~\ref{sec:AGN}.

\section{Discussion}
\label{sec:discussion}

\subsection{Comparison to Previous Work}
\label{sec:comparison_french2018}

Several previous studies have modeled the SFH of TDE host galaxies. \citet{French_2017} use a combination of photometry and spectroscopy to model the SFHs of eight TDE host galaxies using the method of \citet{French_2018}. However, there are some differences in the methods used in this work and in \cite{French_2017}. For example, optical, FUV, and NUV photometry is employed in \cite{French_2017} while no photometry is used in this work. Instead of modeling the entire spectrum, \citet{French_2017} use the Lick indices to parameterize SFH-sensitive features. In contrast, \textsc{Bagpipes} allows us to fully model the optical spectrum, which ensures that this work is more robust against contamination in the UV photometry from the TDE itself. However, the use of photometry by \citet{French_2017} means that those results consider a physically larger region of the galaxy than the inner $\sim$kiloparsec traced by the spectrum. Secondly, \cite{French_2017} consider that PSB galaxies may have two bursts and allows for such in their SFH fitting, which optimizes the method for PSB galaxies. By only allowing for one burst in our model, we are better able to fit the non-PSB galaxies and can more accurately recover the burst mass fraction. Finally, \cite{French_2017} model the burst with an exponential function while we model the burst as a double power law function. We used the double power law to model the burst in lieu of the exponential function because the double power law is better able to fit current star formation that is not burst-like. Results from \cite{French_2017} show that there is a trend towards older burst ages as the burst mass fraction of the galaxies increases, which we find generally recreated in Figure~\ref{fig:age_vs_massfrac} for galaxies with a burst mass fraction $>1\%$.

\cite{Pursiainen_2025} study stellar populations at sub-kpc scale around the nuclei of 20 TDE host galaxies using MUSE integral field spectroscopy. They use the spectral synthesis code \textsc{Starlight} to determine the stellar population present around the central black hole. This model stellar population is constructed out of simple stellar populations, which allow them to find the age(s) of the stars around the black hole. They find that the youngest stellar populations in the nuclear region of these galaxies is $\sim$1 Gyr old (see their Figure 6). This finding agrees with our result that TDEs are more likely to happen in galaxies where a burst of star formation happened $\sim$1 Gyr ago. 

\cite{Newsome_2025} present spectra taken by the Space Telescope Imaging Spectrograph (STIS) aboard the Hubble Space Telescope of the inner 0.2" of four TDE host galaxies, corresponding to a nuclear region on the order of 10--100 parsecs. One of the galaxies studied is the host galaxy of ASASSN-14li. They use \textsc{Bagpipes} to extract the hosts' SFH and follow the same SFH fitting setup from \cite{French_2017}. In the host of ASASSN-14li, they find evidence for a radial age gradient in the stellar distribution around the central black hole where younger stars are located at smaller radii.

\subsection{Caveats and Limitations}
\label{sec:caveats_limits}

\subsubsection{Non-uniform Sample}
\label{sec:nonuniformsample}

The selection of galaxies in this sample is imperfect. Many TDE samples are plagued by the difficulty of distinguishing rare TDEs from relatively common AGN, leading to the exclusion of nuclear transients from hosts with AGN signatures (i.e. favoring purity over completeness). The non-uniformity of the sample selection in this paper also means that the spectra for the TDE host galaxies come from multiple different sources and do not cover the same regions of each galaxy. There are ultimately many selection functions at play in constructing this sample, and more TDE observations from larger-scale transient surveys are needed to rectify this.  

\begin{figure}
\begin{center}
    \includegraphics[width=0.95\linewidth]{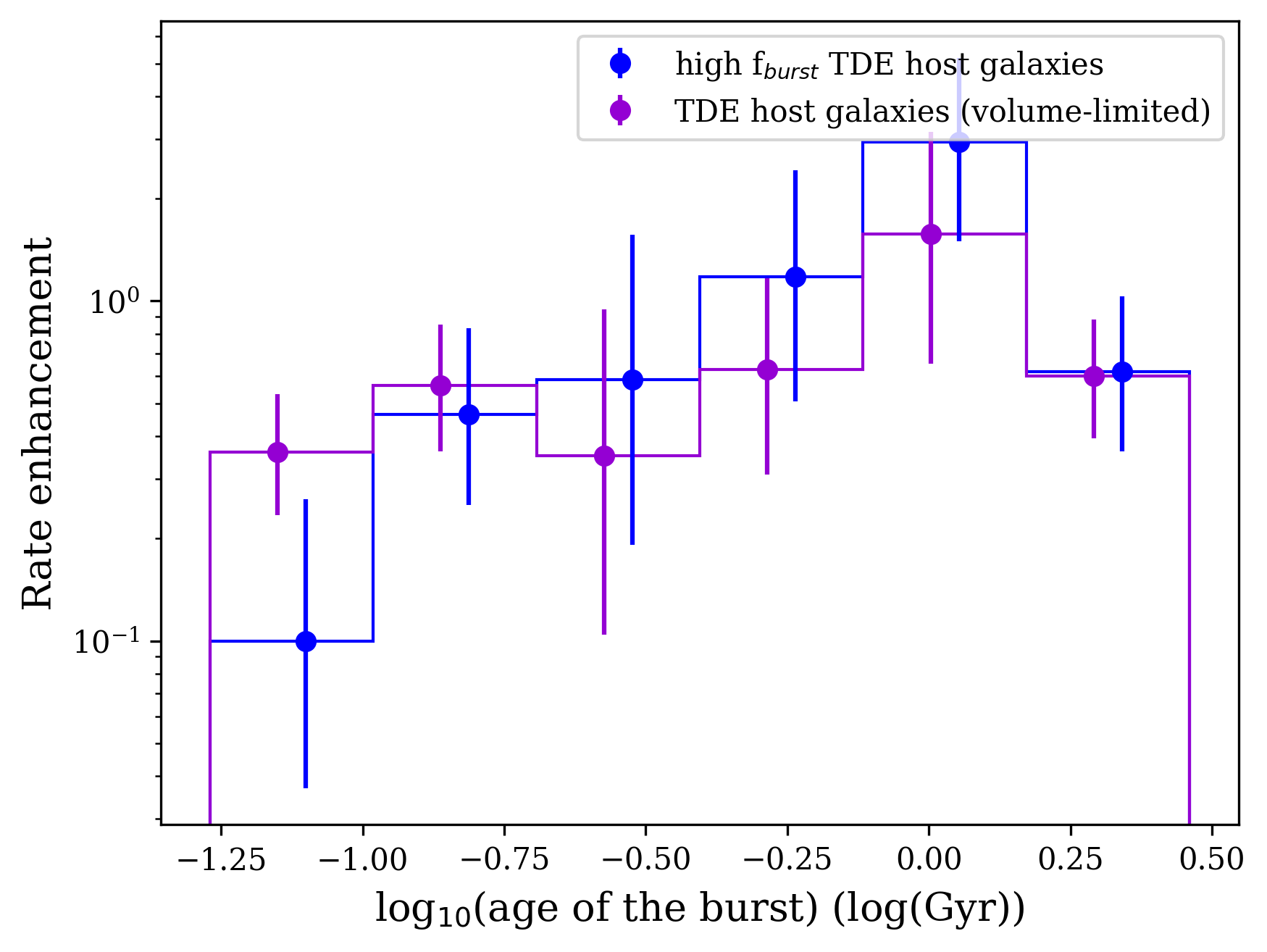}
\end{center}
\caption{The TDE rate as a function of burst age for a volume-limited subset of TDE host galaxies normalized by a volume-limited subset of the control sample. There is no burst mass fraction cutoff implemented for this subset. Also shown is the DTD for galaxies we determine to have experienced a significant burst. The DTD for the volume-limited subset of TDE host galaxies shows similarities to the DTD of galaxies that have experienced a significant burst of star formation, namely a peak in the TDE rate at a burst age of $\sim$1 Gyr, though the rate at early burst ages shows more of a plateau.}
\label{fig:TDErate_op4_volumelim}
\end{figure}

To test some sources of potential bias in our sample due to non-uniformity, we show the TDE rate as a function of time since burst in a volume-limited subset of TDE host galaxies in Figure~\ref{fig:TDErate_op4_volumelim}. This rate is normalized by a volume-limited subset of the control sample. The galaxies used to calculate this DTD are within $0.0 < z < 0.1$ and $9.6<\log_{10}(M_{*}/M_{\odot})<11.2$ (see Figure~\ref{fig:stelmass_vs_z_TDE_control}) and includes galaxies with both high and low burst mass fractions. Also shown is the “fiducial” rate of TDE host galaxies that have experienced a significant burst of star formation from the left panel of Figure~\ref{fig:finalrate_results}. The volume-limited subset has a less pronounced peak at $\sim$1 Gyr but still matches the overall shape of the fiducial DTD. The bins with the youngest burst age in the volume-limited DTD are higher compared to the fiducial DTD because they include some star forming galaxies with low burst mass fractions. See also Appendix~\ref{appendix:c} where we construct a DTD using only host galaxies of TDEs that displayed broad lines, a more stringent method of TDE classification.

\subsubsection{Ambiguous TDE Classification}
\label{sec:F01004_bias}

The classification of F01004 as a TDE has been disputed, and the conclusion that two different TDEs occurred in the galaxy is not certain. However, the results of this paper would not change significantly if we only included it once in our sample or removed it entirely. According to the \textsc{Bagpipes} results, we do not count F01004 as a galaxy that has experienced a recent burst of star formation, because even though it has a very large burst mass fraction, the falling slope index is very small, indicating that the recent component of star formation was not actually burst-like. Additionally, it is classified as a star forming galaxy according to its H$\alpha$ equivalent width and H$\delta$ index. Thus, F01004 is not included in either observational DTD. If we did not double-count F01004 when calculating the PSB and QBS overrepresentation values, the values would increase, but only by small amounts within the error bars already established. Thus, our results are not affected by our choice of interpretation for F01004.

\subsubsection{Limitations to \textsc{Bagpipes}}
\label{sec:bagpipes_limitations}

There is a lower limit to the age of the burst that \textsc{Bagpipes} can return due to the lifetimes of the youngest stars, which means there are no data points at the earliest burst times to compare to the models. For example, \textsc{Bagpipes} finds no galaxies with burst ages younger than 50 Myr (Figure~\ref{fig:age_vs_massfrac}). However, in the case of both DTDs, the galaxies that \textsc{Bagpipes} finds with the youngest bursts are control galaxies, not PSB TDE hosts, suggesting that even though \textsc{Bagpipes} may struggle with fitting galaxies with young bursts, there are still comparatively more non-TDE hosts than TDE hosts with extremely young bursts. 

\subsubsection{Potential TDE Emission Contamination}
\label{sec:tdeplateauemission}

TDEs have been found to have long-lived UV-bright plateaus from disk emission in some cases \citep{vanVelzen_2019, Mummery_2024}, and theoretical models suggest that these plateaus can last for decades or even centuries \citep{Alush_2025}. If this flux is significant relative to the host contribution, it could affect the stellar population synthesis fitting. By only including spectra taken at least a year after the TDE (or prior to the TDE), we avoid contamination from the early-time TDE emission. We can guard against late-time contamination by using late-time light curves of TDEs to estimate the potential contribution of the TDE plateau to the galaxy spectra (see Figure E1 of \citealt{Mummery_2024}). More specifically, we estimate the late-time $g$-band plateau flux and compare that to the average galaxy flux in the spectra used in this analysis. 20 TDEs presented in \cite{Mummery_2024} with late-time light curves are also in our TDE sample. We find that the median plateau contribution in the $g$-band is 5\%. This is smaller than the calibration uncertainty included in the spectral fit conducted by \textsc{Bagpipes}. Additionally, \cite{Newsome_2025} found the late-time TDE disk contribution to be $\sim$10\% when fitting a spectrum of ASASSN-14li that covers the innermost portion of the nucleus, on the order of 10 pc. Because the galaxy spectra used in this paper were taken with larger apertures than the aperture used to take spectra for \cite{Newsome_2025}, we are more confident that the potential disk plateau contribution is 5\% or less. However, the inclusion of TDE contamination would bias the burst ages of our host galaxies younger, meaning that the true burst ages would be even older than reported by \textsc{Bagpipes} here. If the true ages are older, this would make our DTDs even more discrepant with the dynamical models discussed in Section~\ref{sec:analysis}.

\subsection{Potential Biases Against Young Systems}
\label{sec:bias}

\subsubsection{Dust}
\label{sec:dust}

One proposed solution to the PSB overrepresentation among TDE host galaxies is that TDEs in heavily dust-obscured galaxies might not be observable in optical/UV wavelengths \citep{Roth_2021, Masterson_2024}. \cite{Masterson_2024} select TDEs from IR transients, most of which had no transient optical counterpart. These obscured TDEs do not show a PSB preference, suggesting the true PSB overrepresentation is lower than previously reported. Furthermore, \cite{Reynolds_2022} find a high rate of obscured nuclear transients in luminous IR galaxies, especially in starburst/ULIRG galaxies. This suggests that there might be an obscured population of transients in dusty galaxies. \cite{Roth_2021} attempt to predict the host galaxy properties of detected TDEs. They predict that most TDEs in star forming galaxies cannot be detected by ZTF, and that this obscuration can partially but not completely explain the preference for TDE host galaxies to lie in the Green Valley. Others have concluded that nuclear obscuration cannot explain the observed TDE host galaxy preference \citep{Dodd_2023}. \cite{Dodd_2023} do not find an excess of IR transients in starburst galaxies, concluding that we are not missing TDEs that occur in starforming galaxies due to dust obscuration. 

\begin{figure*}
\begin{center}
\includegraphics[width=0.49\linewidth]{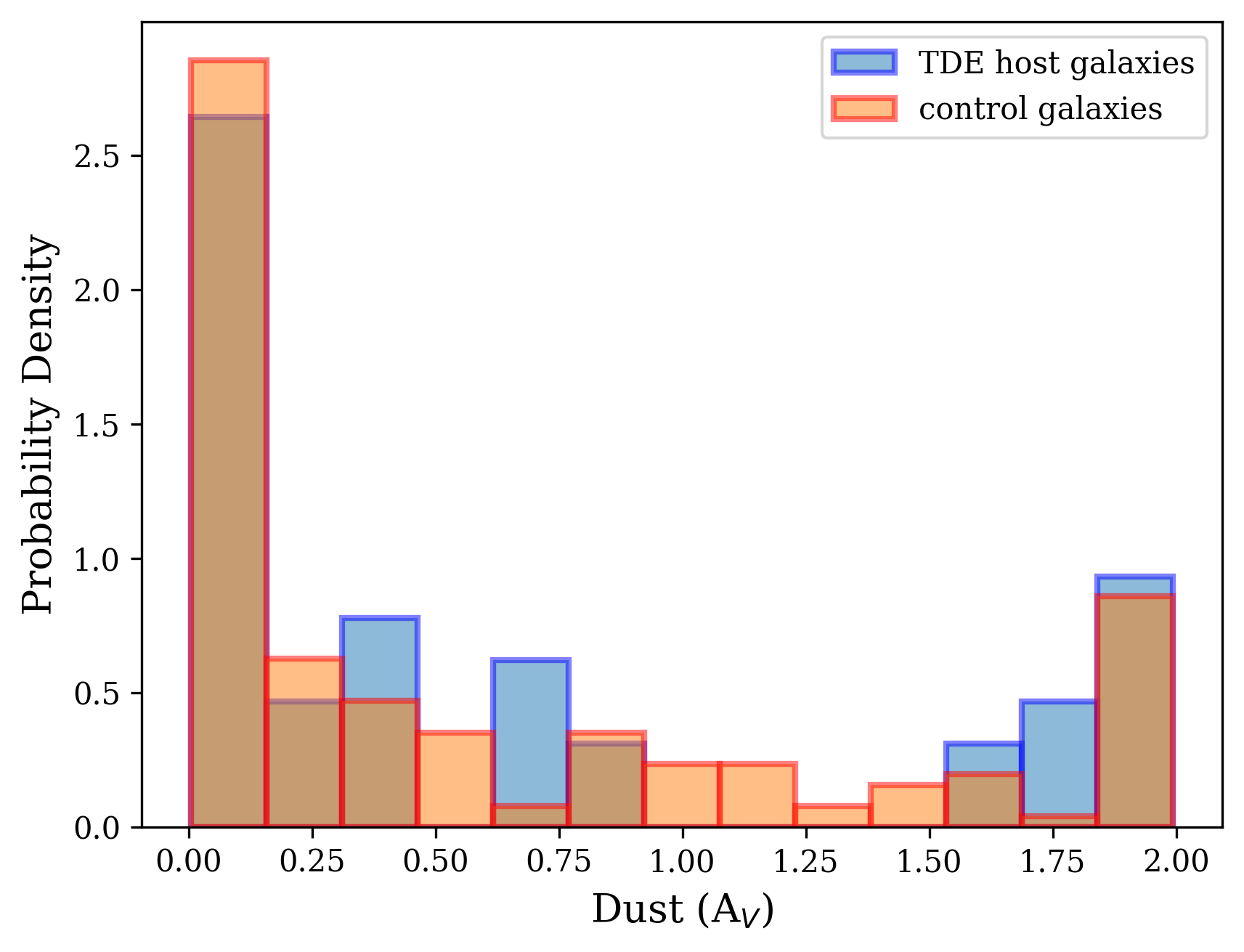}
\includegraphics[width=0.49\linewidth]{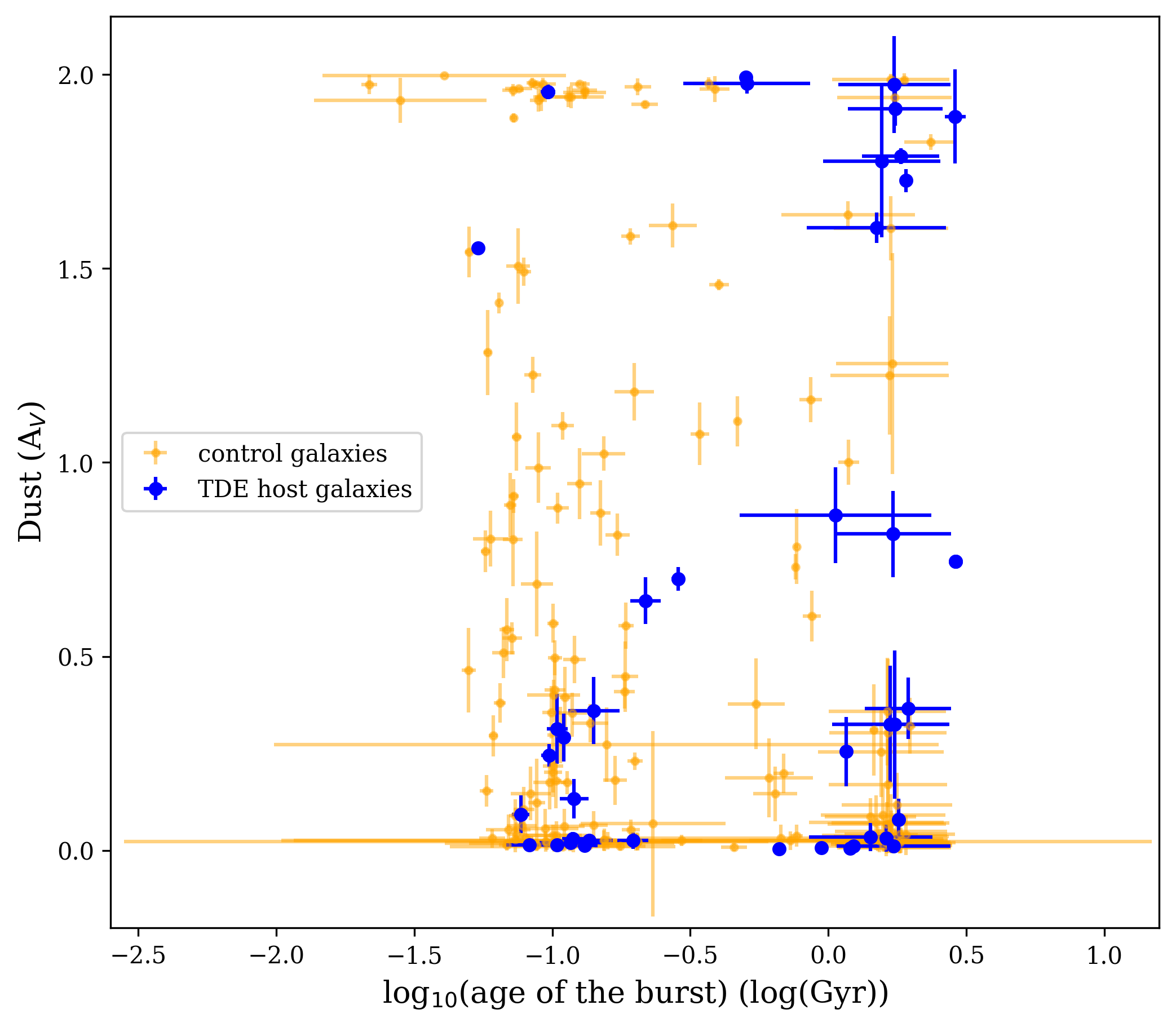}
\end{center}
\caption{{\it Left:} Dust attenuation for the TDE host galaxy sample and the sample of control galaxies. The distributions do not show any statistically significant differences, showing that TDE host galaxies are not preferentially dust-free, which could lead to biases. {\it Right:} Dust attenuation versus burst age for the TDE host galaxy sample and the sample of control galaxies. Dust and burst age are slightly negatively correlated for the control sample and slightly positively correlated for the TDE host galaxy sample. KS tests of the $A_V$ values for galaxies with young and old bursts show that the $A_V$ distributions of TDE host galaxies and the control sample galaxies does not depend on burst age, which indicates that we are not missing a large population of TDE hosts with young burst ages and high $A_V$. Of note are the large error bars on some of the control galaxies. In these cases, \textsc{Bagpipes'} results indicate that these galaxies have not had a burst. \textsc{Bagpipes} is forced to return values for such a “burst” anyway, but because the burst is so small/did not happen, the error bars on the burst age are very large.}
\label{fig:dust}
\end{figure*}

As part of our spectral fitting with \textsc{Bagpipes}, we revisit the issue of preferential obscuration by constraining the dust attenuation $A_V$. Values of $A_V$ as calculated by \textsc{Bagpipes} are shown in the left panel of Figure~\ref{fig:dust}, a probability density histogram which shows that the distribution of dust attenuation values among the TDE host galaxy sample and the control sample is roughly the same. A KS test confirms this; the resulting p-value of 0.64 indicates that we cannot distinguish these distributions from one another. 

The right panel of Figure~\ref{fig:dust} shows the values of $A_V$ versus the burst age for the TDE host galaxies and the control galaxies. We investigate whether there is evidence for a “missing” population of dusty TDE host galaxies at young burst ages. First, we divided both galaxy samples into “young burst” and “old burst” subsets, split at a burst age of $10^{-0.5}$ Gyr, the approximate median age bin. Then, we performed two-sample KS tests on dust attenuation values of the “young burst” and “old burst” TDE host galaxies, and the “young burst” control galaxies and “young burst” TDE host galaxies. None of the resulting p-values are significant,\footnote{The KS test comparing dust attenuation values of the “young burst” and “old burst” TDE host galaxies returns a p-value of 0.057. The KS test comparing dust attenuation values of the “young burst” control galaxies and “young burst” TDE host galaxies returns a p-value of 0.194.} which suggests dust is not disproportionally obscuring TDE host galaxies with young bursts of star formation. There are no significant shifts that we would expect if we are missing high $A_V$ hosts with young burst ages in the TDE sample. In particular, we do not see evidence that the TDE hosts with young burst ages are less dusty than control sample galaxies with young burst ages. 

We must add a caveat that the dust in the nucleus may not be coupled with galaxy-scale dust (which \textsc{Bagpipes} is measuring). However, one would expect the dust in the nucleus to be more strongly correlated with dust on a galaxy-wide scale than with the PSB age. If nuclear dust and galaxy-wide dust are \textit{not} correlated, nuclear dust and galaxy-wide PSB age are also likely not correlated. If this is not true, then the \textsc{Bagpipes} results may not be an effective measure of dust on relevant scales.

To investigate whether the PSB/QBS TDE hosts have a different distribution of dust attenuation compared to the PSB/QBS control sample, which could cause bias in the DTD in the right panel of Figure~\ref{fig:finalrate_results}, we consider trends in $A_V$ versus burst age for the PSB/QBS TDE host galaxies and the PSB/QBS control galaxies. We do not measure a distinguishable difference in the $A_V$ distribution, and we do not measure a significant correlation between $A_V$ and burst age for either the PSB/QBS TDE hosts or PSB/QBS control sample. 

\begin{figure*}
\begin{center}
\includegraphics[width=0.49\linewidth]{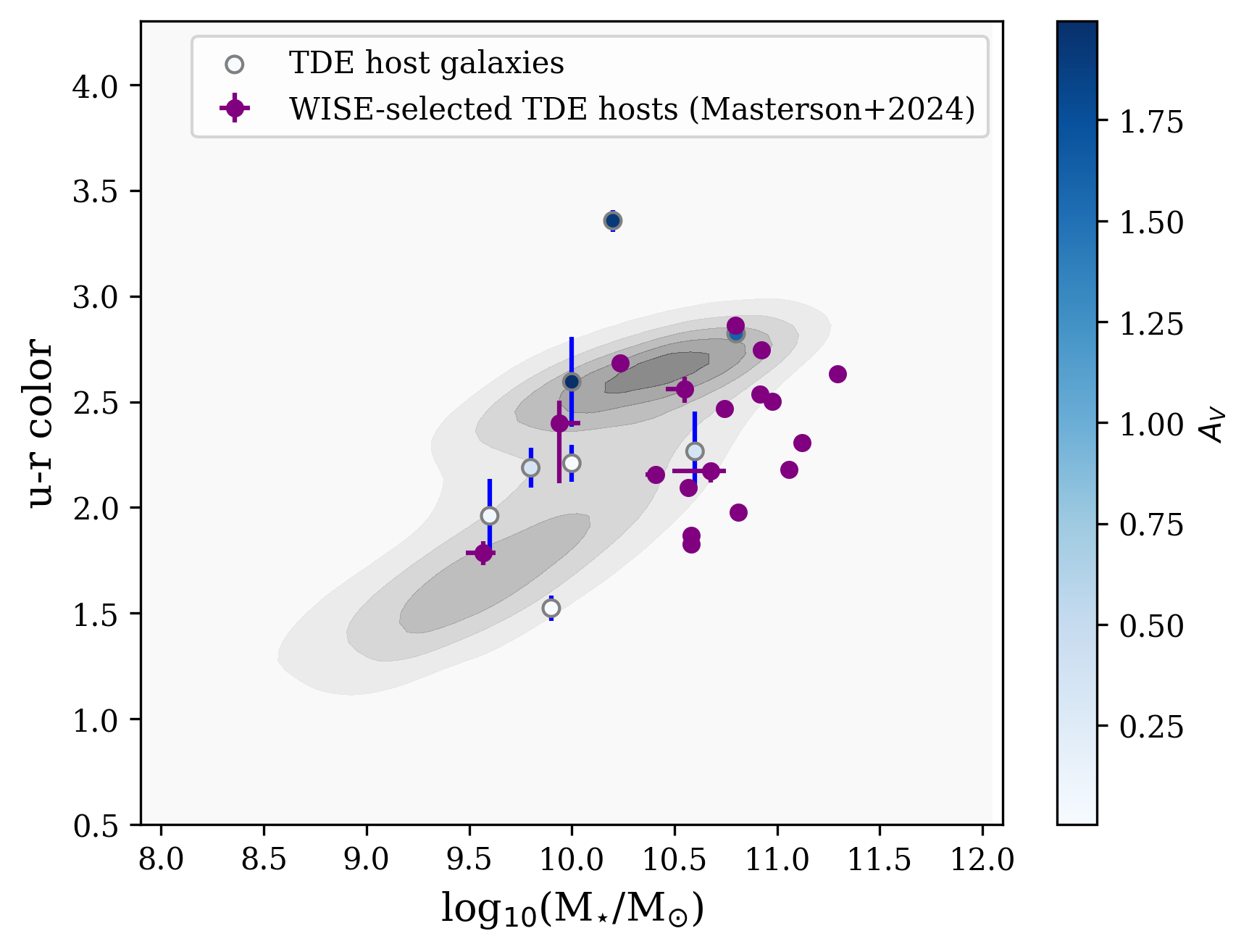}
\includegraphics[width=0.49\linewidth]{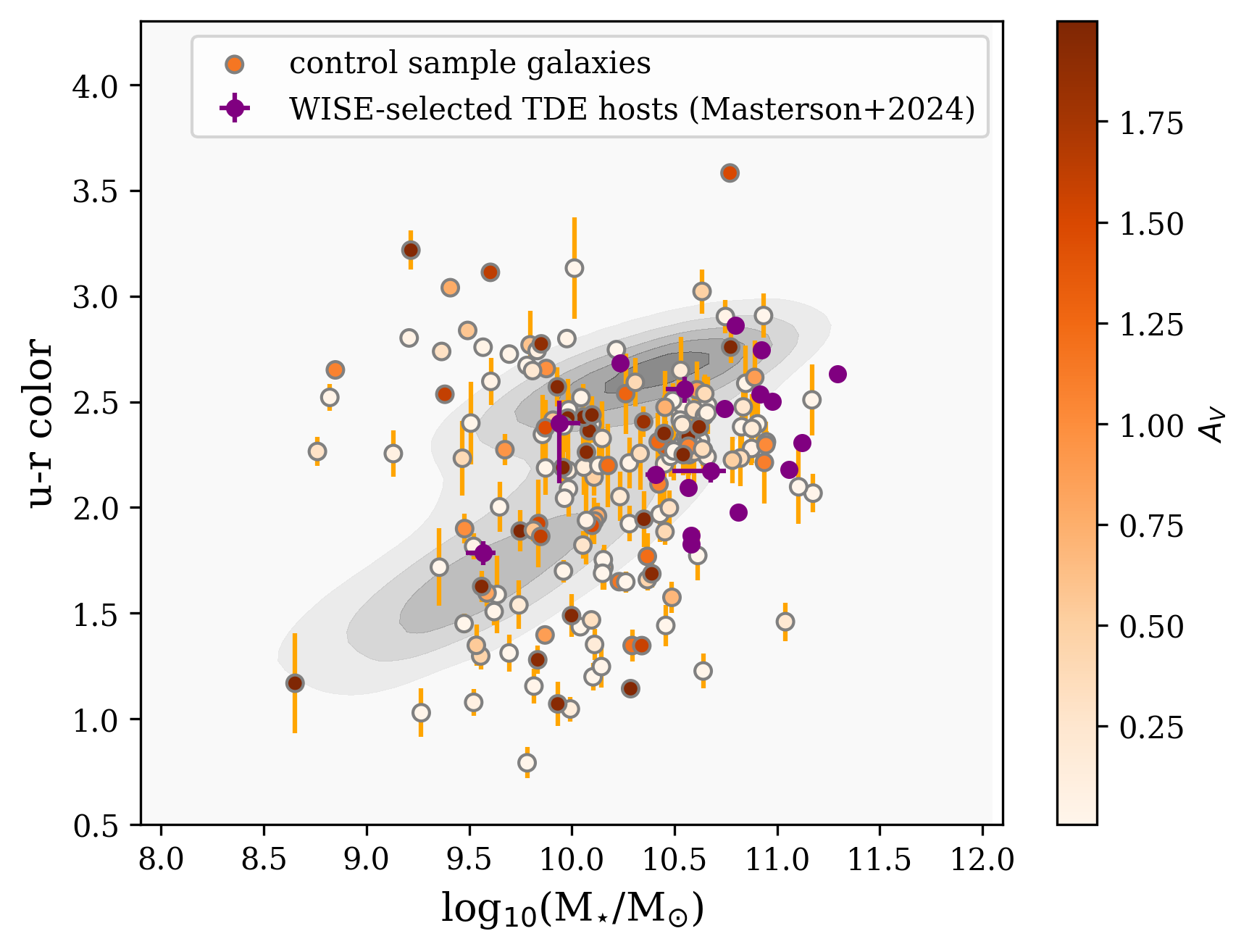}
\end{center}
\caption{Rest-frame $u-r$ colors vs stellar mass values of TDE host galaxies ({\it left}), control sample galaxies ({\it right}), and WISE-selected TDE host galaxies from \cite{Masterson_2024}. We assume the stellar mass values of the TDE host galaxies and control sample galaxies have a characteristic uncertainty of 0.1 dex. The TDE host galaxies and control sample galaxies are colored according to their $A_V$ dust attenuation values from \textsc{Bagpipes}. The $u-r$ colors and stellar mass values of SDSS galaxies at $z<0.05$ are shown in gray contours. There is no apparent correlation between dust attenuation and position in color-mass space for the control sample galaxies, but the TDE host galaxies tend to get dustier as they get redder, and lack blue and high $A_V$ galaxies. Both samples overlap with the IR-selected galaxies from \cite{Masterson_2024}.}
\label{fig:color}
\end{figure*}

Dust attenuation will affect the color of the host galaxy. \cite{Masterson_2024} find a population of IR transient host galaxies to be primarily red, rather than in the Green Valley. It is unclear whether these hosts are red due to dust or older stellar ages. However, if there is sufficient dust in a galaxy to redden the host and obscure the TDE, this level of dust obscuration would be visible in our \textsc{Bagpipes} results. Figure~\ref{fig:color} shows the rest-frame $u-r$ colors versus stellar masses of the TDE host galaxies and control sample galaxies, alongside WISE-selected TDE host galaxies from \cite{Masterson_2024} and SDSS galaxies at $z<0.05$, showing the red cloud and the blue sequence. The $u-r$ colors of the TDE hosts shown are from \cite{Yao_2023}; only eight TDE hosts overlap between both samples. The $u-r$ colors of the control galaxies are calculated from the $u$ and $r$ magnitudes reported by \textsc{Bagpipes}, which models the spectra at the relevant wavelengths even if the original spectra does not cover the required bandpasses. 

The TDE host galaxies lie both below and above the Green Valley, but some of the galaxies are quite red, even more so than the IR-selected galaxies from \cite{Masterson_2024}. The dust attenuation of the control sample galaxies, represented as the shading of the points, does not correlate with their color or stellar mass. The control sample galaxies have a larger range in color than the TDE host galaxies do, consisting of galaxies with a range of dust properties at a range of ages. The large number of TDE hosts with burst ages of 1 Gyr is a contributor to the smaller range of intrinsic color in the TDE hosts. Thus, the dust attenuation values clearly correlate with the color of the TDE hosts for this sample. 

The lack of blue, high $A_V$ galaxies in the TDE sample contrasted with the presence of blue, high $A_V$ galaxies in the control sample (see Figure~\ref{fig:color}) could indicate that we are potentially missing a population of dusty and blue TDE host galaxies, which would host younger stellar populations. We have investigated this by looking for a statistically significant difference in the $A_V$ distributions of the TDE hosts with young burst ages and the control galaxies with young burst ages (see Figure~\ref{fig:dust}), which we do not find. This suggests that the apparent color-dust trend in TDE host galaxies may not significantly bias our results. Furthermore, if this sample were missing a population of dusty and blue (and thus young) TDE host galaxies, they would display blue colors in the IR-selected sample of TDE host galaxies \citep{Masterson_2024}; however, the colors of this sample are largely green/red.

It remains unclear whether the population of IR (or radio) TDEs occurs in preferentially dust obscured galaxies. Constructing a homogeneous sample of TDEs including those that may be obscured by dust will be necessary to determine whether the dust-obscured TDEs primarily contribute to the young end of the DTD and what the effect on the overall shape of the DTD may be.

\subsubsection{AGN}
\label{sec:AGN}

We should also consider whether selection against AGN activity when classifying TDEs could cause this preference towards host galaxies with older bursts. Galaxy mergers can be a trigger of AGN activity. \cite{Ellison_2025} find that the peak of this post-merger AGN activity occurs within the first 0.16 Gyr after coalescence, and an enhancement of AGN activity as traced by broad line AGN and mid-IR selected AGN extends out to 1.76 Gyr after coalescence. \cite{Goulding_2018} also find that galaxies in the process of merging are more likely to host luminous AGN.

AGN activity and thus AGN variability is more likely to occur in star forming galaxies than in PSB galaxies; \cite{Florez_2020} and \cite{Zhuang_2021} find that AGN activity is positively correlated with the star formation rate and gas content of galaxies. X-ray tracers also show that AGN activity is found in PSB galaxies at rates in between that of star forming galaxies (where AGN activity is more common) and quiescent galaxies (where AGN activity is less common) \citep{Almaini_2025}. AGN activity in star forming or merging galaxies may mask TDEs in young systems, but these young star-forming systems may not necessarily be bursty. 

Additionally, variable AGN are a small part of the young star forming galaxy population. \cite{Baldassare_2020} find that the AGN optical variability rate in systems with $\log_{10}(M_{\star}/M_{\odot}) \sim 10$ is about 1\%. Thus, it is unlikely that the exclusion of TDEs in galaxies with AGN activity is hiding enough young systems to significantly bias our results. Given the low rate of variable AGN, the only way that TDE selection bias against AGN could result in a significant bias in our DTD would be if the AGN itself has a strong influence on the TDE rate, which we test in Section~\ref{sec:agndisk}.

\subsection{Trends in Black Hole Mass}
\label{sec:mass}

If SMBH binaries are the cause of the PSB overrepresentation among TDE host galaxies, another way to test this would be to look for additional signatures of a SMBH binary system. \cite{Melchor_2025} find that the TDE rate arising from SMBH binaries increases as a function of the mass of the disrupting black hole up to $10^8 M_{\odot}$. This predicts that TDEs arising from a SMBH binary may preferentially occur in high mass galaxies. 

\begin{figure*}
\begin{center}
\includegraphics[width=0.49\linewidth]{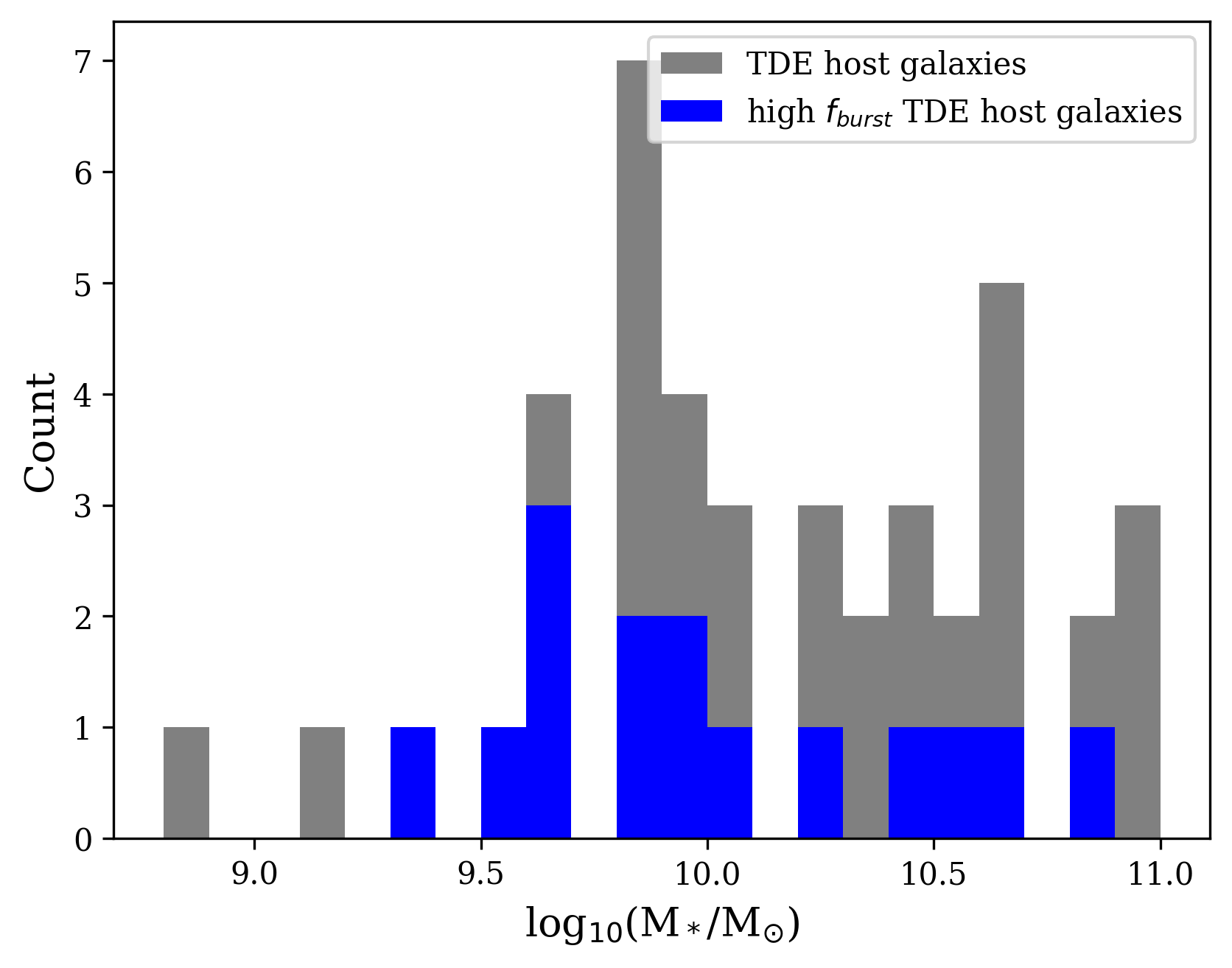}
\includegraphics[width=0.49\linewidth]{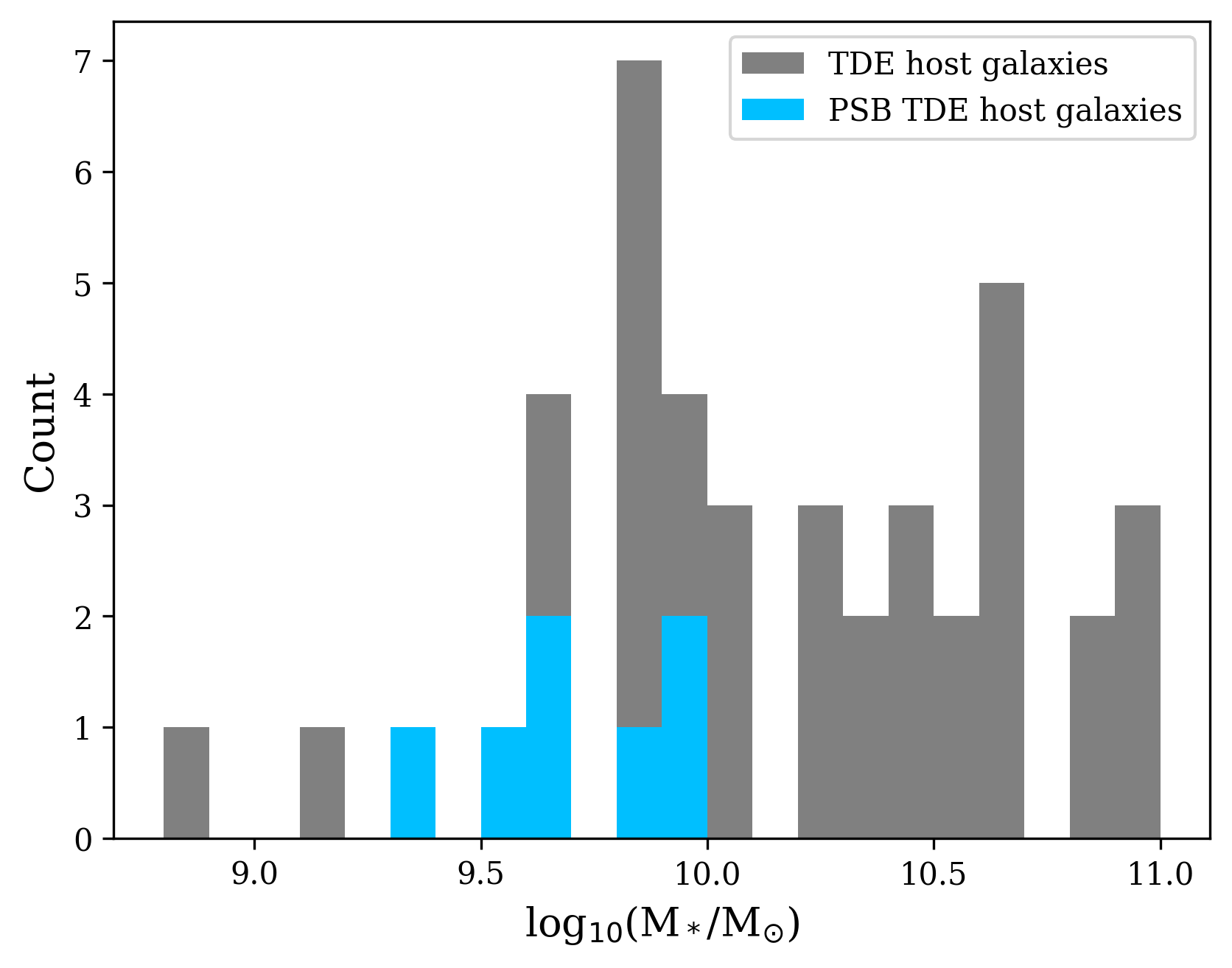}
\end{center}
\caption{Histograms of stellar masses of the TDE host galaxies and ({\it left}) the subset of TDE host galaxies that have experienced a significant burst of star formation and ({\it right}) the subset of PSB TDE host galaxies. A KS test determines that the population of PSB TDE host galaxies is significantly different compared to the parent TDE host galaxy sample, where the PSB TDE host galaxies have stellar masses at the lower end of the stellar mass range of the entire TDE host galaxy sample. The stellar masses of TDE host galaxies that have experienced a significant burst of star formation are not significantly different from the stellar masses of the parent TDE host galaxy sample.}
\label{fig:stelmass}
\end{figure*}

In the left panel of Figure~\ref{fig:stelmass} we show the distribution of stellar masses of our entire TDE host galaxy sample and the subset of TDE host galaxies that have experienced a significant burst. The bin size is determined by the average uncertainty in $\log_{10}(M_{\star}/M_{\odot})$, which we assume to be 0.1 dex. A KS test comparing the two samples returns a p-value of 0.87, indicating there is no statistically significant difference in these two populations. In the right panel of Figure~\ref{fig:stelmass} we show the distribution of stellar masses among our entire TDE host galaxy sample and the subset of TDE host galaxies that are classified as PSB. A KS test comparing the two samples returns a p-value of 0.038, indicating there is a statistically significant difference in these two populations, where PSB TDE hosts have stellar masses at the lower end of the stellar mass range of the entire TDE host galaxy sample.

\begin{figure}
\begin{center}
    \includegraphics[width=0.95\linewidth]{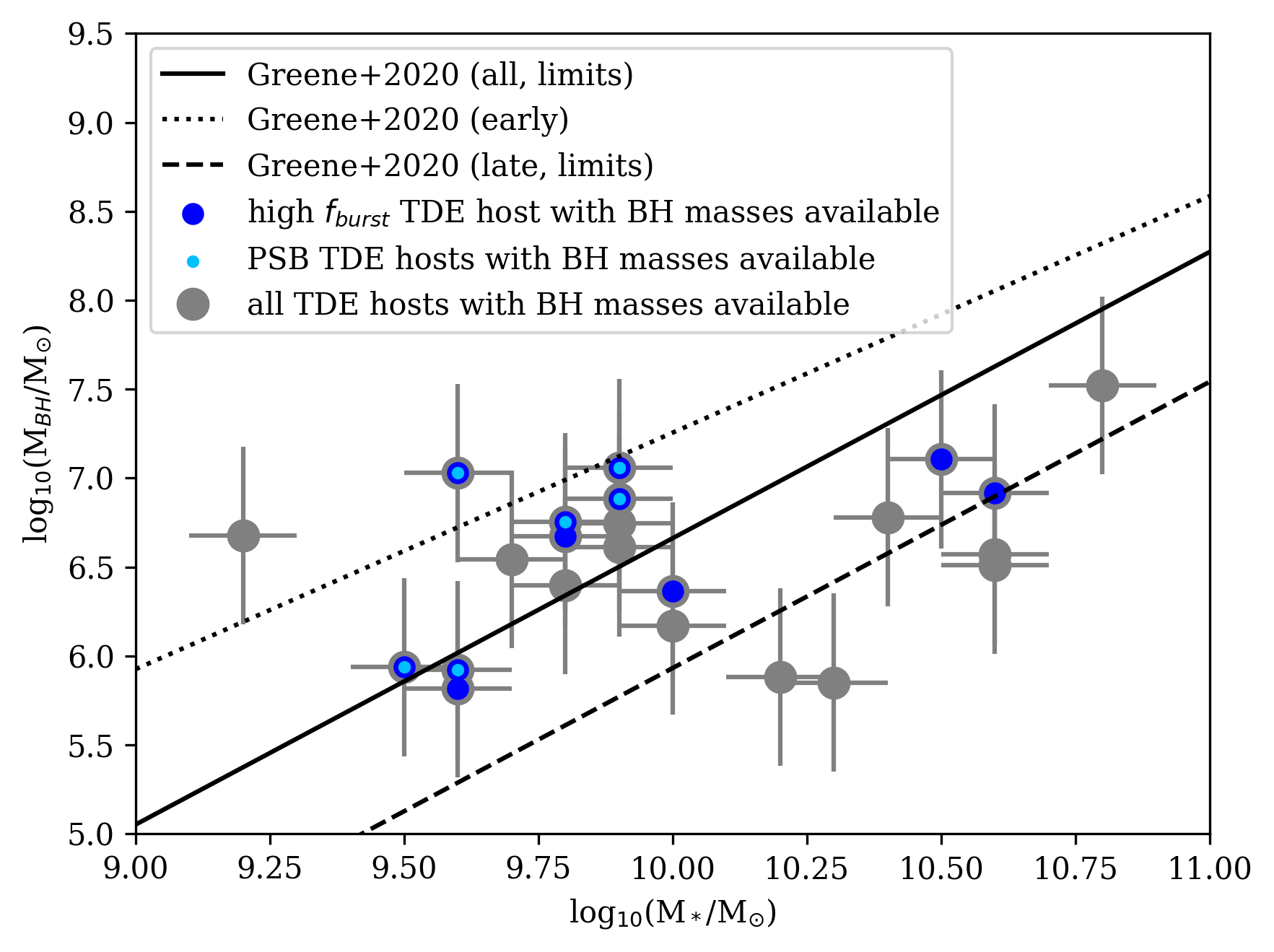}
\end{center}
\caption{Stellar mass versus black hole mass for the TDE host galaxies with black hole masses available. The solid, dotted, and dashed black lines are $M_{BH}$--$M_{stel}$ scaling relationships from \cite{Greene_2020}. The error bars on the stellar mass values are 0.1 dex, the assumed standard deviation of the sample values. The subset of TDE host galaxies that have experienced a significant burst of star formation and the subset of PSB TDE host galaxies are both marked in dark blue and light blue respectively. A KS test determines that the black hole masses for the population of all TDE host galaxies and the population of high f$_{burst}$ TDE host galaxies are not significantly different, and that those of the population of all TDE host galaxies and the population of PSB TDE host galaxies are not significantly different.}
\label{fig:BHmass}
\end{figure}

We can provide a more direct comparison by using black hole masses (see Section~\ref{sec:anisotropy} for an explanation of the source of the black hole masses), though there is data available for fewer galaxies. In Figure~\ref{fig:BHmass} we show the distribution of black hole masses among our TDE host galaxy sample (for the subset of galaxies where such data was available) versus the TDE host galaxies' stellar mass. The TDE host galaxies that have experienced a significant burst of star formation and/or are PSB are marked. A KS test comparing the TDE host galaxy sample and the sample of TDE hosts that have experienced a significant burst of star formation returns a p-value of 0.868, indicating there is no statistically significant difference in these two populations. A KS test comparing the TDE host galaxy sample and the PSB TDE host galaxy sample returns a p-value of 0.607, indicating there is no statistically significant difference in these two populations. 

These results show that PSB TDE host galaxies and TDE host galaxies that have experienced a significant burst of star formation do not have preferentially high black hole masses, as predicted in \cite{Melchor_2025} if the TDEs arose from a SMBH. In fact, one of our tests shows that PSB TDE host galaxies have lower masses. This does not rule out the SMBH binary explanation for the PSB overrepresentation, but may indicate that SMBH binaries are not primarily responsible for the overrepresentation. More observations are needed to test this theory.

Radial anisotropic models (Section~\ref{sec:anisotropy}) for the PSB preference also predict that TDE host galaxies should prefer high black hole masses (see e.g. Figure 11 in \citealt{Stone_2018}) because radial biases relax away more quickly in low mass galaxies. The analysis here further disfavors the radial anisotropy theory, or at least models with high values of the anisotropy factor $\beta$. Figure~\ref{fig:anisotropyplot} compares radial anisotropy models for disruptions around black holes of different masses and finds no agreement with data at high mass black holes and only partial agreement with data at low mass black holes.

\subsection{TDE Rate versus Redshift}
\label{sec:redshift}

Numerous efforts have been made to theoretically calculate the TDE rate as a function of redshift. \cite{Kochanek_2016} calculates the volumetric TDE rate as a function of redshift and finds that the volumetric TDE rate declines as redshift increases (see their Figure 12). \cite{Polkas_2024} calculate the volumetric TDE rate as a function of stellar mass, black hole mass, and redshift from simulations, and find minimal change in the TDE rate at $z < 1$, but there is a more substantial decrease in the rate for $1 < z < 2$ (see their Figure 7). As shown in our Figure~\ref{fig:massfrac}, our results suggest that TDEs are overrepresented in galaxies with large ($>$10\%) burst mass fractions. Galaxies with large burst mass fractions should increase with redshift because the rate of galaxy mergers over time increases before reaching a peak at $1 < z < 2$ \citep{Bertone_2009}. If the burst mass fraction has a strong effect on the TDE rate, we predict a TDE rate that declines with redshift more slowly than predicted in \cite{Polkas_2024}. Similarly, \cite{Kochanek_2016} predicts that TDEs driven by mergers will decline with redshift less slowly than TDEs not driven by mergers. If it is the case that galaxy mergers lead to an increased TDE rate, then the TDE rate may not be directly linked to a burst of star formation; rather, the merger would be the root cause of both the starburst and the increased TDE rate. This prediction could be tested in near future TDE samples from the Rubin Observatory or {\it ULTRASAT}, which will likely be able to detect a significant fraction of TDEs occurring at $z \approx 1$ \citep{Shvartzvald_2024}.

\subsection{Implications for TDE Rate Enhancement Mechanisms}
\label{sec:implications}

The models from \cite{Stone_2018} (Figure~\ref{fig:stellaroverdensityplot} for stellar overdensity models and Figure~\ref{fig:anisotropyplot} for radial anisotropy models), \cite{Teboul_2025} (Figure~\ref{fig:stellaroverdensityplot} for stellar overdensity models and Figure~\ref{fig:anisotropyplot} for radial anisotropy models), and \cite{Bortolas_2022} (Figure~\ref{fig:stellaroverdensityplot} for stellar overdensity models) all decline as burst age increases. Although some of the models agree with the observational DTDs at some age bins (for example, the lower panel of Figure~\ref{fig:stellaroverdensityplot}, the middle left panel of Figure~\ref{fig:anisotropyplot}, or the right panel of Figure~\ref{fig:melchorplot}), the disagreement between the overall trend of the models and the overall trend of the data suggests that these models do not correctly diagnose a unique source of PSB overrepresentation. Piecing together partial agreement from different models such as the SMBH binary model from \cite{Melchor_2024} at old burst ages with the stellar overdensity model from \cite{Bortolas_2022} or the radial anisotropy model from \cite{Teboul_2025} (at lower SMBH masses) at intermediate burst ages suggests that a combination of theoretical models may be necessary to explain the overrepresentation. 

The models from \cite{Wang_2024} (Figure~\ref{fig:wangplot} for transitioning AGN disk models) predict that a peak in the TDE rate should occur at early burst ages. This peak agrees with the peak in the youngest age bin of the PSB/QBS DTD, however, the magnitude of the model does not agree with the data, and the overall trend of the model (declining with burst age after the peak) does not match the overall trend of either DTD (increasing with burst age). Together, these pieces of evidence suggest that this model too does not correctly diagnose a unique source of the PSB overrepresentation.

Once the SMBH binary models from \cite{Melchor_2024} are shifted to older burst ages to account for the black hole coalescence time (Figure~\ref{fig:melchorplot}), the peak of the models does overlap with the peak of the observational DTDs. Unfortunately, the model no longer covers younger burst ages, so in order to explain the full range of ages in our DTDs with this model, we would need a range of coalescence times, not a single characteristic time. There are number of other factors that will perturb the overall shape of the model. \cite{Mockler_2023} and \cite{Melchor_2024} consider the parameter space of different binary mass ratios, stellar density profiles, and eccentricities when generating various different models for SMBH binary TDE rates. Overall, the theory of SMBH binaries leading to an increase in the TDE rate at burst ages of $\sim$1 Gyr in PSB/QBS galaxies seems promising for explaining the late peak in the DTD, but this mechanism may have to act in concert with others that can overproduce TDEs in galaxies with young and intermediate burst ages.

If dust obscuration prevents the observation of TDEs in galaxies with young bursts, the true DTD would become more flat. Accounting for TDEs obscured by dust, if they occur preferentially at young burst ages, would bring the DTDs closer to agreement with all the models discussed in this paper to varying degrees. However, models would still have to address the increase in TDE rates at older burst ages. To go from an observed DTD with a positive slope to a negative slope, one would have to at least double the sample of TDEs in galaxies with young bursts when accounting for preferential obscuration.

We quantify this disagreement by assuming one of the theoretical DTDs, which we take as a representative example of a declining DTD, is correct and then using inverse transform sampling to draw theoretical TDE host burst ages from that distribution. We fit a functional form to the $\gamma = 2.25$, strong scattering model from \cite{Teboul_2025} (red dashed line in the upper right panel of Figure~\ref{fig:stellaroverdensityplot}). When we draw 15 burst ages from this model (representing 15 potential high f$_{burst}$ TDE hosts) and compare this burst age distribution to 60 burst ages from a flat distribution (representing 60 potential control galaxies) using the Anderson-Darling 2-sample test, we find that $\sim$50\% of the time, the sample of burst ages from fake TDE hosts and the sample of burst ages from fake control galaxies are drawn from different populations (p-value $< 0.01$) \textit{and} the median TDE host burst age is \textit{younger} than the median control galaxy burst age. We find the opposite scenario (median TDE host burst age is \textit{older} than the median control galaxy burst age when p-value $< 0.01$) to be true 0\% of the time out of 1000 tests.

\section{Conclusion}
\label{sec:conclusion}

In this paper, we collected a sample of TDE host galaxies and analyzed their spectra to determine if and when a recent burst of star formation occurred. We then calculated the rate of TDEs as a function of burst age for two subsets of the TDE host galaxy population, both designed to isolate galaxies that fit the description of “post-starburst”. We compared these observational DTDs to theoretical DTDs \citep{Stone_2018, Bortolas_2022, Wang_2024, Melchor_2024, Teboul_2025} that have been calculated from a range of dynamical hypotheses to explain the overrepresentation of TDEs among PSB galaxies. Our main conclusions are the following:

\begin{itemize}
  \item The DTD for the subset of TDE host galaxies that experienced a significant burst of star formation features a TDE rate increasing with burst age, until it reaches a peak at $\sim$1 Gyr (Figure~\ref{fig:finalrate_results}). The DTD for the subset of TDE host galaxies that are PSB or QBS is flatter. We find that the distribution of burst ages for TDE host galaxies with high burst mass fractions can be distinguished from the distribution of burst ages for control galaxies, while the distribution of burst ages for PSB/QBS TDE host galaxies cannot be distinguished from the distribution of burst ages for PSB/QBS control galaxies.
  
  \item The theoretical models discussed in this paper largely predict a DTD that decreases as the age of the burst of star formation increases. However, the observational DTD using TDE host galaxies with high burst mass fractions indicates that TDEs are more common in galaxies with older bursts. The observational DTD using PSB/QBS TDE hosts is more consistent with a flat shape, in contrast with expectations from a falling DTD predicted by many models. Thus, none of the theoretical models on their own provide an overwhelmingly convincing explanation for the TDE overrepresentation in PSB galaxies. The behavior of both DTDs perhaps indicates that a variety of mechanisms may be at play to cause the overrepresentation of TDEs in PSB galaxies: one mechanism with a short DTD that produces TDEs immediately after a burst, and one mechanism that delays TDEs until at least 1 Gyr after the burst. 
  
  \item The TDE rate enhancement in galaxies with large burst mass fractions (Figure~\ref{fig:massfrac}) confirms that galaxies that have experienced a large burst of star formation are indeed overrepresented among TDE host galaxies. Specifically, galaxies with a burst mass fraction above 10\% experience TDE overrepresentation of almost 10 times the average optical TDE rate. This effect may act in combination with a time-dependent DTD at $\sim$1 Gyr to produce the overrepresentation seen in PSB galaxies. 
  
  \item We confirm that PSB and QBS galaxies are both overrepresented in our sample, regardless of what the comparison sample is. However, they are overrepresented to varying degrees. PSB and QBS galaxies are overrepresented among TDE host galaxies by factors of 83 and 14 respectively when compared to the population of galaxies in SDSS DR8, and they are overrepresented among TDE host galaxies by factors of 12 and 8.6 respectively when compared to the stellar mass-matched and redshift-matched expanded control sample.
  
  \item Dust does not appear to contribute to our result that the TDE rate increases with burst age. This is apparent in the similar distributions of \textsc{Bagpipes} dust attenuation measurements for the control sample and the TDE host galaxy sample; dust does not affect the TDE host galaxies more or less than the control galaxies. Additionally, there is no statistically significant difference between the distributions of $A_V$ for the TDE host galaxies with young bursts and the control sample galaxies with young bursts. However, a correlation between $A_V$ and $u-r$ color appears in the TDE hosts but not the control sample galaxies, suggesting that we may be missing a population of dusty but blue (and thus with young stellar populations) TDE host galaxies. Further analysis of the host galaxies of TDEs discovered in multi-wavelength searches will be needed to determine the full effect of dust obscuration on the DTD. 
  
  \item The distributions of stellar mass and black hole mass among the TDE host galaxies do not show any preference for one TDE rate enhancement mechanism over another. While SMBH binary effects or radially anisotropic orbits could cause an enhanced TDE rate in high $M_{BH}$ systems, we do not see evidence that the PSB/QBS or high $f_{burst}$ hosts are in higher stellar mass or black hole mass systems.
\end{itemize}

Many of the results in this work are limited by the small number of TDEs with well-characterized hosts studied to date. With a larger sample of TDEs and host galaxy spectra, the procedure in this paper could be repeated and improved. For example, it would be more straightforward to accurately calculate the TDE rate, rather than just the TDE rate enhancement, with a uniformly selected, volume-complete sample. In the near future, the Legacy Survey of Space and Time (LSST) and the {\it ULTRASAT} all-sky survey will each likely expand the TDE sample by 1-2 orders of magnitude \citep{Bricman_2020, Shvartzvald_2024}, allowing the estimates in this paper to be repeated without the hindrance of small number statistics. While LSST and {\it ULTRASAT} will discover so many transients that our capacity to obtain follow-up spectra will be greatly diminished, efforts are being made to prioritize follow-up of particularly interesting transients \citep{Frohmaier_2025}, from which we anticipate a significant increase in the sample of well-studied TDEs.

\section{Acknowledgments}
\label{sec:acknowledgments}

We thank the anonymous reviewer, whose comments and suggestions improved this work.

The authors thank Thomas Wevers, Erica Hammerstein, and Clive Tadhunter for providing TDE host galaxy spectra via private communication. The authors thank Megan Masterson for providing colors and stellar masses of TDE host galaxies via private communication. M.S. thanks Gautham Narayan and Xin Liu for useful discussion and comments.

M.S., K.D.F., and N.E. acknowledge support from NSF grant AST 22-06164. M.S. acknowledges support from the Illinois Space Grant Consortium. N.E. acknowledges support from the Center for Astrophysical Surveys Graduate Fellowship. N.C.S. gratefully acknowledges support from the Binational Science Foundation (grant No. 2020397) and the Israel Science Foundation (Individual Research Grant No. 2414/23). O.T. gratefully acknowledges the support of the Benoziyo Fellowship. M.E.V. acknowledges support from NSF grant NSF-2307375.

The material contained in this document is based upon work supported by a National Aeronautics and Space Administration (NASA) grant or cooperative agreement. Any opinions, findings, conclusions, or recommendations expressed in this material are those of the author and do not necessarily reflect the views of NASA. This work was supported through a NASA grant awarded to the Illinois/NASA Space Grant Consortium.

Funding for SDSS-III has been provided by the Alfred P. Sloan Foundation, the Participating Institutions, the National Science Foundation, and the U.S. Department of Energy Office of Science. The SDSS-III web site is http://www.sdss3.org/.

SDSS-III is managed by the Astrophysical Research Consortium for the Participating Institutions of the SDSS-III Collaboration including the University of Arizona, the Brazilian Participation Group, Brookhaven National Laboratory, Carnegie Mellon University, University of Florida, the French Participation Group, the German Participation Group, Harvard University, the Instituto de Astrofisica de Canarias, the Michigan State/Notre Dame/JINA Participation Group, Johns Hopkins University, Lawrence Berkeley National Laboratory, Max Planck Institute for Astrophysics, Max Planck Institute for Extraterrestrial Physics, New Mexico State University, New York University, Ohio State University, Pennsylvania State University, University of Portsmouth, Princeton University, the Spanish Participation Group, University of Tokyo, University of Utah, Vanderbilt University, University of Virginia, University of Washington, and Yale University.

\software{Astropy \citep{astropy2013, astropy2018}, Matplotlib \citep{matplotlib}, NumPy \citep{numpy}, \textsc{Bagpipes} \citep{Carnall_2018, Carnall_2019}, pyLick \citep{pyLick}, SciPy \citep{2020SciPy-NMeth}}

\bibliography{Bibliography}

\begin{appendices}

\section{Alternative SFH Models}
\label{appendix:a}

\setcounter{figure}{0}
\renewcommand{\thefigure}{\Alph{section}\arabic{figure}}

We choose to model the burst with a double power law because it was the most physical option that was able to accommodate the greatest variety in SFHs. We tried modeling the burst as a delta function ($SFR(t) \propto \delta(t)$), an exponential function ($SFR(t) \propto e^{-t/\tau}$), a delta function with an additional constant star formation component ($SFR(t) \propto \delta(t) + C$), and a double power law ($SFR(t) \propto [(t/\tau)^{\alpha} + (t/\tau)^{-\beta}]^{-1}$). For each of these options, the old stellar component was always modeled as a delayed exponential function ($SFR(t) \propto t \times e^{-t/\tau}$). The exponential function burst and the delta function burst with an additional constant star formation component were both attempts to make the burst more realistic than a delta function alone. The constant star formation component tried to allow for galaxies that were still actively forming stars, regardless of whether or not a burst occurred. However, the results from when the burst was modeled as a delta function with an additional constant star formation component were almost identical to when the burst was modeled only as a delta function, so we excluded both of those options. Modeling the burst as a double power law function has an advantage over modeling the burst as an exponential function. The steepness of the decline of the double power law function is a free parameter in \textsc{Bagpipes}, which allows us to identify star forming galaxies that have a constant but non-bursty recent SFH. 

We compared the results between \textsc{Bagpipes} runs where the burst was modeled as an exponential function versus when it was modeled as a double power law function. The total number of galaxies with a high burst mass fraction ($> 1\%$) when the burst was modeled as an exponential function was 21, while the total number of galaxies with a high burst mass fraction when the burst was modeled as an double power law function was 16 (counting F01004 only once in both cases). One galaxy had a high burst mass fraction when the burst was modeled as a double power law function but not when the burst was modeled as an exponential function. Six galaxies had a high burst mass fraction when the burst was modeled as an exponential function but not when the burst was modeled as a double power law function.

\begin{figure}
\begin{center}
    \includegraphics[width=0.49\linewidth]{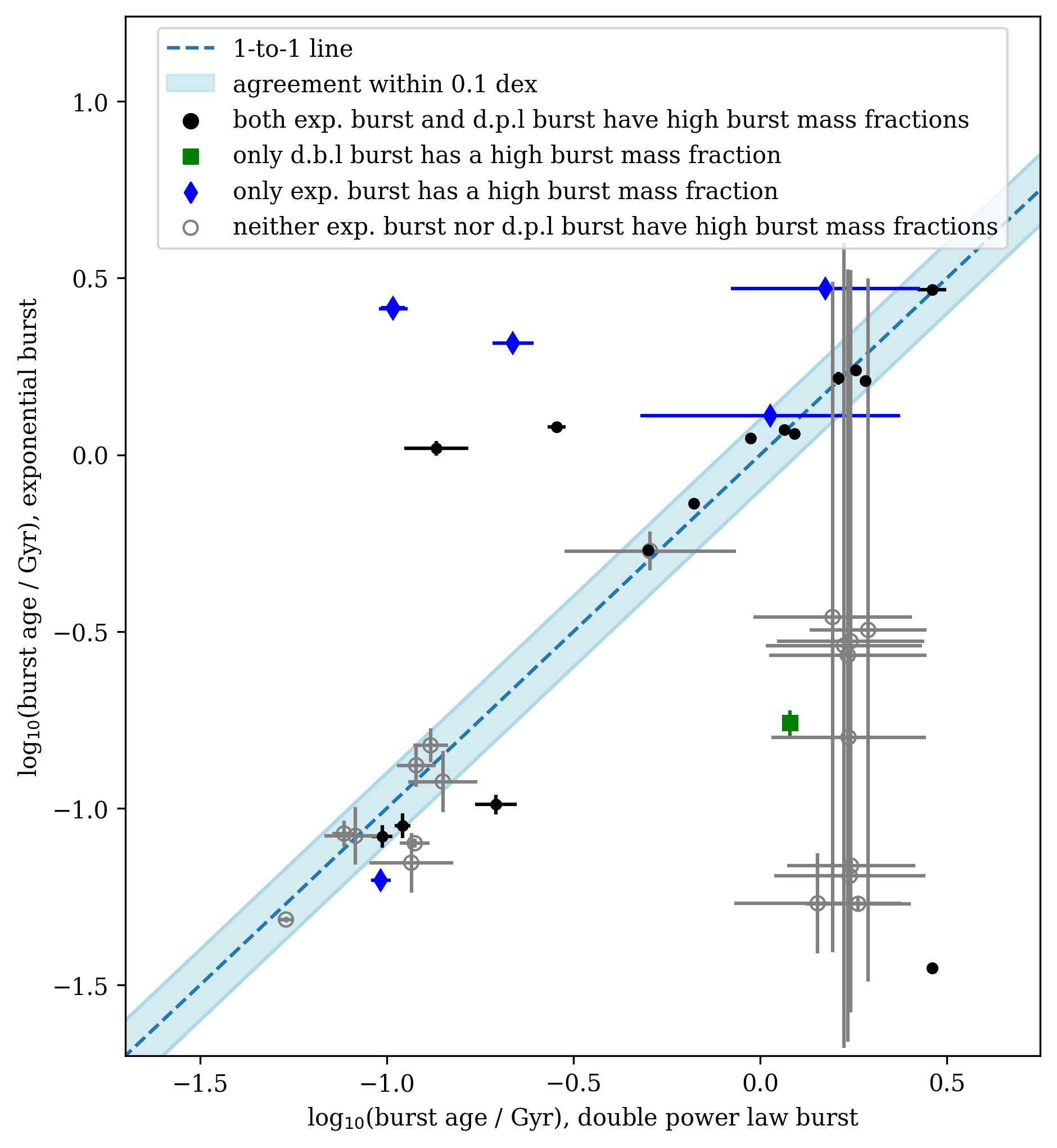}
\end{center}
\caption{The age of the burst modeled as an exponential function versus the age of the burst modeled as a double power law function, for all TDE hosts in the sample. The differently-colored markers represent burst mass fraction classifications in one or both functional forms, and the dashed blue line represents a 1-to-1 match between burst ages. Points within the blue shaded region have burst ages that agree within 0.1 dex of each other. The error bars are the standard deviation of the \textsc{Bagpipes} posterior (for some galaxies, the error bars are smaller than the marker size). In cases with large error bars, \textsc{Bagpipes'} results indicate that these galaxies have not had a burst. \textsc{Bagpipes} is forced to return values for such a “burst” anyway, but because the burst is so small/did not happen, the error bars on the burst age are very large.}
\label{fig:age_vs_age}
\end{figure}

To compare the burst ages between the two functional forms of the burst, see Figure~\ref{fig:age_vs_age}, which plots the burst ages of all the TDE host galaxies in the sample when the burst was modeled as an exponential function on the y-axis and when the burst was modeled as a double power law function on the x-axis. The points are colored by their burst mass fraction classification. The points included in the final high $f_{burst}$ DTD are the black points and the green point, with the exception of the black point in the lower right corner, which represents F01004. This galaxy was removed from the final high $f_{burst}$ DTD due to the cut on the falling slope index $\alpha$, intended to remove galaxies with non-zero constant star formation from consideration as galaxies that have experienced a significant burst of star formation. 

Our pivot from the burst with an exponential form to the burst with a double power law form was motivated by the exponential burst model's inability to fit more ongoing/continuous star formation. The galaxies that ended up in the high $f_{burst}$ DTD when modeled with a double power law burst are relatively close to the 1-to-1 line. 11/15 high $f_{burst}$ TDE hosts have burst ages that agree within 0.1 dex between the two functional forms, two hosts have burst ages that are older with an exponential burst, and two hosts have burst ages that are older with a double power law burst. Thus, the overall trend of the DTD (rate increasing as a function of burst age until a peak at $\sim$1 Gyr) does not change.

\section{\textsc{Bagpipes} Fit Instructions and Results}
\label{appendix:b}

\setcounter{table}{0}
\renewcommand{\thetable}{\Alph{section}\arabic{table}}

Below is example code that displays the fit instructions provided to \textsc{Bagpipes} to retrieve SFH parameters and other parameters from host galaxies' SDSS spectra. The fit instructions for galaxy spectra from other sources followed the same structure. Table~\ref{table:BagpipesOp4} describes the TDE host galaxies' \textsc{Bagpipes} fitting results.

\begin{lstlisting}[language=Python]
def fit_instructions(index):
    delayed = {}                               # Old stellar component: delayed exponential
    delayed["age"] = (3., 15.)                 # Time since SF began: Gyr
    delayed["age_prior"] = "log_10"
    delayed["tau"] = 1.0
    delayed["massformed"] = (1., 15.)          # Vary log_10(M*/M_solar) between 1 and 15
    delayed["metallicity"] = (0.1, 2.0)
    delayed["metallicity_prior"] = "Gaussian"
    delayed["metallicity_prior_mu"] = metallicity(SDSSgalarray['logmass'][index])[1]
    delayed["metallicity_prior_sigma"] = metallicity(SDSSgalarray['logmass'][index])[2]
    
    dblplaw = {}                               # New stellar component: double power law
    dblplaw["alpha"] = (0.1, 1000)             # Falling slope index, taken from Carnall+2019a
    dblplaw["alpha_prior"] = "log_10"
    dblplaw["beta"] = 250                      # Rising slope index, taken from Wild+2020
    dblplaw["tau"] = (cosmo.age(SDSSgalarray['z'][index]).value - 3, cosmo.age(SDSSgalarray['z'][index]).value)                             # Age of Universe at turnover: Gyr
    dblplaw["tau_prior"] = "log_10"
    dblplaw["massformed"] = (1., 15.)          # Vary log_10(M*/M_solar) between 1 and 15
    dblplaw["metallicity"] = "delayed:metallicity"
    
    nebular = {}
    nebular["logU"] = -3.
    dust = {}
    dust["type"] = "Calzetti"
    dust["Av"] = (0., 2.)
    
    fit_instructions = {}
    fit_instructions["redshift"] = (0.0001,0.9)
    fit_instructions["redshift_prior"] = "Gaussian" 
    fit_instructions["redshift_prior_mu"] = SDSSgalarray['z'][index] 
    fit_instructions["redshift_prior_sigma"] = 40*SDSSgalarray['zErr'][index]
    fit_instructions["t_bc"] = 0.01
    fit_instructions["delayed"] = delayed
    fit_instructions["dblplaw"] = dblplaw
    fit_instructions["nebular"] = nebular
    fit_instructions["dust"] = dust
    fit_instructions["veldisp"] = (30., 300.)  # km/s
    fit_instructions["veldisp_prior"] = "log_10"
    
    calib = {}
    calib["type"] = "polynomial_bayesian"
    calib["0"] = (0.5, 1.5) 
    calib["0_prior"] = "Gaussian"
    calib["0_prior_mu"] = 1.0
    calib["0_prior_sigma"] = 0.25
    calib["1"] = (-0.5, 0.5) 
    calib["1_prior"] = "Gaussian"
    calib["1_prior_mu"] = 0.
    calib["1_prior_sigma"] = 0.25
    calib["2"] = (-0.5, 0.5)
    calib["2_prior"] = "Gaussian"
    calib["2_prior_mu"] = 0.
    calib["2_prior_sigma"] = 0.25
    fit_instructions["calib"] = calib
    
    noise = {}
    noise["type"] = "white_scaled"
    noise["scaling"] = (1., 10.)
    noise["scaling_prior"] = "log_10"
    fit_instructions["noise"] = noise
    
    return(fit_instructions)
\end{lstlisting}

\begin{table}
{\scriptsize
\begin{tabular}{cDDDDDDDDD}

TDE name & \multicolumn2c{$t_{burst}$} & \multicolumn2c{$\sigma_t$} & \multicolumn2c{$M_{*, burst}$} & \multicolumn2c{$\sigma_{M_{*, burst}}$} & \multicolumn2c{$M_{*, old}$} & \multicolumn2c{$\sigma_{M_{*, old}}$} & \multicolumn2c{$M_{*, burst}/M_{*, tot}$} & \multicolumn2c{$\alpha$} & \multicolumn2c{$\sigma_{\alpha}$} \\

 & \multicolumn2c{Gyr} & \multicolumn2c{Gyr} & \multicolumn2c{log$_{10} (M_{\odot})$} & \multicolumn2c{log$_{10} (M_{\odot})$} & \multicolumn2c{log$_{10} (M_{\odot})$} & \multicolumn2c{log$_{10} (M_{\odot})$} & \multicolumn2c{} & \multicolumn2c{} & \multicolumn2c{} \\

\hline
\decimals
AT 2023clx & 0.131 & 0.014 & 6.989 & 0.059 & 9.274 & 0.045 & 0.005 & 895 & 90 \\
AT 2022dyt & 0.077 & 0.006 & 8.245 & 0.027 & 10.425 & 0.03 & 0.007 & 966 & 37 \\
AT 2022bdw & 0.11 & 0.005 & 8.127 & 0.032 & 9.921 & 0.023 & 0.016 & 916 & 48 \\
AT 2021nwa & 0.116 & 0.03 & 7.331 & 0.087 & 10.173 & 0.023 & 0.001 & 785 & 130 \\
AT 2020wey & 1.802 & 0.047 & 9.11 & 0.052 & 3.84 & 1.75 & 1.0 & 60 & 2 \\
AT 2020vwl & 0.141 & 0.031 & 7.551 & 0.074 & 9.893 & 0.037 & 0.005 & 862 & 103 \\
AT 2020ohl & 0.104 & 0.008 & 7.443 & 0.031 & 10.195 & 0.021 & 0.002 & 948 & 49 \\
AT 2019gte & 0.082 & 0.016 & 8.018 & 0.052 & 11.066 & 0.032 & 0.001 & 877 & 101 \\
AT 2019azh & 0.663 & 0.012 & 9.248 & 0.021 & 9.228 & 0.063 & 0.512 & 135 & 3 \\
AT 2018mac & 0.218 & 0.028 & 7.849 & 0.053 & 10.058 & 0.026 & 0.006 & 730 & 114 \\
ASASSN-18zj & 0.285 & 0.016 & 8.468 & 0.022 & 10.347 & 0.015 & 0.013 & 819 & 88 \\
AT 2018dyk & 0.12 & 0.014 & 7.928 & 0.049 & 10.377 & 0.029 & 0.004 & 876 & 90 \\
ASASSN-14li & 1.236 & 0.006 & 9.202 & 0.014 & 3.711 & 1.197 & 1.0 & 86 & 1 \\
AT 2023mhs & 0.096 & 0.006 & 8.38 & 0.017 & 10.4 & 0.011 & 0.009 & 891 & 37 \\
RBS 1032 & 1.914 & 0.016 & 9.52 & 0.012 & 4.506 & 1.301 & 1.0 & 60 & 1 \\
SDSS J0952 & 1.677 & 0.817 & 4.341 & 2.053 & 10.275 & 0.076 & 0.000001 & 16 & 213 \\
SDSS J1342 & 1.425 & 0.74 & 6.422 & 2.386 & 9.756 & 0.057 & 0.0005 & 121 & 221 \\
SDSS J0748 & 0.054 & 0.002 & 8.284 & 0.012 & 10.652 & 0.012 & 0.004 & 957 & 19 \\
AT 2018dyb & 1.736 & 0.818 & 4.026 & 1.826 & 10.578 & 0.071 & 0.0000003 & 14 & 208 \\
AT 2021ehb & 1.752 & 0.699 & 3.732 & 1.411 & 10.226 & 0.01 & 0.0000003 & 13 & 154 \\
AT 2022gri & 1.829 & 0.599 & 4.498 & 1.131 & 10.753 & 0.009 & 0.0000006 & 3 & 65 \\
AT 2021yzv & 1.495 & 0.879 & 4.742 & 1.864 & 11.218 & 0.012 & 0.0000003 & 5 & 150 \\
ASASSN-15oi & 0.104 & 0.009 & 7.646 & 0.05 & 10.151 & 0.042 & 0.003 & 929 & 50 \\
RX J1242-A & 1.745 & 0.8 & 3.883 & 1.791 & 10.631 & 0.08 & 0.0000002 & 17 & 188 \\
RX J1624 & 0.507 & 0.271 & 7.719 & 0.797 & 10.134 & 0.053 & 0.004 & 511 & 228 \\
TDE2 & 2.883 & 0.254 & 10.788 & 0.047 & 6.482 & 2.709 & 1.0 & 25 & 150 \\
XMM J0740 & 0.136 & 0.027 & 6.472 & 0.052 & 8.088 & 0.047 & 0.024 & 801 & 100 \\
PS1-10jh & 1.562 & 0.772 & 3.361 & 1.228 & 10.375 & 0.034 & 0.0000001 & 36 & 211 \\
PTF09axc & 1.161 & 0.023 & 9.957 & 0.063 & 4.682 & 2.004 & 1.0 & 85 & 4 \\
PTF09djl & 1.619 & 0.062 & 10.101 & 0.065 & 3.56 & 1.662 & 1.0 & 531 & 230 \\
AT 2018fyk & 0.119 & 0.011 & 8.138 & 0.03 & 10.254 & 0.016 & 0.008 & 744 & 76 \\
AT 2019ahk & 1.202 & 0.024 & 8.57 & 0.019 & 9.614 & 0.011 & 0.083 & 678 & 152 \\
iPTF16fnl & 0.943 & 0.017 & 8.864 & 0.017 & 9.222 & 0.036 & 0.305 & 107 & 2 \\
AT 2019dsg & 1.946 & 0.707 & 6.55 & 2.814 & 10.361 & 0.08 & 0.0002 & 84 & 221 \\
iPTF15af & 1.724 & 0.828 & 3.938 & 1.724 & 10.405 & 0.052 & 0.0000003 & 18 & 210 \\
ASASSN-14ae & 0.196 & 0.025 & 7.888 & 0.038 & 9.632 & 0.023 & 0.018 & 740 & 133 \\
AT 2018bsi & 1.063 & 0.859 & 5.133 & 2.469 & 10.525 & 0.064 & 0.000004 & 87 & 271 \\
AT 2018hco & 1.716 & 0.84 & 3.987 & 1.717 & 10.233 & 0.036 & 0.0000006 & 11 & 205 \\
AT 2019qiz & 0.502 & 0.018 & 8.828 & 0.067 & 7.805 & 1.098 & 0.913 & 782 & 147 \\
F01004 & 2.889 & 0.056 & 10.128 & 0.008 & 9.948 & 0.038 & 0.603 & 0.112 & 0.008 \\
AT 2020nov & 0.097 & 0.006 & 8.578 & 0.027 & 10.294 & 0.019 & 0.019 & 897 & 66 \\
\end{tabular}
}
\caption{\textsc{Bagpipes} results when modeling the burst with a double power law function. The columns are the name of the TDE, the time ($t_{burst}$) since a burst of star formation in the host galaxy, the standard deviation of $t_{burst}$, the logarithm of the stellar mass formed in the burst ($M_{\star, burst}$), the standard deviation of $M_{\star, burst}$, the logarithm of the stellar mass formed in the old stellar component ($M_{\star, old}$), the standard deviation of $M_{\star, old}$, the burst mass fraction (stellar mass formed in the burst divided by total stellar mass formed), the falling slope index $\alpha$, and the standard deviation of $\alpha$. The values in the $t_{burst}$, $M_{\star, burst}$, $M_{\star, old}$, and $\alpha$ columns are the median values of the posteriors returned by \textsc{Bagpipes}.}
\label{table:BagpipesOp4}
\end{table}

\section{DTD of TDEs with Broad Lines}
\label{appendix:c}

\setcounter{figure}{0}
\renewcommand{\thefigure}{\Alph{section}\arabic{figure}}

\begin{figure}
\begin{center}
    \includegraphics[width=0.49\linewidth]{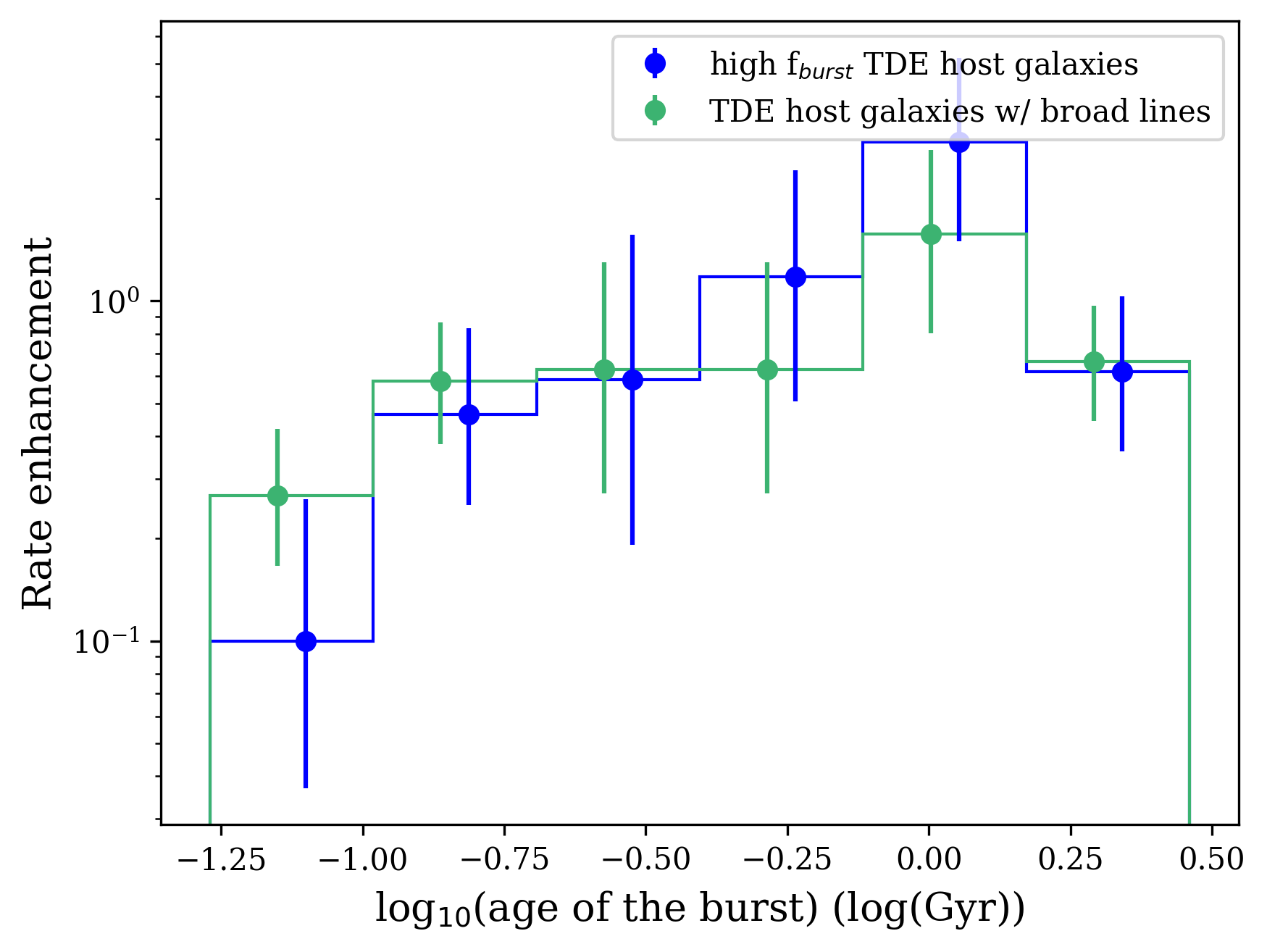}
\end{center}
\caption{The TDE rate for galaxies whose TDEs display broad lines in their spectra as a function of time since burst, where the rate is normalized by the time since burst in the control sample. Also shown is the fiducial TDE rate for galaxies we determine to have experienced a significant burst. The TDE rate for TDEs with broad lines shows the same pattern as the fiducial rate (lower rates at early burst ages, peaking at a burst age of $\sim$1 Gyr), but with a less dramatic increase. Confirming the similarities between these two DTDs is important because PSB galaxies have also been shown to be more overrepresented among the host galaxies of TDEs with broad lines in their spectrum compared with more general optical and X-ray selections \citep{Graur_2018, French_2020}.}
\label{fig:TDErate_op4_BL}
\end{figure}

Another way to ensure that the non-uniform nature of our sample is not biasing our results is to create a DTD using only TDEs with broad H or He lines in their spectra, a more stringent method of classifying TDEs. 28/41 TDE host galaxies in our sample display broad lines in the spectra of the TDE itself. 12/41 TDE host galaxies have experienced a significant burst of star formation and display broad lines in the spectra of the TDE. The TDE rate as a function of time since burst is shown in Figure~\ref{fig:TDErate_op4_BL}, for the subset of TDE host galaxies where the TDE displayed broad lines. These rates are normalized by the control sample. Also shown is the “fiducial” rate of TDE host galaxies that have experienced a significant burst of star formation from the left panel of Figure~\ref{fig:finalrate_results}. The subset with broad lines is similar to the fiducial rate, with a less pronounced peak at 1 Gyr. 

\section{Informational Tables}
\label{appendix:d}

\setcounter{table}{0}
\renewcommand{\thetable}{\Alph{section}\arabic{table}}

Table~\ref{table:finalsample} describes the TDE host galaxies' coordinates, redshift, stellar mass, stellar mass source, spectrum source, spectrum observing information, and discovery citation. Table~\ref{table:broadlinestable} describes the H$\alpha$, H$\delta$, broad line, and galaxy label information for the TDE host galaxies.

\begin{sidewaystable}
{\scriptsize
\begin{tabular}{cDDDDccDc}

TDE name & \multicolumn2c{RA} & \multicolumn2c{Dec} & \multicolumn2c{$z$} & \multicolumn2c{$M_{\star}$} & $M_{\star}$ source & Spectrum source & \multicolumn2c{Slit width} & TDE discovery citation \\

 & \multicolumn2c{deg} & \multicolumn2c{deg} & \multicolumn2c{} & \multicolumn2c{log$_{10} (M_{\odot})$} & & & \multicolumn2c{arcsec} & \\

\hline
\decimals
AT 2023clx & 175.0391 & 15.327303 & 0.0111 & 8.8 & SDSS & SDSS & 3.0 & \citet{Jiazheng_2023} \\
AT 2022dyt & 150.03335 & 26.460692 & 0.072 & 10.6 & SDSS & SDSS & 3.0 & \citet{Somalwar_2022} \\
AT 2022bdw & 126.29314 & 18.582651 & 0.0378 & 10.2 & SDSS & SDSS & 3.0 & \citet{Arcavi_2022} \\
AT 2021nwa & 238.46366 & 55.588842 & 0.0468 & 10.0 & SDSS & SDSS & 3.0 & \citet{Yao_2021} \\
AT 2020wey & 136.35777 & 61.80255 & 0.0274 & 9.6 & SDSS & SDSS & 3.0 & \citet{Arcavi_2020} \\
AT 2020vwl & 232.65754 & 26.982471 & 0.0325 & 9.8 & SDSS & SDSS & 3.0 & \citet{Hammerstein_2021TNS} \\
AT 2020ohl & 255.9023 & 62.025617 & 0.0167 & 10.5 & SDSS & SDSS & 3.0 & \citet{Hinkle_2022} \\
AT 2019gte & 196.66416 & -1.5484348 & 0.0856 & 11.0 & SDSS & SDSS & 3.0 & \citet{Swann_2019} \\
AT 2019azh & 123.32063 & 22.648304 & 0.0222 & 9.9 & SDSS & SDSS & 3.0 & \citet{Brimacombe_2019, vanVelzen_2019} \\
AT 2018mac & 185.18771 & 49.551287 & 0.0284 & 10.0 & SDSS & SDSS & 3.0 & \citet{Arcavi_2022_2018mac} \\
ASASSN-18zj & 151.71195 & 1.6927759 & 0.0457 & 9.6 & SDSS & SDSS & 3.0 & \citet{Brimacombe_2018, Dong_2018} \\
AT 2018dyk & 233.2834 & 44.535656 & 0.0367 & 10.9 & SDSS & SDSS & 3.0 & \citet{Arcavi_2018} \\
ASASSN-14li & 192.06346 & 17.774016 & 0.0206 & 9.5 & SDSS & SDSS & 3.0 & \citet{Jose_2014} \\
AT 2023mhs & 205.81524 & 19.250264 & 0.0482 & 9.8 & SDSS & SDSS & 3.0 & \citet{Sollerman_2023} \\
RBS 1032 & 176.86121 & 49.716067 & 0.026 & 9.3 & SDSS & SDSS & 3.0 & \citet{Maksym_2014} \\
SDSS J0952 & 148.03985 & 21.720364 & 0.0795 & 10.9 & SDSS & SDSS & 3.0 & \citet{Komossa_2008} \\
SDSS J1342 & 205.68507 & 5.5155945 & 0.0365 & 9.8 & SDSS & SDSS & 3.0 & \citet{Wang_2012} \\
SDSS J0748 & 117.08614 & 47.203973 & 0.0616 & 10.2 & SDSS & SDSS & 3.0 & \citet{Wang_2011} \\
AT 2018dyb & 242.7448917 & -60.9231 & 0.018 & 10.4 & * & WISeREP & n/a & \citet{Pan_2018, Leloudas_2019} \\
AT 2021ehb & 46.9492423 & 40.3113166 & 0.017 & 10.2 & \citet{Yao_2023} & WISeREP & n/a & \citet{Gezari_2021TNS} \\
AT 2022gri & 109.5865345 & 33.9948899 & 0.028 & 10.6 & * & WISeREP & n/a & \citet{Yao_2022} \\
AT 2021yzv & 105.277487 & 40.8251838 & 0.286 & 10.8 & \citet{Yao_2023} & WISeREP & n/a & \citet{Chu_2022} \\
ASASSN-15oi & 309.785659 & -30.75658 & 0.0484 & 9.9 & \citet{French_2020} & [1] & 1.65 & \citet{Brimacombe_2015, Holoien_2016} \\
RX J1242-A & 190.65375 & -11.3263889 & 0.05 & 10.3 & \citet{Wevers_2019} & [1] & 1.5 & \citet{Komossa_1999} \\
RX J1624 & 246.2360833 & 75.9155806 & 0.0636 & 10.4 & \citet{Wevers_2019} & [1] & 1.7 & \citet{Grupe_1999} \\
TDE2 & 350.9525833 & -1.1362056 & 0.2515 & 10.6 & \citet{French_2020} & [1] & 1.5 & \citet{vanVelzen_2011} \\
XMM J0740 & 115.0337083 & -85.6586944 & 0.0173 & 10.8 & * & [1] & 4.8 & \citet{Saxton_2017} \\
PS1-10jh & 242.3678333 & 53.6733306 & 0.1696 & 9.2 & [1] & [1] & 1.0 & \citet{Gezari_2012} \\
PTF09axc & 223.3045 & 22.2422972 & 0.1146 & 9.8 & [1] & [1] & 1.0 & \citet{Arcavi_2014} \\
PTF09djl & 248.4832083 & 30.2379583 & 0.184 & 9.9 & [1] & [1] & 1.0 & \citet{Arcavi_2014} \\
AT 2018fyk & 342.56723 & -44.86457 & 0.06 & 9.7 & WISeREP & [2] & 2.0 & \citet{Wevers_2019_2018fyk} \\
AT 2019ahk & 105.0481083 & -66.04003889 & 0.0262 & 10.5 & * & [2] & 2.0 & \citet{Holoien_2019} \\
iPTF16fnl & 7.4875417 & 32.8936778 & 0.0163 & 9.6 & [1] & [2] & 2.0 & \citet{Gezari_2016} \\
AT 2019dsg & 314.2623917 & 14.20440556 & 0.0512 & 10.6 & [4] & [2] & 2.0 & \citet{Cannizzaro_2021} \\
iPTF15af & 132.11726 & 22.059315 & 0.079 & 10.3 & SDSS & [2] & 2.0 & \citet{French_2016} \\
ASASSN-14ae & 167.16716 & 34.097847 & 0.0436 & 9.8 & SDSS & [2] & 2.0 & \citet{Holoien_2014} \\
AT 2018bsi & 123.860919 & 45.592208 & 0.051 & 10.6 & [3] & [3] & 2.0 & [3], \citet{Dahiwale_2020} \\
AT 2018hco & 16.8901468 & 23.4761884 & 0.088 & 9.9 & [3] & [3] & 2.5 & [3], \citet{Reynolds_2018} \\
AT 2019qiz & 71.657851 & -10.2263679 & 0.0151 & 10.0 & [3] & [3] & 2.5 & [3], \citet{Siebert_2019} \\
F01004 & 15.7133333 & -22.3641667 & 0.1178 & 9.8 & \citet{French_2020} & [5] & 1.2 & \citet{Heikkila_2016, Sun_2024} \\
AT 2020nov & 254.554042 & 2.117511 & 0.0826 & 10.4 & \citet{Earl_2025} & MOST & n/a & \citet{Dahiwale_2020b} \\

\end{tabular}
}
\caption{TDE and associated host galaxy information}
\tablecomments{* indicates that the stellar mass was calculated from the 2MASS K-band magnitude. Galaxies with spectra from WISeREP or MOST did not have available slit dimensions. [1] \citet{Graur_2018} [2] MUSE (Thomas Wevers) [3] \citet{Hammerstein_2021} [4] \citet{Hammerstein_2023} [5] \citet{Tadhunter_2021}}
\label{table:finalsample}
\end{sidewaystable}

\begin{table}
\begin{tabular}{cDDDDcc}

TDE name & \multicolumn2c{H$\alpha$} & \multicolumn2c{H$\alpha$ uncertainty} & \multicolumn2c{H$\delta$} & \multicolumn2c{H$\delta$ uncertainty} & Broad lines & Galaxy label \\

 & \multicolumn2c{\AA} & \multicolumn2c{\AA} & \multicolumn2c{\AA} & \multicolumn2c{\AA} & & \\

\hline
\decimals
AT 2023clx & 3.37 & 0.35 & 0.52 & 0.39 & 1 & SF \\
AT 2022dyt & 4.36 & 0.31 & 1.22 & 0.66 & 1 & SF \\
AT 2022bdw & 9.41 & 0.33 & 1.61 & 0.56 & 1 & SF \\
AT 2021nwa & 0.64 & 0.19 & 0.55 & 1.0 & 1 & Quiescent \\
AT 2020wey & 0.08 & 0.15 & 2.92 & 0.69 & 1 & QBS \\
AT 2020vwl & 0.04 & 0.2 & 0.2 & 1.22 & 1 & Quiescent \\
AT 2020ohl & 0.68 & 0.06 & -0.91 & 0.27 & 0 & Quiescent \\
AT 2019gte & 0.88 & 0.12 & -2.21 & 0.67 & 0 & Quiescent \\
AT 2019azh & 1.73 & 0.08 & 7.67 & 0.17 & 1 & PSB \\
AT 2018mac & 0.23 & 0.04 & 2.15 & 0.53 & 1 & QBS \\
ASASSN-18zj & 0.35 & 0.14 & 5.43 & 0.41 & 1 & PSB \\
AT 2018dyk & 2.17 & 0.11 & -0.27 & 0.48 & 1 & Quiescent \\
ASASSN-14li & 0.6 & 0.5 & 5.7 & 0.6 & 1 & PSB \\
AT 2023mhs & 13.92 & 0.32 & 2.38 & 0.47 & 1 & SF \\
RBS 1032 & 0.5 & 0.4 & 4.1 & 0.4 & 0 & PSB \\
SDSS J0952 & 27.8 & 0.4 & -2.0 & 1.1 & 0 & SF \\
SDSS J1342 & 15.1 & 0.5 & -1.0 & 1.3 & 0 & SF \\
SDSS J0748 & 11.4 & 1.0 & 1.2 & 0.8 & 1 & SF \\
AT 2018dyb & 1.29 & 0.32 & -0.57 & 1.58 & 1 & Quiescent \\
AT 2021ehb & 0.76 & 0.07 & -1.95 & 0.92 & 0 & Quiescent \\
AT 2022gri & 0.64 & 0.08 & -1.12 & 0.91 & 0 & Quiescent \\
AT 2021yzv & 1.32 & 0.11 & -1.6 & 0.44 & 0 & Quiescent \\
ASASSN-15oi & -0.1 & 0.3 & 1.9 & 0.7 & 1 & QBS \\
RX J1242-A & -1.1 & 0.8 & 0.9 & 1.2 & 0 & Quiescent \\
RX J1624 & -0.6 & 1.3 & -1.1 & 2.1 & 0 & Quiescent \\
TDE2 & 4.5 & 0.5 & 3.7 & 0.6 & 1 & SF \\
XMM J0740 & 0.3 & 0.6 & 0.4 & 0.4 & 0 & Quiescent \\
PS1-10jh & 0.5 & 0.7 & 1.7 & 0.8 & 1 & QBS \\
PTF09axc & 1.1 & 0.7 & 4.9 & 0.4 & 1 & PSB \\
PTF09djl & 0.3 & 0.7 & 4.7 & 0.5 & 1 & PSB \\
AT 2018fyk & 1.26 & 0.07 & * & * & 1 & -- \\
AT 2019ahk & 27.97 & 0.8 & * & * & 0 & SF \\
iPTF16fnl & -0.8 & 0.6 & 5.8 & 0.3 & 1 & PSB \\
AT 2019dsg & 11.53 & 1.42 & * & * & 1 & SF \\
iPTF15af & 1.7 & 0.3 & 1.3 & 1.9 & 1 & QBS \\
ASASSN-14ae & 0.7 & 0.4 & 3.4 & 0.8 & 1 & QBS \\
AT 2018bsi & 5.25 & 0.69 & -0.73 & 0.47 & 1 & SF \\
AT 2018hco & 1.97 & 0.54 & 0.33 & 1.09 & 1 & Quiescent \\
AT 2019qiz & 0.02 & 0.92 & 0.87 & 1.11 & 1 & Quiescent \\
F01004 & 47.0 & 0.2 & -0.2 & 0.8 & 0 & SF \\
AT 2020nov & 16.45 & 0.63 & 3.58 & 0.63 & 1 & SF \\
\end{tabular}
\caption{The columns are the name of the TDE, the H$\alpha$ equivalent line width, the uncertainty on the H$\alpha$ equivalent line width, the H$\delta$ equivalent line width, the uncertainty on the H$\delta$ equivalent line width, whether or not broad lines appear in the spectrum, and the galaxy type label. A broad lines value of 1 indicates that H and/or He broad lines were observed in the spectrum, and includes TDEs classified as TDE-H, TDE-He, and TDE-H+He. A broad lines value of 0 does not necessarily mean that there are no broad lines, only that no broad lines were observed in the galaxy spectrum and identified in a published paper or AstroNote. An asterisk * in the H$\delta$ and H$\delta$ uncertainty columns indicates that those values could not calculated because the galaxy spectra did not extend to blue enough wavelengths.}
\label{table:broadlinestable}
\end{table}

\end{appendices}

\end{document}